\documentclass{SciPost}
\hypersetup{
    colorlinks,
    linkcolor={red!50!black},
    citecolor={blue!50!black},
    urlcolor={blue!80!black}
}
\usepackage[bitstream-charter]{mathdesign}
\usepackage{wrapfig}
\DeclareSymbolFont{usualmathcal}{OMS}{cmsy}{m}{n}
\DeclareSymbolFontAlphabet{\mathcal}{usualmathcal}
\fancypagestyle{SPstyle}{
\fancyhf{}
\lhead{\colorbox{scipostblue}{\bf \color{white} ~Prepared for SciPost Physics
}}
\rhead{{\bf \color{scipostdeepblue} 
}}

\fancyfoot[C]{\textbf{\thepage}}
}

\newcommand{\scri}{\mathcal{I}}
\newcommand{\sech}{\operatorname{sech}}

\newcommand{\bea}{\begin{eqnarray}}
\newcommand{\eea}{\end{eqnarray}}
\newcommand{\be}{\begin{equation}}
\newcommand{\ee}{\end{equation}}
\def\nn{\nonumber}

\begin{document}

\pagestyle{SPstyle}

\begin{center}{\Large \textbf{\color{scipostdeepblue}{
Conservation law of super-Lorentz charges across spatial infinity\\
}}}\end{center}

\begin{center}\textbf{
Geoffrey Comp\`ere\textsuperscript{$\star$} and
S\'ebastien Robert\textsuperscript{$\dagger$}
}\end{center}

\begin{center}
Universit\'e Libre de Bruxelles, International Solvay Institutes and Brussels Laboratory of the Universe (BLU-ULB), C.P. 231, B-1050 Bruxelles, Belgium%
\\[\baselineskip]
$\star$ \href{mailto:email1}{\small geoffrey.compere@ulb.be}\,,\quad
$\dagger$ \href{mailto:email2}{\small sebastien.robert@ulb.be}
\end{center}

\section*{\color{scipostdeepblue}{Abstract}}
\textbf{\boldmath{%
Under assumptions compatible with generic gravitational scattering, the vacuum relativistic gravitational field is entirely determined at leading order in the large radius expansion at spatial infinity by its supermomentum, its dual supermomentum and its global supertranslation frame. At subleading order, the gravitational field is determined by three additional sets of charges: the super-Lorentz charges, the leading mixed tail/ peeling-breaking  charges and the leading peeling-breaking charges. In this work we provide a proper-supertranslation-invariant definition of these charges in terms of asymptotic Bondi-Sachs fields as well as a corresponding proper supertranslation and logarithmic translation invariant definition of these charges in terms of  Beig-Schmidt fields. Using the properties of homogeneous and inhomogeneous solutions to relevant wave equations over the boundary de Sitter spacetime at spatial infinity, we derive the conservation law of super-Lorentz charges between the future and past of spatial infinity. We obtain that the super-Lorentz aspects are non-locally defined from the Bondi-Sachs fields. 
}}

\vspace{\baselineskip}



\vspace{10pt}
\noindent\rule{\textwidth}{1pt}
\tableofcontents
\noindent\rule{\textwidth}{1pt}
\vspace{10pt}


\section{Introduction}
\label{sec:intro}

Conservation laws are of fundamental importance in physics. Through Noether's theorems they are related to global symmetries and after quantization they lead to Ward identities. In asymptotically flat spacetimes, a special role is played by the $\text{Diff}(S^2)$ extension of the Lorentz group $SO(3,1)$ which is canonically associated to the super-Lorentz charges  \cite{Campiglia:2015yka,Compere:2018ylh,Flanagan:2019vbl,Campiglia:2020qvc,Compere:2020lrt}\footnote{In its initial formulation, the superrotations \cite{Barnich:2009se,Barnich:2010ojg,Barnich:2010eb,Barnich:2011mi} are defined as meromorphic functions with poles over the $2$-sphere. In this paper all quantities will be smooth over the $2$-sphere. The super-Lorentz transformations are split into superrotations and superboosts \cite{Compere:2018ylh}.}. The conservation of the total super-Lorentz flux between future and past null infinity leads to the subleading soft graviton theorem \cite{Cachazo:2014fwa,Saha:2019tub} as a Ward identity \cite{Kapec:2014opa,Campiglia:2014yka,Adamo:2014yya,Campiglia:2015kxa,Freidel:2021dfs,Agrawal:2023zea}. 

Several partial proofs of the conservation of super-Lorentz charges and their relationship to the subleading soft graviton theorem have appeared in the literature but all such proofs are unsatisfactory. For example, the framework set up in \cite{Freidel:2021dfs} enforces restrictive boundary conditions on the news tensor which are incompatible with the existence of tails (e.g. terms $u^{-1}$ in the shear as $u \to \pm \infty $), while they are generic in scattering \cite{Laddha:2018vbn}. A logarithmically divergent definition of superrotations has been conjectured to obey a conservation law at spatial infinity even though it is divergent \cite{Agrawal:2023zea}. A Hamiltonian generator has been derived for super-Lorentz charges at spatial infinity  \cite{Fiorucci:2024ndw} but it has not been related to fields at null infinity. Moreover, the boundary conditions used for the derivation in \cite{Fiorucci:2024ndw} do not allow for possible violations of peeling, which however generically occur in the presence of incoming/outgoing particles and radiation \cite{PhysRevD.19.3495,Damour:1985cm,ChristodoulouKlainerman+1994,Friedrich:2017cjg,Saha:2019tub,Kehrberger:2021uvf,Kehrberger:2021vhp,Kehrberger:2021azo,Masaood:2022bvi,Kehrberger:2024clh,Kehrberger:2024aak,Schneider:2025tek}. The main objective of this paper is to prove the conservation of super-Lorentz charges across spatial infinity with a definition that allows to read off the charges as a limit of codimension one fields  defined at null infinities and under a set of assumptions which is compatible with currently known properties of gravitational scattering and binary black hole mergers. 

\begin{wrapfigure}[]{r}{.33\textwidth}\vspace{-0.6cm}
\includegraphics[width=\linewidth]{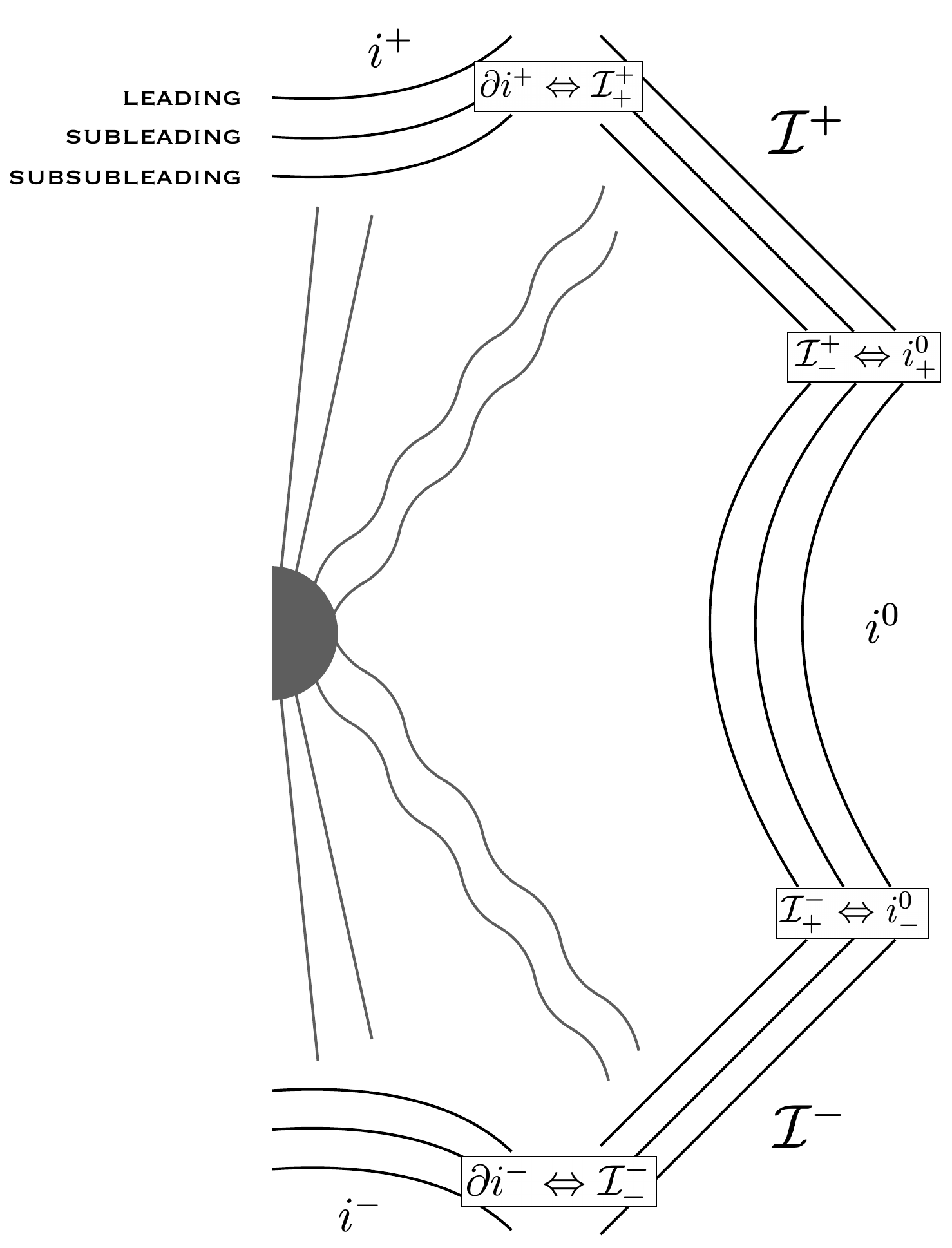}
\caption{Gravitational onion structure of asymptotically flat spacetimes.}
\label{fig:onion}\vspace{-0.6cm}
\end{wrapfigure}
Following the methodology of \cite{Compere:2023qoa,Boschetti:2026gfd,Compere:2026jmk} (see also earlier work \cite{1972JMP....13..956G,Ashtekar:1978zz,beig1982einstein,chrusciel1995gravitational,Friedrich:2002ru,Ashtekar:2008jw,Troessaert:2017jcm,Henneaux:2019yax,Prabhu:2021cgk,chakraborty2022supertranslations,Capone:2022gme,Solanki:2024oci}), asymptotically flat spacetimes are defined from asymptotic fall-off conditions at their five infinities: future and past null infinities $\mathcal I^\pm$, spatial infinity $i^0$ and future and past timelike infinities $i^\pm$, together with fall-off conditions on the codimension 1 fields defined at null infinities that ensure that a matching can be performed at the four corners between the boundaries. The leading, subleading and higher subleading behavior of the metric field can be defined from asymptotic expansions around each respective boundary. The matching conditions at the four corners mix the respective leading and subleading components of the two corresponding boundaries in a nested pattern that will be detailed in the article, as usual in matched asymptotic expansion schemes. This is the gravitational onion structure, see Figure \ref{fig:onion}. For alternative geometric approaches, see e.g. \cite{Ashtekar:2023wfn,Ashtekar:2023zul,Borthwick:2024skd,Marajh:2025yhg}.

Our larger aim is to identify all matching conditions of the leading and subleading fields at the corners $\mathcal I^+_-$ and $\mathcal I^-_+$, and deduce from the evolution of the codimension one fields at $i^0$ implied by Einstein's equations how the ``corner fields'' defined in the matching regions $\mathcal I^+_-$ and $\mathcal I^-_+$ are related to one another. The structure at leading order is well-understood: a single BMS frame can be defined from matching conditions \cite{Strominger:2013jfa,Compere:2023qoa}, and the asymptotic Bondi mass aspect is antipodally identified between $\mathcal I^+_-$ and $\mathcal I^-_+$, which leads to the leading soft graviton theorem \cite{He:2014laa}. The structure at subleading order has however only been partially understood so far. Two identities have been described \cite{Compere:2026jmk,Boschetti:2026gfd,Boschetti:2026ogm}, which are related to the conservation of tails and the leading law of violation of peeling that leads to the logarithmic correction to the classical soft graviton theorem \cite{Laddha:2018myi}. However, the identity at spatial infinity leading to the subleading soft graviton theorem has not been derived in this general setting. 

In this work, we derive the conservation law of the super-Lorentz charge across spatial infinity and also provide a comprehensive derivation of the results of our Letter \cite{Compere:2026jmk}. This provides a complete derivation of all conservation laws at spatial infinity at leading and subleading order in the large radius asymptotic expansion given our hypotheses. Our results bear consequences on the classical soft graviton theorems and on the conditions for violation of peeling which we will contrast with the literature \cite{1974SvA....18...17Z,Laddha:2018myi,Laddha:2018vbn,Sahoo:2018lxl,Saha:2019tub,Sahoo:2021ctw,PhysRevD.46.4304,1987Natur.327..123B}. Our hypotheses will be described in the main text and can be summarized as follows: 
\begin{enumerate}
    \item[H1] The metric field admits a polyhomogeneous expansion in Bondi-Sachs coordinates at $\scri^+$ and $\scri^-$ of the form \eqref{BondiSachsMetric}-\eqref{hABdef}-\eqref{BondihExp}. 
    \item[H2] The shear admits the expansion \eqref{CABasymptoticsscri} at $\mathcal I^+_-$ and at $\mathcal I^-_+$, which defines the asymptotic tails $C^{\pm (1)}_{AB}$ at spatial infinity. 
    \item[H3] The metric field admits a polyhomogeneous expansion in Beig-Schmidt coordinates at $i^0$ of the form \eqref{BSExpansion}. 
\end{enumerate}

Hypotheses 1 and 2 are compatible with known properties of the asymptotic gravitational field in the context of gravitational scattering \cite{PhysRevD.19.3495,Damour:1985cm,ChristodoulouKlainerman+1994,Cachazo:2014fwa,Friedrich:2017cjg,Laddha:2018myi,Laddha:2018vbn,Saha:2019tub,Satishchandran:2019pyc,Godazgar_2019,Kol_2019,Donnay:2020lur,Blanchet:2020ngx,Freidel:2021dfs,Kehrberger:2021uvf,Kehrberger:2021vhp,Kehrberger:2021azo,Masaood:2022bvi,Blanchet:2023pce ,Agrawal:2023zea,Choi:2024ajz,Geiller:2024ryw,Kehrberger:2024clh,Kehrberger:2024aak,Schneider:2025tek,1985FoPh...15..605W}. 
We advocate as in \cite{Compere:2026jmk} that Hypothesis 3 is a direct consequence of the two other assumptions by consistency with the matching conditions. If other fields are turned on in the Beig-Schmidt expansion, we expect that Einstein's equations will constrain them to obey free wave equations and since there is no field at the corners to match them given the Bondi-Sachs expansion, these fields are zero. The fields at spatial infinity that are considered constitute the minimal set required for the matching with null infinity.

In what follows, we will first set our boundary conditions at null infinity in Einstein's theory coupled to generic matter in Section \ref{sec:null} following \cite{Compere:2026jmk}, which generalizes earlier work  \cite{Flanagan:2015pxa,1985FoPh...15..605W,chrusciel1995gravitational,Geiller:2022vto,Geiller:2024ryw}. Note that we do not derive these boundary conditions from classes of initial data, see e.g. \cite{Kroon:1998tu,Friedrich:1998xmu,ValienteKroon:2002gb,Hintz:2017xxu,Kehrberger:2024clh,Fang:2024ivh,Marajh:2025yhg} for work in that direction. We expose the definition of the leading peeling-breaking charge and the leading mixed tail/peeling-breaking charge following the work of  \cite{Compere:2026jmk,Boschetti:2026ogm,Boschetti:2026gfd}. We then present our definition of super-Lorentz charges at $\mathcal I^+_-$. In Section \ref{sec:spatial} we derive the metric expansion at spatial infinity and deduce the antipodal relationships of the leading and subleading fields between $i^0_+$ and $i^0_-$ using the results on harmonics on $dS_3$ derived in \cite{Compere:2025bnf}.  We perform the matching between past and future null infinity and spatial infinity in Section \ref{sec:matching}. After rewriting the corner fields as conserved charges in Section \ref{sec:charges}, we discuss relevant features of gravity at timelike infinity in Section \ref{sec:timelike}.  Our results are compared with some earlier results on gravitational scattering  in Section \ref{sec:comp}. We conclude in Section \ref{sec:ccl}.

\vspace{10pt}
\noindent {\bf Note added} This version matches  with the published version, which corrects earlier errors and typos. It is consistent with version 4 on the arXiv of \cite{Compere:2026jmk} which also matches the published version of that preprint. We also inform the reader that version 3 on the arXiv of \cite{Compere:2025bnf} takes precedence over its Erratum correction as some further minor typos were corrected in the arXiv version 3.

\section{Null infinities $\mathcal I^+$ and $\mathcal I^-$}
\label{sec:null}
In this section, we start by defining the solution space we are considering at future null infinity $\mathcal I^+$ and past null infinity $\mathcal I^-$. This solution space was already presented in \cite{Compere:2026jmk}, generalizing earlier work \cite{Flanagan:2015pxa,1985FoPh...15..605W,chrusciel1995gravitational,Geiller:2022vto,Geiller:2024ryw}. Then, we define a complete set of asymptotic charges at the two  corners $\scri^+_-$, $\scri^-_+$, up to subleading order in the asymptotic expansion at large radius.

\subsection{Bondi-Sachs expansion}
\label{sec:2}

At future null infinity, we work in Bondi-Sachs gauge where the metric takes the form 
\cite{bondi1962gravitational,sachs1961gravitational,Sachs1962BMS,bondi1960gravitational,sachs1962gravitational}:
\begin{equation}
    ds^2=\frac{V e^{2\beta}}{r} du^2 - 2 e^{2\beta}du dr + r^2 h_{AB} (dx^A-U^Adu)(dx^B-U^Bdu),
\label{BondiSachsMetric}
\end{equation}
where
\begin{equation}
    h_{AB}:=\gamma_{AB}\sqrt{1+\frac{\mathcal C^{CD}\mathcal C_{CD}}{2r^2}}+ \frac{\mathcal C_{AB}}{r}.\label{hABdef}
\end{equation}
Here calligraphic $\mathcal C_{AB}(u,r,x^A)$ is the traceless part (with respect to the metric $\gamma_{AB}$) of $h_{AB}$. We fix the super-Lorentz frame by choosing  $\gamma_{AB}$ as the round sphere metric. We denote as $\nabla_A$ its Levi-Civita connection, $\epsilon_{AB}$ its volume form and $\int_{S^2}d^2\Omega=\int_{S^2}d^2 x \sqrt{\gamma} =4\pi$ its measure. It is possible to enlarge the phase space to include the action of $\text{Diff}(S^2)$ generators \cite{Campiglia:2015yka,Compere:2018ylh,Geiller:2024amx}, but this extension is not required to define the $\text{Diff}(S^2)$ charges so we shall not consider it here. 

The choice of Bondi-Sachs gauge fixes the logarithmic translation frame to outgoing radiation gauge frame \cite{Boschetti:2025tru}. 

In the large radius limit, we consider spacetimes that admit a polyhomogeneous expansion of the form \cite{1985FoPh...15..605W,chrusciel1995gravitational,Geiller:2022vto}\footnote{Here $V$ is only expanded one order beyond zeroth order. The second order coefficients won't be useful for any computations made in this paper.}
\begin{subequations}\label{BondihExp}
\begin{align}
    V(u,r,x^A)&:=-r+2m+{o}(r^{0})\label{BondiVExp},\\
    \beta (u,r,x^A)&:=\frac{\beta_1}{r}+\frac{\beta_2}{r^2}+o(r^{-2})\label{BondiBExp},\\
    U^A(u,r,x^A)&:=\frac{U_2^A}{r^2}+\frac{\log r \, U_{3,1}^{A}+U_3^A}{r^3}+o(r^{-3})\label{BondiUExp},\\
    \mathcal C_{AB}(u,r,x^A) &:= C_{AB}+\frac{1}{r}D_{AB}+o(r^{-1}),  
\end{align}    
\end{subequations}
where all coefficients in the $r$ expansion are $u-$dependent tensor fields over the sphere. By construction, calligraphic $\mathcal C_{AB}$, non-calligraphic $C_{AB}$ and $D_{AB}$ are traceless with respect to the spherical metric. In this super-Lorentz frame, the news tensor is $N_{AB}:= \partial_u C_{AB}$ \cite{Barnich:2009se,Compere:2018ylh,Campiglia:2020qvc,rignonbret2024centerlessbmschargealgebra}. This expansion was recently studied in \cite{Geiller:2024ryw} for the vacuum Einstein equations and in \cite{Compere:2026jmk} for non-vacuum Einstein equations. The most generic stress-energy tensors compatible with the expansion \eqref{BondihExp} admit an expansion of the form :
\begin{align}
    T_{uu}=& \frac{T^{(0)}_{uu}}{r^2}+\frac{T^{(1)}_{uu}}{r^3}+o(r^{-3}),\label{Tuu}\\
    T_{ur}=& \frac{T^{(0)}_{ur}}{r^3}+o(r^{-3}),\\
     T_{rr}=&\frac{T^{(0)}_{rr}}{r^3}+\frac{T^{(1)}_{rr}}{r^4}+o(r^{-4}),\\
    T_{uA}=&\frac{T^{(0)}_{uA}}{r}+\frac{T^{(1)}_{uA}+T^{(1,log)}_{uA}\log r}{r^2}+o(r^{-2}),\\
     T_{rA}=&\frac{T^{(0)}_{rA}}{r^2}+\frac{T^{(1)}_{rA}}{r^3}+o(r^{-3}),\\     T_{AB}=&T^{(0)}\gamma_{AB}+\frac{T^{(1)}_{\langle AB\rangle}+T^{(1)}\gamma_{AB}}{r}+o(r^{-1})\label{TAB},
\end{align}
where all coefficients are $u$-dependent tensor fields over the sphere. The divergence-free condition with respect to the four-dimensional metric imposes
\begin{align}
    &T_{ur}^{(0)} = - \nabla^A T^{(0)}_{rA}-\frac{1}{2} \nabla^2 T^{(0)}_{rr},\\
    &T^{(0)}= -\frac{1}{2} \partial_uT_{rr}^{(0)},\\
    &T^{(0)}_{uA}=-\partial_u T^{(0)}_{rA}-\frac{1}{2}  \nabla_A \partial_uT^{(0)}_{rr},\\
    &T^{(1)}= -\frac{1}{4} (\nabla^2+2)T_{rr}^{(0)}+ T_{rr}^{(0)}\partial_u\beta_1-\frac{1}{2}\partial_uT_{rr}^{(1)} ,\\
    &T^{(1,log)}_{uA}= \frac{1}{2} \nabla^B (\partial_u T_{rr}^{(0)}C_{AB})-\frac{1}{4} \nabla_A (\nabla^2 +2)T_{rr}^{(0)}+\nabla^B T^{(1)}_{\langle A B\rangle } -\partial_u T_{rA}^{(1)}-\frac{1}{2}\nabla_A \partial_u T_{rr}^{(1)}.
\end{align}

Then, Einstein's equations imply the following algebraic identities \cite{Geiller:2024bgf,Flanagan:2015pxa,Madler_2016,Geiller:2024ryw,Compere:2026jmk}: 
\begin{subequations}
\begin{align}
    \beta_1&=-2 \pi T_{rr}^{(0)},\\
    \beta_2&= -\frac{1}{32}C_{AB}C^{AB}-\pi T_{rr}^{(1)},\\
    U_2^A&= -\frac{1}{2} \nabla_B C^{AB}-6 \pi \nabla^A  T_{rr}^{(0)}-8\pi T_{rB}^{(0)}\gamma^{AB},\\
    U_{3,1}^A &=-\frac{2}{3}\nabla_B D^{AB}-\frac{8\pi}{3}\nabla^AT_{rr}^{(1)} -\frac{16\pi}{3}T_{rB}^{(1)}\gamma^{AB},
\end{align}    
\end{subequations}
and the following flux-balance laws: 
\begin{align}
      \partial_u m &= -\frac{1}{8} N_{AB}N^{AB} + \frac{1}{4} \nabla_A \nabla_B N^{AB}+3 \pi (\nabla^2+\frac{2}{3})\partial_uT_{rr}^{(0)}+4\pi \nabla^A \partial_uT_{rA}^{(0)}-4 \pi T_{uu}^{(0)}\label{FluxBalancedLawsm},\\
      \partial_u \mathcal{P}_A &=\nabla^B \mathcal{M}_{AB} + \frac{1}{2} C_{A}^B \nabla^C N_{BC}-48\pi^2 \partial_uT_{rr}^{(0)}\nabla_AT_{rr}^{(0)}-24 \pi^2 T_{rr}^{(0)}\partial_u\nabla_AT_{rr}^{(0)}-64 \pi^2 \partial_u T^{(0)}_{rr} T^{(0)}_{rA} \nn\\&-32\pi^2 T^{(0)}_{rr}\partial_u  T^{(0)}_{rA}-4\pi C_{AB}\partial_u\nabla^B T_{rr}^{(0)}-6 \pi N_{AB}\nabla^B T_{rr}^{(0)}-2 \pi \partial_u(T_{rr}^{(0)} \nabla^B  C_{AB}) \nn\\& 
      - 8\pi \partial_u (C_{A}^B T_{rB}^{(0)})-\pi \nabla_A(\nabla^2+2)T_{rr}^{(0)}+4\pi \nabla_A \nabla^B T_{rB}^{(0)}-4\pi (\nabla^2-1) T_{rA}^{(0)}\nn\\&+\frac{8}{3} \pi \partial_u T_{rA}^{(1)}-8\pi T_{uA}^{(1)}+\frac{7}{3}\pi \partial_u \nabla_A T_{rr}^{(1)} +4\pi \nabla^B T_{\langle A B \rangle}^{(1)}\label{FluxBalancedLawsN},\\
      \partial_ u D_{AB} &=-8\pi T_{\langle AB \rangle }^{(1)}-4\pi C_{AB} \partial_u T^{(0)}_{rr}+4\pi \nabla_{\langle A}\nabla_{B\rangle}T^{(0)}_{rr}\label{FluxBalancedLawsD},
\end{align}
where $\mathcal{M}_{AB}$ is the covariant mass tensor and $\mathcal P_A$ the covariant Bondi angular momentum aspect defined in \cite{Geiller:2024ryw}:
\begin{align}
    \mathcal{M}_{AB}&:=m \gamma_{AB}+\frac{1}{16} \partial_u \left(C_{CD}C^{CD}\right)\gamma_{AB} + \frac{1}{2} \nabla_{[A} \nabla^C C_{B]C} - \frac{1}{8} \epsilon^{CD} C_{DE} N^{E}_C \epsilon_{AB}\nn\\&=\mathcal{M}\gamma_{AB}+\tilde{\mathcal{M}}\epsilon_{AB},\label{defMAB}\\
    \mathcal P_A &:=\frac{3}{32} \partial_A\left(C_{B C} C^{B C}\right)-\frac{4}{3}\nabla^BD_{AB}-\frac{3}{2}\Big(U_{3\,A}-\frac{1}{2} C_{AB} \nabla_C C^{B C}\Big).
\end{align}
Here $\mathcal{M}:=m+\frac{1}{16} \partial_u \left(C_{CD}C^{CD}\right)$ is the covariant mass aspect and 
\begin{align}
\tilde{\mathcal{M}}:=\frac{1}{4}\epsilon^{CD} \nabla_C \nabla^E C_{DE}- \frac{1}{8} \epsilon^{CD} C_{DE} N^{EC}    
\end{align}
is the covariant dual mass aspect. The set $\{ C_{AB}, D_{AB}, m ,\mathcal P_A \}$ of the shear $C_{AB}$, the $D_{AB}$ field, the mass aspect $m$ and the angular momentum aspect $\mathcal P_A$ is the complete set of Bondi-Sachs fields up to subleading order. 

The Weyl scalars associated to the expansion \eqref{BondihExp} were derived in \cite{Geiller:2024ryw} using vacuum Einstein's equations. We now extend this computation by including the stress-energy tensor \eqref{Tuu}-\eqref{TAB}. 
We define the outgoing Bondi tetrad as\footnote{ Our conventions differ from \cite{Geiller:2024ryw}: their $m^A$ is our $\bar m^A$, their $\ell^\mu$ is our $\sqrt{2}\ell^\mu$ and their $n^\mu$ is our $\frac{1}{\sqrt{2}}n^\mu$.} 
\begin{equation}\label{tetrad}
\ell^\mu\partial_\mu:=\frac{1}{\sqrt{2}}\partial_r,\;\; n^\mu\partial_\mu:=\sqrt{2}\,e^{-2\beta}\Big(\partial_u+\frac{V}{2r}\partial_r+U^A\partial_A\Big),\;\; m^A \partial_A:=\frac{1}{r}\sqrt{\frac{h_{\theta\theta}}{2 h}}\big(\frac{\sqrt{h}-i\,h_{\theta \phi}}{h_{\theta\theta}}\, \partial_\theta + i\,\partial_\phi \big).
\end{equation}
These are defined such that $\ell_\mu n^\mu=-1$, $m_\mu\bar m^\mu=1 $ and all other contractions are vanishing. Note that  $\ell_\mu dx^\mu = -\frac{1}{\sqrt{2}}e^{2\beta}du$. We denote $m=r^{-1}m_1^A(x^B) \partial_A + r^{-2} m_2^A(u,x^B) \partial_A+o(r^{-2})$ with 
\begin{equation}
    m_1^A \partial_A:=\frac{1}{\sqrt{2}}\big( \partial_\theta + \frac{i}{\sin\theta}\partial_\phi \big),\qquad m_2^A=\bar m_1^A\sigma_2+2m_1^A \epsilon_2
\end{equation}
and $\sigma_2=-\frac{1}{2}C_{AB}m_1^A m_1^B$, $\epsilon_2=\frac{1}{4}(\sigma_2-\bar \sigma_2)$. 

The Weyl scalars are given by 
\begin{align}
 &\Psi_0\vert_{\scri^+}:=-W_{\ell m\ell m}\vert_{\scri^+} = \frac{1}{2 r^4}(D_{AB}+2\pi T^{(0)}_{rr}C_{AB}) m^A_1 m^B_1 +o(\frac{1}{r^4}),\label{Weyl1}\\
& \Psi_1\vert_{\scri^+}:=-W_{\ell n\ell m}\vert_{\scri^+}=\frac{1}{\sqrt{2}}\frac{1}{r^3}4\pi (\nabla_AT_{rr}^{(0)}+T_{rA}^{(0)})m^A_1\nn\\&\qquad\qquad\qquad\qquad+\frac{1}{\sqrt{2}}\frac{\log r}{r^4}(\nabla^BD_{AB}+8\pi T_{rA}^{(1)}+4\pi \nabla_A T_{rr}^{(1)})m^A_1\nn\\&\qquad\qquad\qquad\qquad+\frac{1}{\sqrt{2}}\frac{1}{r^4}\Big(\mathcal{P}_A+32\pi^2T_{rr}^{(0)}T_{rA}^{(0)}+24\pi^2T_{rr}^{(0)}\nabla_A T_{rr}^{(0)}+2\pi T_{rr}^{(0)}\nabla^BC_{AB}\nn\\&\qquad\qquad\qquad\qquad-\frac{20}{3}\pi T_{rA}^{(1)}-\frac{7}{3}\pi\nabla_A T_{rr}^{(1)}+8 \pi C_{A}^{B}\nabla_BT_{rr}^{(0)}+12 \pi C_{A}^{B}T_{rB}^{(0)}\Big)m^A_1\nn\\&\qquad\qquad\qquad\qquad +\frac{1}{\sqrt{2}}\frac{1}{r^4} 4\pi (\nabla_AT_{rr}^{(0)}+T_{rA}^{(0)})m^A_2+o(\frac{1}{r^4}),\\
&\Psi_2\vert_{\scri^+}:=-W_{\ell m\bar m n}\vert_{\scri^+}=\frac{1}{r^2}\frac{2\pi}{3}\partial_u T^{(0)}_{rr}\nn\\&\qquad\qquad\qquad\qquad+\frac{1}{r^3}\Big(\mathcal{M}+\frac{4\pi}{3} (\frac{1}{2}\partial_{u}T^{(1)}_{rr}-\nabla^A T^{(0)}_{rA}-\nabla^2 T^{(0)}_{rr}+2 \pi T^{(0)}_{rr}\partial_uT^{(0)}_{rr})\nn\\&\qquad\qquad\qquad\qquad+i(\tilde{\mathcal{ M}}+4\pi \epsilon^{AB}\nabla_A T^{(0)}_{rB})\Big)+o(\frac{1}{r^3}),\\
&\Psi_3\vert_{\scri^+}:=-W_{n\bar m n\ell}\vert_{\scri^+}=\sqrt{2}\frac{1}{r^2}(\frac{1}{2}\nabla^B N_{AB}+4\pi \partial_u T^{(0)}_{rA}+2\pi \partial_u \nabla_AT^{(0)}_{rr})\bar m^A_1+o(\frac{1}{r^2}),\\
&\Psi_4\vert_{\scri^+} :=-W_{n\bar m n \bar m}\vert_{\scri^+}= \frac{1}{r}\partial_u N_{AB}\bar m^A_1 \bar m^B_1 +o(\frac{1}{r}).
\end{align}
All expansions of the metric field and asymptotic fields at $\mathcal I^-$ are formally identical to the expansions at $\mathcal I^+$ with $u$ replaced by $-v$. The angular coordinates are identical between $\mathcal I^+$ and $\mathcal I^-$ at leading order in the asymptotic expansion (they are not related by the antipodal map). 
We introduce the notation \(u^+ := u\) and \(u^- := -v\) so that expansions at \(\scri^+\) and \(\scri^-\) can be written in a unified manner. To distinguish the coefficients in the expansion \eqref{BondihExp} at \(\scri^+\) and \(\scri^-\), we assign an additional superscript \(+\) to all quantities at \(\scri^+\), for instance \(C^+_{AB} := C_{AB}\), while quantities at \(\scri^-\) carry a superscript \(-\). In particular, the news tensor at \(\scri^-\) is defined by \(N^-_{AB} := -\partial_v C_{AB}\).

We denote the limit \(u \to -\infty\) along \(\scri^+\) by \(\scri^+_-\), and the limit \(v \to +\infty\) along \(\scri^-\) by \(\scri^-_+\). The points \(\scri^+_-\) and \(\scri^-_+\) will be referred to as the ``corners'' of spacetime at spatial infinity, corresponding to the limit \(u^\pm \to -\infty\).

At $\mathcal I^-$, the outgoing Bondi tetrad can be expressed in incoming Bondi coordinates as 
\begin{align}
\ell^\mu \partial_\mu = - (n^\mu\partial_\mu )\vert_{u \mapsto -v},\qquad n^\mu \partial_\mu = -(\ell^\mu \partial_\mu)\vert_{u \mapsto -v}    \label{democratic}
\end{align}
where we can formally take the vectors \eqref{tetrad} by replacing $u$ with $-v$. With our choices of factors of $\sqrt{2}$ in the definition of $n^\mu$ and $\ell^\mu$, the factor 2 originating from the change of coordinates at leading order $v=u+2r$ is exactly cancelled and the ``democratic'' relations \eqref{democratic} between $\mathcal I^+$ and $\mathcal I^-$ are obeyed. The angular tetrad is unchanged, $m^A\partial_A = (m^A\partial_A)\vert_{u \mapsto -v}$ as the coordinates on the sphere are identical between $\scri^+$ and $\scri^-$. The roles of $\Psi_k$ and $\Psi_{4-k}$, $k=0,...,4$ become inverted with some minus signs. We therefore have
 \begin{align}
 &\Psi_0\vert_{\scri^-}:=-W_{\ell m\ell m}\vert_{\scri^-} =-\frac{1}{r}\partial_v N_{AB}^- m^A_1  m^B_1 +o(\frac{1}{r}) \\
& \Psi_1\vert_{\scri^-}:=-W_{\ell n\ell m}\vert_{\scri^-}= -\frac{\sqrt{2}}{r^2}(\frac{1}{2}\nabla^BN_{AB}^--4\pi  \partial_v T^{-(0)}_{rA}-2\pi \partial_v \nabla_AT^{-(0)}_{rr}) m^A_1+o(\frac{1}{r^2}),\\
&\Psi_2\vert_{\scri^-}:=-W_{\ell m\bar m n}\vert_{\scri^-}=-\frac{1}{r^2}\frac{2\pi}{3}\partial_v T^{-(0)}_{rr}\nn\\&\qquad\qquad\qquad\qquad+\frac{1}{r^3}\Big(\mathcal{M}^--\frac{4\pi}{3} (\frac{1}{2}\partial_{v}T^{-(1)}_{rr}+\nabla^A T^{-(0)}_{rA}+\nabla^2 T^{-(0)}_{rr}+2 \pi T^{-(0)}_{rr}\partial_vT^{-(0)}_{rr})\nn\\&\qquad\qquad\qquad\qquad-i(\tilde{\mathcal{ M}}^-+4\pi \epsilon^{AB}\nabla_A T^{-(0)}_{rB})\Big)+o(\frac{1}{r^3}),\\
&\Psi_3\vert_{\scri^-}:=-W_{n\bar m n\ell}\vert_{\scri^-}=
-\frac{1}{\sqrt{2}}\frac{1}{r^3}4\pi (\nabla_AT_{rr}^{-(0)}+T_{rA}^{-(0)})\bar m^A_1\nn\\&\qquad\qquad\qquad\qquad-\frac{1}{\sqrt{2}}\frac{\log r}{r^4}(\nabla^BD^-_{AB}+8\pi T_{rA}^{-(1)}+4\pi \nabla_A T_{rr}^{-(1)})\bar m^A_1\nn\\&\qquad\qquad\qquad\qquad-\frac{1}{\sqrt{2}}\frac{1}{r^4}\Big(\mathcal{P}_A^-+32\pi^2T_{rr}^{-(0)}T_{rA}^{-(0)}+24\pi^2T_{rr}^{-(0)}\nabla_A T_{rr}^{-(0)}+2\pi T_{rr}^{-(0)}\nabla^BC_{AB}^-\nn\\&\qquad\qquad\qquad\qquad-\frac{20}{3}\pi T_{rA}^{-(1)}-\frac{7}{3}\pi\nabla_A T_{rr}^{-(1)}+8 \pi C_{A}^{-\,B}\nabla_BT_{rr}^{-(0)}+12 \pi C_{A}^{-\,B}T_{rB}^{-(0)}\Big)\bar m^A_1\nn\\&\qquad\qquad\qquad\qquad -\frac{1}{\sqrt{2}}\frac{1}{r^4} 4\pi (\nabla_AT_{rr}^{-(0)}+T_{rA}^{-(0)})\bar m^A_2+o(\frac{1}{r^4}),\\
&\Psi_4\vert_{\scri^-} :=-W_{n\bar m n \bar m}\vert_{\scri^-}=\frac{1}{r^4}\frac{1}{2}(D_{AB}^-+2\pi T^{-(0)}_{rr}C_{AB}^-) \bar m^A_1 \bar m^B_1 +o(\frac{1}{r^4}).\label{Weyl2}
\end{align}

\subsection{Hypotheses in the approach to spatial infinity}
\label{sec:hypo}

In this paper, we expand the shear and all Bondi aspects around spatial infinity as a polyhomogeneous expansion in $-u^\pm \to + \infty$. The choice of expanding in $-u^\pm$ instead of in $u^\pm$ is motivated by the idea to expand fields at spatial infinity in terms of a quantity reaching $+\infty$. We now assume the following fall-off conditions on the Bondi shear as $-u^\pm\to +\infty$:
\begin{align}
    C_{AB}^\pm(u^\pm ,x^A) & = C_{AB}^{\pm (0)}(x^A)-\frac{C_{AB}^{\pm (1)}(x^A)}{u^\pm}+o(\frac{1}{u^\pm})\label{CABasymptoticsscri},
\end{align}
It is convenient to decompose the asymptotic shear $ C_{AB}^{\pm (0)}(x^A)$ using two scalars: 
\begin{equation}
   C_{AB}^{\pm (0)}(x^A)=\left(-2 \nabla_A \nabla_B + \gamma_{AB} \nabla^2\right)C^{(0)\pm}(x^A)+\epsilon_{C(A}\nabla_{B)}\nabla^C \Psi^{(0)\pm}(x^A)\label{CABODecomp} .
\end{equation}
The scalar $C^{(0)\pm}$ encodes the electric parity piece $C_{AB}^{\pm(E)(0)}:=\left(-2 \nabla_A \nabla_B + \gamma_{AB} \nabla^2\right)C^{(0)\pm}$ and $\Psi^{(0)\pm}$ encodes the magnetic parity piece $C_{AB}^{\pm(B)(0)}:=\epsilon_{C(A}\nabla_{B)}\nabla^C \Psi^{(0)\pm}$. Here and throughout this article we use the unit weighted symmetrization $v_{(A}w_{B)}=\frac{1}{2}(v_A w_B + v_B w_A)$. Because the differential operator acting on $C^{(0)\pm}$ and $\Psi^{(0)\pm}$ annihilates $\ell\le1$ harmonics, we will take $C^{(0)\pm}$ and $\Psi^{(0)\pm}$ to have no $\ell=0,1$ harmonics. The decomposition \eqref{CABODecomp} is valid for any symmetric traceless tensor over the sphere.  

Differentiating with respect to $u^\pm$ and assuming that the asymptotic expansion is stable under differentiation we obtain the asymptotic behavior of the news at spatial infinity: 
\begin{align}
    N_{AB}^\pm(u^\pm ,x^A) & = \frac{C_{AB}^{\pm (1)}(x^A)}{(u^\pm)^2}+o(\frac{1}{(u^\pm)^2})\label{NABasymptoticsscri},
\end{align}
The aspect $C_{AB}^{\pm (1)}(x^A)$ is therefore the leading tail defined at spatial infinity. 

The ansatz \eqref{CABasymptoticsscri} is compatible with generic gravitational scattering \cite{Saha:2019tub} where $C_{AB}^{\pm (B)(0)}$ vanishes. Following \cite{Compere:2026jmk}, we consider a more general case where this magnetic piece might not vanish, motivated by the solutions provided in \cite{Satishchandran:2019pyc} where this quantity is non-vanishing. For simplicity, we assume that the stress-energy tensors on $\scri^\pm$ are compactly supported. While still including massless particles, gravitational waves and massive fields, this excludes massless matter fields. We could relax this assumption by assuming suitably defined boundary conditions for the matter stress-energy tensor in the limit to the corners but this analysis is not presented here.

Given these assumptions, the flux balance laws 
\eqref{FluxBalancedLawsm}-\eqref{FluxBalancedLawsD} imply the following relations: 
\begin{subequations}\label{falloffIplus}
\begin{align}
D^\pm_{AB}(u^\pm,x^A)&= D_{AB}^{(0)\pm}(x^A), \\
m^\pm(u^\pm,x^A) & = m^{\pm (0)}(x^A) -\frac{m^{\pm (1)}(x^A)}{u^\pm}+o\big( \frac{1}{u^\pm}\big)\label{mapectasymptotics}, \\
\mathcal{P}^\pm_A(u^\pm,x^A) & =- N_A^{\pm(-1)}(x^A) u^\pm+N_A^{\pm (\text{log})}(x^A)\log(-u^\pm)+ N_A^{\pm (0)}(x^A)+o(1),
\end{align}
\end{subequations}
where 
\begin{align}
m^{\pm (1)} &:=\frac{1}{4} \nabla^A \nabla^B C^{\pm (1)}_{AB},\label{EvolutionConstrainm1}\\
N_A^{\pm (-1)}&:= -\nabla^B \mathcal M_{AB}^{(0)}=-\nabla_A m^{(0)}-\epsilon_{AB}\nabla^B \widetilde{\mathcal M}^{(0)}, \\ 
    N_A^{\pm (\text{log})} &:=-\frac{1}{2} \nabla_C \nabla_{\langle A} \nabla_{B\rangle} C^{\pm (1)\;BC}\label{NALog}. 
\end{align}
The dual mass aspect is expanded as 
\begin{align}\label{exptildeM}
 \tilde {\mathcal M}^\pm(u^\pm,x^A) = \tilde {\mathcal M}^{\pm(0)}(x^A) + O\big( \frac{1}{u^\pm}\big) 
\end{align}
where $\tilde {\mathcal M}^{\pm(0)} :=\frac{1}{4}\epsilon^{CD} \nabla_C \nabla^E C_{DE}^{\pm(0)}=\frac{1}{4}\epsilon^{CD} \nabla_C \nabla^E C_{DE}^{\pm(B)(0)}$. 
A useful identity is
\begin{equation}
    \nabla_C \nabla_{\langle A} \nabla_{B\rangle} C^{\pm (1)\;BC}=\frac{1}{2} \Big(\nabla_A \nabla^B\nabla^C C^{\pm (1)}_{BC}+\epsilon_{AB}\nabla^B \epsilon^{DE}\nabla_D\nabla^C C^{\pm (1)}_{CE}\Big).\label{propC}
\end{equation}
Note that one convention of this paper differs from \cite{Compere:2023qoa}: the global sign of $m^{\pm(1)}$ and $C^{\pm (1)}_{AB}$ is opposite to the one defined in \cite{Compere:2023qoa}.

\subsection{BMS and generalized BMS frame}

Infinitesimal diffeomorphisms that preserve the Bondi-Sachs gauge \eqref{BondiSachsMetric}-\eqref{hABdef} obey the conditions
\begin{equation}
    \mathcal{L}_{\xi} g_{rr} = 0,\quad  \mathcal{L}_{\xi} g_{rA} = 0,\quad  \mathcal{L}_{\xi}\partial_r \det(g_{AB}/r^2)=0.
\label{gaugeInvariant}
\end{equation}
The general solution can be expressed as 
\begin{equation}
\begin{aligned}
\xi_{T, Y}= & {\left[T(x^C)+\frac{u}{2} \nabla_A Y^A(x^C)+o\left(r^0\right)\right] \partial_u }  +\left[Y^A(x^C)-\frac{1}{r} \nabla^A T(x^C)-\frac{u}{2r}\nabla^A\nabla^BY_B+o\left(r^{-1}\right)\right] \partial_A \\
& +\left[-\frac{r+u}{2} \nabla_A Y^A(x^C)+\frac{1}{2} \nabla^2 T(x^C)+o\left(r^0\right)\right] \partial_r
\end{aligned}
\end{equation}
where $T(x^A)$ is an arbitrary function and $Y^A(x^A)$ is a vector that obeys the conformal Killing equation over the round sphere. Indeed, under a gauge transformation of parameter $Y^A$, the metric over the sphere transforms as 
\begin{equation}
    \delta_Y (\gamma_{AB})=\nabla_A Y_B + \nabla_B Y_A - \gamma_{AB} \nabla_C Y^C. 
\end{equation}
We distinguish the standard BMS and the generalized BMS algebra. The standard BMS algebra is defined from generators $\xi_{T,Y}$ where $Y^A$  obeys $\delta_Y (\gamma_{AB})=0$. The generators $\xi_{T=0,Y}$ then form the Lorentz algebra under the Lie bracket. The generalized BMS algebra is generated by arbitrary smooth vectors $Y^A(x^B)$ and scalars $T(x^A)$. The $\ell=0,1$ harmonics of $T$ are the four translations $T^{(\mu)}$, $\mu=0,1,2,3$ and the $\ell \geq 2$ harmonics of $T$ are the proper supertranslations. 

We can decompose an arbitrary (co-)vector $Y_A(x^A)$ in vector harmonics on $S^2$ as 
\begin{equation}
    Y_A(x^C) = \sum_{\ell \ge 1}\sum_{m=-\ell}^\ell\left( k_{\ell m} \mathring V^{(E)\ell m}_A(x^C) + j_{\ell m} \mathring V^{(B)\ell m}_A(x^C)\right) 
\label{YADecomposition},
\end{equation}
where the vector harmonics are built from scalar spherical harmonics on the sphere $Y_{\ell m}$ as follows 
\begin{align}
    \mathring V^{(B)\ell m}_A(x^C)=\frac{1}{\sqrt{\ell(\ell+1)}} \epsilon_{AB} \nabla^B Y_{\ell m}(x^C),\label{VBHarmonics}\\
    \mathring V^{(E)\ell m}_A(x^C)=\frac{1}{\sqrt{\ell(\ell+1)}} \nabla_A  Y_{\ell m}(x^C).\label{VEHarmonics}
\end{align}

In particular, when $\ell = 1$, the parameters $j_{1m}$ label rotations (which correspond to Killing vectors on the sphere) and parameters $k_{1m}$ label boosts (which correspond to proper conformal Killing vectors on the sphere). The higher harmonics $\ell \ge 2$ correspond to superrotations and superboosts, which we call super-Lorentz transformations. Here, we fixed the super-Lorentz frame by requiring $\gamma_{AB}$ to be the round-sphere metric. 

Under parity $\Upsilon(\theta,\phi)=(\pi-\theta,\phi+\pi)$, the vector harmonics transform as 
\begin{align}
    \Upsilon^* \mathring V^{(B)\ell m}_A(x^C)  =  (-1)^{\ell +1} \mathring V^{(B)\ell m}_A(x^C)
\label{BParity} , \\
    \Upsilon^* \mathring V^{(E)\ell m}_A(x^C) =  (-1)^{\ell } \mathring V^{(E)\ell m}_A(x^C).\label{Eparity}
\end{align}
We will refer as magnetic-parity (B-parity) vector, vectors that are linear combinations of $\mathring V^{(B)\ell m}_A$ and electric-parity (E-parity) vectors, vectors that are linear combinations of $\mathring V^{(E)\ell m}_A$. 

Fields vary non-trivially under BMS transformations. Let us display the transformation laws of the asymptotic fields in the absence of matter only. Under an arbitrary supertranslation $T(x^A)$, the fields in the Bondi-Sachs expansion transform as (see \cite{Geiller:2024ryw}):
\begin{align}
    \delta_T C_{AB} &= - 2 \nabla_{A} \nabla_B T + \gamma_{AB} \nabla^2 T + T N_{AB},\label{detlaCAB}\\
   \delta_T m &= \frac{1}{4} \bigg( N_{AB} \nabla^A \nabla^B T + 2 \nabla_C N^{CB} \nabla_B T \bigg) + T \partial_u m ,\label{deltam}\\
    \delta_T D_{AB} &=T \partial_u D_{AB},\\
    \delta_T \mathcal{P}_A &= 3  \mathcal{M}_{AB}\nabla^B T + T \partial_u \mathcal{P}_A.\label{deltaPA}
\end{align}
We do not detail the action of Lorentz transformations on the fields here.

At the corner $\mathcal I^+_-$, under the hypotheses provided in Sec. \ref{sec:hypo}, the boundary fields transform as 
\begin{align}
    \delta_T C^{+(0)}&=\Pi_{\ell \geq 2}T\label{detlaC0}\\
    \delta_T N_A^{+(0)} &= 3  \mathcal{M}_{AB}^{(0)}\nabla^B T + T \nabla^B \mathcal{M}_{AB}^{(0)}, \label{deltaNA}
\end{align}
with the other variations vanishing: $\delta_T m^{+(0)}= 0$, $\delta_T C^{+(B)(0)}_{AB} = \delta_T D_{AB}^{+(0)} =\delta_T C^{+(1)}_{AB}=0$. Here $\Pi_{\ell \geq 2}$ is a projector onto $\ell \geq 2$ harmonics since $C^{+(0)}$ is defined without $\ell=0,1$ harmonics.

\subsection{Asymptotic set of charges at subleading order}
\label{sec:dressedBondi}

We now aim to define a complete list of charges in the limit $u^\pm\to-\infty$ that are defined as integrals of asymptotic Bondi-Sachs fields over the sphere.
We see from the expansions \eqref{CABasymptoticsscri}-\eqref{exptildeM} that there are three independent asymptotic quantities at leading order: $C^{\pm(0)}$, $m^{\pm(0)}$ and $\tilde {\mathcal M}^{\pm(0)}$ and three independent asymptotic quantities at subleading order : $N_A^{\pm(0)}$, $C_{AB}^{\pm(1)}$ and $D_{AB}^{\pm(0)}$. Each field defined over the sphere at spatial infinity defines a charge by integration over the sphere with a test rank $r=0,1,2$ tensor (scalar/vector/tensor) that corresponds to the tensorial nature of the field over the sphere. We will therefore refer to fields at spatial infinity and charges at spatial infinity on an equal footing. 

The leading order charge associated with $m^{\pm (0)}$ is well documented in the literature (see e.g. \cite{Strominger:2013jfa,Troessaert:2017jcm,Prabhu:2019fsp,Magdy:2021rmi,Compere:2023qoa}): its $\ell=0,1$ harmonics define the total spacetime momenta $\mathcal P^\mu := (E,P_i)$ and its $\ell \geq 2$ harmonics are the proper supermomenta at $\scri^+_-/\scri^-_+$. We refer to the charge associated with $m^{\pm (0)}$ as the supermomenta.
The charge associated with $C^{\pm (0)}$ determines the supertranslation frame at $\scri^+_-/\scri^-_+$ \cite{Strominger:2013jfa}. It allows to build proper supertranslation invariant quantities at the corners (see below). The dual supertranslation charges associated with $\tilde{\mathcal M}^{\pm (0)}$ are defined from the expansion of the dual mass aspect \eqref{exptildeM} \cite{Kol_2019,Godazgar_2019N}. The $\ell \geq 2$ harmonics can be non-vanishing in physically motivated scenarios \cite{Satishchandran:2019pyc} while the $\ell =0,1$ modes correspond to NUT charges \cite{1981JMP....22.2612R,1982JMP....23.2168A} which are discarded in our analysis. Both the supermomenta and dual supermomenta are supertranslation invariant. 

At subleading order, the leading peeling-breaking aspects $\mathcal D_A^{\pm (0)}(x^A)$ and $\mathcal N_A^{ \text{(log)}\pm(0)}(x^A)$ are defined as \cite{Compere:2026jmk,Boschetti:2026ogm,Boschetti:2026gfd}
\begin{align}
\mathcal D_A^{\pm (0)} & := \nabla^B D_{AB}^{\pm(0)} - \frac{1}{2} \nabla_C \nabla_{\langle A} \nabla_{B\rangle} C^{\pm (1) BC} \pm 6 P_i \mathcal{M}^{\pm(0)}_{AB}\nabla^B n_i- 2 (E\mp P_i n_i) \nabla^B\mathcal{M}^{\pm(0)}_{AB} ,\label{defDA}\\ 
\mathcal  N_A^{\text{(log)}\pm (0)}  & := 
    \nabla_C\nabla_{\langle A}\nabla_{B\rangle}
    \mathcal C^{\pm(1)BC} + \mathcal D_A^{\pm(0)},\label{defNAlog}
\end{align}
with $\mathcal{M}^{\pm(0)}_{AB}(x^A):=m^{\pm(0)}\,\gamma_{AB}+\tilde{\mathcal{M}}^{(0)\pm}\,\epsilon_{AB}$ and with the leading tails $C_{AB}^{\pm (1)}(x^A)$ defined from our assumptions \eqref{CABasymptoticsscri}. $\mathcal C_{AB}^{\pm (1)}$ is defined as 
\begin{equation}
    \mathcal C_{AB}^{\pm (1)}:=  C_{AB}^{\pm (1)}+ (-E \pm P_i n_i)C_{AB}^{\pm (B)(0)}. \label{defCAB1}
\end{equation}
The corresponding charges are supertranslation invariant. 

The last remaining charges to be defined are the super-Lorentz charges associated with $\mathcal N_A^{\pm(0)}(x^A)$ and their proper-super\-translation-invariant version  associated with $\mathcal N_A^{\pm(0)\text{inv}}(x^A)$ which we build out of the covariant Bondi angular momentum aspect $\mathcal P^\pm_A(u,x^A)$ as a core result of this paper. 
More precisely, given a definition of dressed Bondi angular momentum aspect $\mathcal{N}^\pm_{A}(u,x^A)$ such that 
\begin{align}\label{condNA}
    \mathcal{N}^\pm_{A}(u,x^A) =\mathcal N_A^{\pm(0)}(x^A) + o((u^\pm)^{0}) ,
\end{align}
as $u^\pm \to -\infty$ and given a smooth vector field $Y^A(x^B)$ generating an infinitesimal  diffeomorphism over the sphere, the super-Lorentz charges are defined as 
\begin{equation}
    Q_{Y}^{\scri^+}(u) = \frac{1}{8\pi} \int_{S^2} d^2 \Omega\,  Y^A \mathcal{N}^+_A 
\label{scri+superrotationscharges},
\end{equation}
at $\scri^+$ and 
\begin{equation}
    Q_{Y}^{\scri^-}(v) =- \frac{1}{8\pi} \int_{S^2} d^2 \Omega\,  (\Upsilon^*Y^A) \mathcal{N}_A^-
\label{scri-superrotationscharges},
\end{equation}
at $\scri^-$.

Several proposals for the dressed Bondi angular momentum aspect $\mathcal{N}_A^\pm$ have been defined in the literature \cite{Barnich:2011mi,Flanagan:2015pxa,Compere:2019gft,Geiller:2024bgf,Freidel:2021qpz,Freidel:2024jyf}. With an appropriate definition, the Lorentz charges ($\ell=1$ harmonics) are finite and conserved across spatial infinity in the past and future of $\scri^\pm$ under the falloff conditions \eqref{CABasymptoticsscri} as derived in \cite{Compere:2023qoa}. However, the super-Lorentz charges have not yet been properly defined because of a logarithmic divergence of $\mathcal{P}^\pm_A$ in the limit $u^\pm \to -\infty$. Such a logarithmic divergence has also been identified in \cite{Geiller:2024ryw,Choi:2024ajz}.

In this work, we propose to define $\mathcal N^{\pm (0)}_A$ such that there is an explicit match between past and future null infinity:
\begin{align}
 \lim_{u \to -\infty}   Q_{Y}^{\scri^+}(u) = \lim_{v \to \infty}  Q_{Y}^{\scri^-}(v) \label{antip}
\end{align}

As we will derive in Section \ref{sec:matching}, our proposal is
\begin{align}\label{curlyN0}
\mathcal{N}_A^{\pm(0)}=&N^{\pm(0)}_A -\frac{1}{4}C^{\pm(B)(0)\,B}_{A}\nabla^CC^{\pm(B)(0)}_{CB}-\frac{1}{16}\nabla_A (C^{\pm(B)(0)\,C}_{D}C^{\pm(B)(0)\,D}_{C})\nn\\&-(\log 2+\frac{3}{2})\nabla^BD_{AB}^{\pm(0)}+\nabla^BD_{AB}^{\pm(E)(0)}\nn\\&+\frac{5}{4}\nabla_C \nabla_{\langle A} \nabla_{B\rangle}\mathcal{C}^{\pm(1)\,BC}-\frac{1}{4}\nabla_A \nabla^B\nabla^C \mathcal{C}^{\pm(1)}_{BC}-\frac{1}{4}\nabla^B\mathcal{C}^{\pm(1)}_{AB}+\frac{1}{4}O^{(q)n=0}_V(\nabla^B\mathcal{C}^{\pm(1)}_{AB})\nn\\&+\frac{7}{4}\nabla_C \nabla_{\langle A} \nabla_{B\rangle}\Big((-E\pm P_i n_i)C^{\pm(B)(0)\,BC}\Big) -\frac{1}{2}\nabla_A \nabla^B\nabla^C \Big((-E\pm P_i n_i)C^{\pm(B)(0)}_{BC}\Big)\nn\\&\pm 2 P_i \mathcal{\tilde{M}}^{\pm(0)}\epsilon_{AB}\nabla^Bn_i\mp \frac{1}{4} P_i\nabla^B n_i C^{\pm(B)(0)}_{BA} +\frac{1}{4}(E\pm3P_in_i)\nabla^B C^{\pm(B)(0)}_{BA}\pm \mathcal  S^{\pm(q)}_A[\mathcal M_{AB}^{-(0)}]
\end{align}
where the operator $O^{(q)n=0}_V$ is defined in Eq. \eqref{defOVq}, $\mathcal S^{\pm(q)}_A[\mathcal M_{AB}^{-(0)}]$ is a non-local quantity over the sphere defined in Eq. \eqref{SqCompact} from an integral over $dS_3$.

We do not provide a definition of $\mathcal N_A^\pm(u,x^A)$ in this paper as there are several quantities that obey \eqref{condNA} and we did not identify a unique definition. The definition of the Bondi angular momentum aspect at spatial infinity is sufficient for our purposes. Following the methods of \cite{Compere:2019gft} (see also \cite{Chen:2021szm,Javadinezhad:2022hhl,Javadinezhad:2022ldc}), the asymptotically-invariant definition of the super-Lorentz aspect under proper supertranslations is 
\begin{align}\label{Ninv0}
    \mathcal{N}_A^{\text{inv}\pm(0)}=&\mathcal N_A^{\pm (0)} -3 \mathcal{M}^{\pm(0)}_{AB}\nabla^BC^{\pm(0)}-C^{\pm(0)}\nabla^B\mathcal{M}^{\pm(0)}_{AB}
\end{align}
where $\mathcal N_A^{\pm (0)}$ has been defined in Eq. \eqref{curlyN0}. As a consequence of Eq. \eqref{detlaC0}-\eqref{deltaNA}, the field $\mathcal{N}_A^{\text{inv}\pm(0)}$ only transforms under ordinary translations as 
\begin{align}
 \delta_T \mathcal{N}_A^{\text{inv}\pm(0)}= 3 \mathcal{M}^{\pm(0)}_{AB}\nabla^B T^{\pm}\vert_{\ell=0,1}+T^{\pm}\vert_{\ell=0,1}\nabla^B\mathcal{M}^{\pm(0)}_{AB}. 
\end{align}
The superscript ``inv'' denotes invariance under proper supertranslations. Ordinary translations act nontrivially, as required for reproducing the asymptotic Poincar\'e algebra. 

Shifting the definition of $\mathcal N^{\pm(0)}_A$ with a term proportional to $\nabla^B D^{\pm(0)}_{AB}$ does not modify the definition of the Lorentz charges due to the fact that $D_{AB}^{\pm(0)}$ does not contain $\ell =0,1$ harmonics: one can integrate by parts inside the spherical integral $Y^A \nabla^B D_{AB}^{\pm(0)}$ to $- \nabla^{\langle A} Y^{B \rangle}D_{AB}^{\pm(0)}$ which is zero by the conformal Killing equation for $Y^A$. 
Therefore, contracting such terms with a conformal Killing vector $Y^A$ and integrating over the sphere is zero by orthogonality of the spherical vector harmonics. The non-linear terms proportional to $C_{AB}^{(B)}$ in Eq \eqref{curlyN0} would bring new contributions to the Lorentz charges. However, since $C_{AB}^{(B)}$ was previously set to zero at spatial infinity, our definition does not affect the established definition of Lorentz charges in \cite{Compere:2023qoa} but allows to furthermore define the super-Lorentz charges. Now, part of $\mathcal{S}_A[\mathcal M^{-(0)}_{AB}]$ originates from contributions proportional to $C_{AB}^{(B)}$ which were discarded in  \cite{Compere:2023qoa} and which could potentially contribute. However, the part of $\mathcal{S}_A[\mathcal M^{-(0)}_{AB}]$ originating from $\sigma$ does not contain $\ell=1$ harmonics\footnote{Indeed, the conserved charge $\oint_{S^2}d^2\Omega\,  S_{ab}\xi^a n^b$ where $\xi^a$ is a conformal Killing vector on $dS_3$ captures the $\ell=1$ harmonics of the source $S_{ab}$ defined in Eq. \eqref{SourceSab} and therefore of $\mathcal{S}_A$. Now, by using the explicit fall-off conditions at $\tau\to \pm \infty$ this charge exactly vanishes.}. The definition of the Lorentz charges is therefore unchanged in the phase space where $k_{ab}^{(B)}=0$. 

We warn the reader that the definition of $\mathcal{N}_A^\pm$ is not unique since one could add to $\mathcal{N}_A^{\pm(0)}$ any term that is quasi-conserved and has the same parity under antipodal matching as $\mathcal{N}_A^{\pm(0)}$. However, since we provide a complete set of charges at spatial infinity, such ambiguities amount to changing the basis of charges (possibly including non-linear redefinitions of the charges). A preferred basis would be one such that the algebra of charges is simplest, but these considerations go beyond our current analysis. 

As usual, we call the (super-)rotation aspect the magnetic-type $\mathcal N^{\pm(0)}_A$ and  the (super-)boost aspect the electric-type $\mathcal N^{\pm(0)}_A$. The (super-)Lorentz charges can therefore be split according to parity as 
\begin{equation}
    \mathcal J_{\ell m}^\pm=  \frac{(-1)^{\delta_{\pm,-}\ell}}{8\pi} \int_{S^2} d^2 \Omega \mathring V^{(B)\ell m}_A\mathcal{N}^{\pm\,A} , \qquad \mathcal K_{\ell m}^\pm=\frac{(-1)^{\delta_{\pm,-}(\ell+1)}}{8\pi} \int_{S^2}  d^2 \Omega \mathring V^{(E)\ell m}_A\mathcal{N}^{\pm\,A}
\label{canonicalCurlyNACharges}.
\end{equation}
The $\ell =1$ harmonics lead to the Lorentz charges ($\mathcal J_{\ell m}^\pm$ are the angular momenta and $\mathcal K_{\ell m}$ are the mass moments) while the higher harmonics lead to the proper super-Lorentz charges. The antipodal matching conditions \eqref{antip} are equivalent to $\mathcal J_{\ell m}^+=\mathcal J_{\ell m}^-$ and $\mathcal K_{\ell m}^+=\mathcal K_{\ell m}^-$.

With these definitions, the complete set of charges is summarized in Table \ref{tab:chargesnull}. 

\begin{table}[!htbp]
    \centering
    \begin{tabular}{|r|c|c|}\hline 
        & Value at $u^\pm \to -\infty$ & Name of the quasi-conserved charge \\ \hline\hline
       Leading & $m^{\pm(0)} $  & Momenta ($\ell=0,1$) \\   
       & & Proper supermomenta ($\ell > 1$) \\ \cline{2-3}
  & $C^{\pm(0)}$  & Supertranslation frame charge ($\ell>1$)\\ \cline{2-3}
  & $\mathcal{\tilde{M}}^{\pm(0)}$  & Dual supertranslation charges ($\ell>1$)\\ \hline  \hline
Subleading & $\mathcal N_A^{\text{inv}\pm(0)}$ & Angular momentum and center-of-mass ($\ell=1$) \\ 
&& Proper super-Lorentz charges ($\ell >1$)\\ \cline{2-3} 
& $\mathcal N_A^{\pm(\log)(0)}$ & Leading mixed tail/peeling-breaking charge ($\ell \geq 2$) \\ \cline{2-3}  
& $\mathcal D_{A}^{\pm(0)} $ & Leading peeling-breaking charge ($\ell \geq 2$)  \\ \hline
    \end{tabular}
    \caption{Complete set of charges at $\mathcal I^\pm$ up to subleading order under our assumptions. The values at $u^\pm \rightarrow -\infty$ are the proper-supertranslation-invariant values defined in the main text except the supertranslation frame that does not admit a proper-supertranslation-invariant definition. The non-vanishing (scalar, vector or tensor) spherical harmonics are indicated in parentheses.}
    \label{tab:chargesnull}
\end{table}

\section{Spatial infinity $i^0$}
\label{sec:spatial}

\subsection{Beig-Schmidt expansion}

Assuming our Hypothesis H3, we consider the following asymptotic form in Beig-Schmidt coordinates $(\rho, \phi^a)$, $a=1,2,3$, with $\phi^a = (\tau, x^A)$ at spatial infinity, 
\begin{align}
      ds^2&= \left(1+\frac{2\sigma}{\rho}+\frac{\sigma^2}{\rho^2}+o(\rho^{-2})\right)d\rho^2 + o(\rho^{-2}) \rho d\rho d\phi^a +\rho^2 \bigg(q_{ab}+\frac{k_{ab}-2\sigma q_{ab}}{\rho} \nonumber \\& +\frac{\log\rho}{\rho^2}  ~ i_{ab}+\frac{j_{ab}}{\rho^2}
     +o(\rho^{-2})\bigg) d\phi^a d\phi^b, \label{BSExpansion}
\end{align}
where the condition $k:=k_{ab} q^{ab}=0$ is imposed. All Beig-Schmidt fields $\{ \sigma,\, k_{ab},\, i_{ab},\, j_{ab}\}$ depend upon $\phi^a$. We advocate as in \cite{Compere:2026jmk} that the class of metrics admitting the expansion \eqref{BondihExp} in Bondi-Sachs coordinates around past and future null infinity precisely admits such a Beig-Schmidt expansion.

The Beig-Schmidt expansion \eqref{BSExpansion} has already been formulated and studied to  second order in the $\rho\to \infty$ expansion in \cite{Compere:2011ve}. The logarithmic Beig-Schmidt field $i_{ab}$ was introduced in \cite{Compere:2011ve} in addition to the original polynomial expansion \cite{beig1982einstein,beig1984integration}.  The leading order metric is Minkowski in hyperboloidal coordinates with $q_{ab}$ the unit metric over $dS_3$,
\begin{align}
 q_{ab}d\phi^a d\phi^b = -d\tau^2+ \cosh^2\tau (d\theta^2+\sin^2\theta d\phi^2).   
\end{align}
Latin indices will be raised and lowered with this metric and $\mathcal D_a$ will denote its metric-compatible covariant derivative. We define as $\Upsilon_{\mathcal H}$ the combined time reversal $\tau \mapsto - \tau$ and antipodal map over the sphere $\Upsilon$. Following \cite{Compere:2025bnf}, we define the $p$-parity (resp. $q$-parity) tensor of rank $r$ on $dS_3$ as the tensor $\Psi_{i_1...i_r}$ satisfying $\Upsilon_{\mathcal H}^*\Psi_{i_1...i_r}=(-1)^{n+1+r}\Psi_{i_1...i_r}$ (resp. $\Upsilon_{\mathcal H}^*\Psi_{i_1...i_r}=(-1)^{n+r}\Psi_{i_1...i_r}$) and the equations 
\begin{equation}
    (\Box+((n+1)^2-1-r))\Psi_{i_1...i_r}=S_{i_1...i_r},\qquad \mathcal{D}^{a}\Psi_{a\,i_2...i_r}=0,\qquad 
    q^{ab}\Psi_{ab\,i_3...i_r}=0\label{generici0equ}
\end{equation}
where $S_{i_1...i_r}$ is a source independent of $\Psi_{i_1...i_r}$ and $n+1\in\mathbb N$. The second equality is only defined for $r\ge1$ and the third for $r\ge2$. 
Such rank $r$ tensors are scalars for $r=0$, transverse vectors for $r=1$ and symmetric divergence-free traceless (SDT) tensors for $r=2$. 

Following the work \cite{Compere:2026jmk}, we consider $k_{ab}$ as a tensor admitting an electric piece and a (traceless) magnetic piece $k^{(B)}_{ab}$ (for a definition, see \cite{Compere:2025bnf}) as 
\begin{equation}
 k_{ab}= -2 (\mathcal D_a \mathcal D_b + q_{ab})\Phi + k_{ab}^{(B)}
\label{kabPhi}
\end{equation}
where $\Phi$ is a scalar field on $dS_3$. The trace condition $k=0$ is equivalent to 
\begin{equation}
    (\mathcal D_c \mathcal D^c + 3)\Phi = 0.
\label{EquationPhi}
\end{equation}
The operator acting on $\Phi$ in Eq. \eqref{kabPhi} admits a non-trivial kernel consisting in the $4$ $q$-parity $\ell=0,1$ harmonics spanned by $\sinh \tau$ and $\cosh\tau n_i$.  In order to provide a unique definition for $\Phi$ we set these harmonics to zero in $\Phi$, which will turn out to be consistent with the similar condition imposed on $C^{(0)}$ after imposing the matching condition, see Eq. \eqref{MatchPhi}.


\subsection{Einstein's equations}
\label{sec:EIN}
Under the assumption of a compact stress-energy tensor at null infinities, Einstein's equations near spatial infinity amount to $G_{\mu\nu}=0$. These equations have been studied in detail in \cite{Compere:2011ve} using a 3+1 decomposition adapted to Beig-Schmidt coordinates. They reduce to a set of equations for each field appearing in the expansion \eqref{BSExpansion}:
\begin{equation}
    (\Box + 3)\sigma = 0,~~~~~ \mathcal{D}^b k_{ab} = 0,~~~~~(\Box - 3)k_{ab} = 0,
\label{BSEquationOrder1}
\end{equation}
\begin{equation}
   i_a^a = 0,~~~~~ \mathcal{D}^b i_{ab} = 0,~~~~~(\Box - 2)i_{ab} = 0\label{iabequation},
\end{equation}
and
\begin{align}
  j_a^a &= 12 \sigma^2+\frac{1}{4} k^{ij} k_{ij}+k_{ij} \mathcal{D}^{i}\mathcal{D}^{j}\sigma+\mathcal{D}_{i}\sigma \mathcal{D}^{i}\sigma\label{jabTrace},\\
  \mathcal{D}^b j_{ab} &=16 \sigma \mathcal{D}_a\sigma+2 \mathcal{D}^i\sigma \mathcal{D}_i\mathcal{D}_a\sigma+ \mathcal{D}^i\mathcal{D}^j\sigma \mathcal{D}_a k_{ij}+\frac{1}{2} k^{ij} \left(\mathcal{D}_i k_{aj}-\frac{1}{2} \mathcal{D}_a k_{ij}+2 \mathcal{D}_a\mathcal{D}_i\mathcal{D}_j\sigma\right)\label{jabDivergence} ,\\
  (\Box - 2)j_{ab} &= 2 i_{ab} + NL_{ab}(\sigma,\sigma)+NL_{ab}(\sigma,k)+NL_{ab}(k,k),\label{jabEOM}
\end{align}
where the non-linear terms $NL_{ab}(\sigma,\sigma)$, $NL_{ab}(\sigma,k)$ and $NL_{ab}(k,k)$ are  
\begin{align}
    NL_{ab}(\sigma,\sigma) :=&q_{ab} \left(6 \mathcal{D}_i\sigma \mathcal{D}^i\sigma-18 \sigma^2\right)+14 \sigma \mathcal{D}_a\mathcal{D}_b\sigma+8 \mathcal{D}_a\sigma \mathcal{D}_b\sigma+2\mathcal{D}^i\sigma \mathcal{D}_i\mathcal{D}_a\mathcal{D}_b\sigma\nn\\&+2\mathcal{D}^i\mathcal{D}_a\sigma \mathcal{D}_i\mathcal{D}_b\sigma,\\
    NL_{ab}(k,k) :=&k_a^i k_{ib}+k^{ij} (\mathcal{D}_i\mathcal{D}_j k_{ab} - \mathcal{D}_i\mathcal{D}_{(a} k_{b)j})-\frac{1}{2} \mathcal{D}_a k^{ij} \mathcal{D}_b k_{ij}+\mathcal{D}^i k_{(a}^j \mathcal{D}_{b)} k_{ij}\nn\\& -\mathcal{D}^i k_a^j \mathcal{D}_j k_{bi}+\mathcal{D}^i k_a^j\mathcal{D}_i k_{bj},\\
    NL_{ab}(\sigma,k) :=& 4 \sigma k_{ab} -2 q_{ab} k^{ij} \mathcal{D}_i\mathcal{D}_j\sigma +4 k^i_{(a} \mathcal{D}^{}_{b)} \mathcal{D}_i\sigma+4 \mathcal{D}^i\sigma (\mathcal{D}_{(a} k_{b)i} - \mathcal{D}_i k_{ab})\nn\\&+ k^{ij}  \mathcal{D}_{(a} \mathcal{D}_{b)} \mathcal{D}_i \mathcal{D}_j\sigma +2 \mathcal{D}_{(a} k^{ij} \mathcal{D}_{b)} \mathcal{D}_i \mathcal{D}_j\sigma +\mathcal{D}^i \mathcal{D}^j\sigma  \mathcal{D}_{(a} \mathcal{D}_{b)} k_{ij}.\label{NLsigmakab}
\end{align}

\subsection{Residual gauge transformations and invariant quantities}
\label{sec:res}

One can show that the form of the metric \eqref{BSExpansion} is invariant under Lorentz transformations, logarithmic translations and supertranslations, which are the residual transformations of Beig-Schmidt gauge under our assumptions on fall-off conditions   \cite{beig1982einstein,Compere:2011ve}. The logarithmic translations are generated by a scalar $H$ that obeys $\mathcal D_a \mathcal D_b H + q_{ab} H = 0$\footnote{We do not consider the extension to logarithmic supertranslations \cite{beig1982einstein,Compere:2011ve,Troessaert:2017jcm,Fuentealba:2022xsz,Compere:2023qoa,Mishra:2025nmd,girelli2026covariantformulationlogarithmicsupertranslations} in this work.}. We denote a basis of such functions as $H_{(\mu)}$, $\mu=0,1,2,3$ which are built from $q$-parity scalar harmonics with $n=1$ and $\ell=0,1$, see Eq. \eqref{generici0equ} and Sec. III.C. of \cite{Compere:2025bnf}. The general solution is a linear combination of these 4 harmonics, 
\begin{equation}\label{HlogSolutions}
H = \sum_{\mu=0,1,2,3} H_{(\mu)}\zeta_{(\mu)}  .
\end{equation}
The diffeomorphisms associated to logarithmic translations are \cite{Compere:2023qoa}
\begin{align}
    \xi^\rho &=H \log\rho+\frac{\log\rho}{\rho}(-\sigma H+\mathcal{D}_c \sigma \mathcal{D}^cH)+\frac{1}{\rho}2 \mathcal{D}_c \sigma \mathcal{D}^cH+o(\rho^{-1}),\\
   \xi^a &=\frac{(1+\log\rho)}{\rho}\mathcal{D}^aH + \frac{\log\rho}{\rho^2} \frac{1}{2}\Big( \mathcal{D}^a(-\sigma H+\mathcal{D}_c \sigma \mathcal{D}^cH)+4\sigma \mathcal{D}^a H-k^{ac}\mathcal{D}_c H\Big)\nn \\& +\frac{1}{\rho^2}\frac{1}{4}\Big(\mathcal{D}^a(-\sigma H+3\mathcal{D}_c \sigma \mathcal{D}^cH)-k^{ac}\mathcal{D}_c H+4 \sigma \mathcal{D}^a H\Big)+o(\rho^{-2}).
\end{align}

Logarithmic translations act on the fields as \cite{Compere:2011ve} 
\begin{align}
\delta_H \sigma & = H , \\
\delta_H k_{ab} &= 0, \\
\delta_H i_{ab} &=\mathcal{D}_c \left( (\mathcal{D}^c H)(\mathcal{D}_a \mathcal{D}_b \sigma + q_{ab} \sigma) \right)
-(\mathcal{D}_{(a} k_{b)c}-\mathcal{D}_ck_{ab})\mathcal{D}^cH , \\
\delta_H j_{ab} &= 
+ 10 \sigma H q_{ab} - 4 \mathcal{D}_{(a}H \mathcal{D}_{b)}\sigma + 2 H \mathcal D_{a} \mathcal D_{b} \sigma + 2 q_{ab} \mathcal D_{c} \sigma \mathcal D^{c} H \nn\\&+ \frac{5}{2}\mathcal D_{c} \left( (\mathcal D_{a}\mathcal D_{b} \sigma + q_{ab} \sigma)\mathcal D^{c} H \right)
+ \frac{1}{2}\mathcal D_{c} (k_{ab}\mathcal D^{c} H) -\frac{1}{2}(\mathcal{D}_{(a} k_{b)c}-\mathcal{D}_ck_{ab})\mathcal{D}^cH .
\end{align}

Since the only part of $\sigma$ that varies under a logarithmic translation is the $\ell=0,1$ $q$-parity harmonic part of $\sigma$, we perform the decomposition 
\begin{align}\label{decompsigma}
\sigma = \sigma^\text{inv} + \eta,
\end{align}
where $\eta$ only contains the $H_{(\mu)}$ spherical $\ell=0,1$ harmonics which  are shifted by logarithmic translations while  $\sigma^\text{inv}$ does not contain any such harmonic. 

The asymptotic matching at $\scri^+_-$ or $\scri^-_+$ with the alternative Bondi-Sachs expansion \eqref{BondihExp} in the limit $u^\pm \to -\infty$ imposes that $\sigma$ converges in the limit $\tau \rightarrow \pm \infty$ up to a logarithmic translation. This change of logarithmic translation frame is unavoidable in order to define the convergence of $\sigma$ at both $\tau \rightarrow \pm \infty$ \cite{Ashtekar:1978zz,Ashtekar:1991aa,Compere:2023qoa}. As demonstrated in \cite{Compere:2023qoa}, the field $\sigma$ vanishing at $\tau \rightarrow \pm \infty$ has 
\begin{align}\label{etaH0}
\eta=\pm H_0, \qquad H_0 :=  -2 E \sinh \tau +2 P^i n_i \cosh\tau   
\end{align}
where $E$ and $P^i$ are the total energy and momentum of the spacetime. In order to perform the matching with Bondi-Sachs fields we will gauge fix the logarithmic transformations at $\mathcal I^+_-$ (resp. $\mathcal I^-_+$) by canceling the divergent $\ell =0,1$ parts of  $\sigma$, which leads to Eq. \eqref{etaH0}. These gauge fixing conditions differ between $\mathcal I^+_-$ and $\mathcal I^-_+$, but since logarithmic translation charges do commute with BMS charges, we can freely gauge fix them in separate ways at distinct boundaries \cite{Compere:2023qoa}. It can be interpreted as the fact that future radiative gauge and past radiative gauge lie in distinct logarithmic translation frames \cite{Boschetti:2026gfd}.

The Beig-Schmidt supertranslations also called spi-supertranslations are generated by an arbitrary scalar $\omega(\phi^a)$ which obeys $(\mathcal D_c \mathcal D^c+3) \omega = 0$. This wave equation admits two sets of solutions, one of them corresponding to supertranslations on $\mathcal I^\pm$ and the other set corresponding to trivial gauge transformations \cite{Troessaert:2017jcm}. We will also gauge fix these trivial gauge transformations below. There is then a one-to-one correspondence between BMS supertranslation generators $\omega(\phi^a)$ and $T(x^A)$ defined at $\scri^\pm$. The associated diffeomorphism 
is given up to subsubleading order as 
\begin{align}
    \xi^\rho &= \omega + \frac{1}{\rho} \Big(-\sigma \omega + \mathcal{D}^a \sigma \mathcal{D}_a \omega\Big)+o(\rho^{-1}),\\
   \nn  \xi^a &= \frac{1}{\rho} \mathcal{D}^a \omega + \frac{1}{\rho^2}\Big(\frac{1}{2}\mathcal{D}^a \mathcal{D}^b \sigma \mathcal{D}_b \omega - \frac{1}{2} \omega \mathcal{D}^a \sigma + \frac{3}{2}\sigma \mathcal{D}^a \omega - \frac{1}{2} k^{ab}\mathcal{D}_b \omega\Big)+o(\rho^{-2}).
\end{align}

Under infinitesimal supertranslations, the Beig-Schmidt fields transform as 
\begin{align}
\delta_{\omega} \sigma & =0, \label{sigmaSTransl}\\
\delta_{\omega} k_{a b} & = 2\left(\mathcal{D}_a \mathcal{D}_b+q_{ab}\right) \omega \label{ktransfo}, \\
\delta_{\omega} i_{a b} & =0, \\
\delta_{\omega} j_{a b} & = k_{c(a} \mathcal{D}_{b)} \mathcal{D}^c \omega +k_{a b} \omega +\mathcal{D}^c \omega \left(\mathcal{D}_c k_{a b}-\mathcal{D}_{(a} k_{b) c}\right) -4 \sigma \omega  q_{ab}  \nn\\
&+4 \mathcal{D}_{(a} \sigma \mathcal{D}_{b)} \omega+\mathcal{D}_a \mathcal{D}_b\left(-\sigma \omega +\mathcal{D}_c \sigma \mathcal{D}^c \omega \right).\label{deltajab}
\end{align}
We do not detail the Lorentz transformation here. 

Given the field transformation laws and the equations of motion, it is possible to construct second order proper supertranslation and logarithmic translation invariant tensors (see \cite{Compere:2026jmk}). This will be useful for matching the data between $\scri^+_-$ and $\scri^-_+$ since a different logarithmic gauge fixing will be made at the corners. The second order invariant tensors can be defined as 
\begin{align}
    I_{ab} :=& i_{ab}-\mathcal{D}_c \left( \mathcal{D}^c \eta~(\mathcal{D}_a \mathcal{D}_b \sigma + q_{ab} \sigma)\right)+(\mathcal{D}_{(a} k_{b)c}-\mathcal{D}_ck_{ab})\mathcal{D}^c\eta \label{defIab},\\
    {J}_{ab}:=& j_{ab}-\frac{5}{2} i_{ab}+\frac{1}{2} q_{ab} \Big(-2\mathcal{D}_g\sigma \mathcal{D}^g\sigma+4 \mathcal{D}^g\sigma \mathcal{D}_g\Phi -8 \sigma \Phi-12 \sigma^2-2 \Phi^2\Big)+\mathcal{D}^g\sigma \mathcal{D}_g\mathcal{D}_a\mathcal{D}_b\Phi\nn\\&+\mathcal{D}^g\Phi \mathcal{D}_g\mathcal{D}_a\mathcal{D}_b\sigma+2 \mathcal{D}^g\mathcal{D}_{(a}\sigma \mathcal{D}_{b)}\mathcal{D}_g\Phi-\Phi \mathcal{D}_a\mathcal{D}_b\sigma-\mathcal{D}^g\mathcal{D}_a\Phi \mathcal{D}_g\mathcal{D}_b\Phi-\sigma \mathcal{D}_a\mathcal{D}_b\Phi\nn\\&+2 \mathcal{D}_a\sigma \mathcal{D}_b\sigma-2 \sigma \mathcal{D}_a\mathcal{D}_b\sigma-2 \Phi \mathcal{D}_a\mathcal{D}_b\Phi-4 \sigma k^{(B)}_{ab} -4 \mathcal{D}_{(a}(k^{(B)}_{b)i}\mathcal{D}^i\sigma )+ 4 \mathcal{D}^i \sigma \mathcal{D}_i k^{(B)}_{ab}\nn\\&+\mathcal{D}^i \mathcal{D}^j \sigma k^{(B)}_{ij}q_{ab}+\Phi k^{(B)}_{ab} +k^{(B)}_{i(a}\mathcal{D}^i \mathcal{D}_{b)}\Phi-\mathcal{D}^i\Phi \mathcal{D}_{(a}k^{(B)}_{b)i}+\mathcal{D}^i\Phi \mathcal{D}_i k^{(B)}_{ab}-4\eta k^{(B)}_{ab} \nn\\& + 2  \mathcal{D}_{(a}(k^{(B)}_{b)i}\mathcal{D}^i\eta ) -\frac{5}{2}\mathcal{D}^i(\mathcal{D}_i\eta  k^{(B)}_{ab})
    +\frac{1}{2} k^{(B)}_{ac}k^{(B)\,c}_b - \frac{1}{8} q_{ab} (k^{(B)}_{cd}k^{(B)\,cd}+\mathcal{D}_ik^{(B)}_{cd}\mathcal{D}^ik^{(B)\,cd}) \nn\\& - \frac{1}{8} k^{(B)\,cd} \mathcal{D}_{(a} \mathcal{D}_{b)} k^{(B)}_{cd} +\frac{3}{8} \mathcal{D}_{(a} k^{(B)\,cd} \mathcal{D}_{b)}  k^{(B)}_{cd} , \label{defJab}
\end{align}
where $\eta$ is the odd piece of $\sigma$ defined in Eq. \eqref{decompsigma} that obeys $\mathcal D_a \mathcal D_b \eta + q_{ab} \eta = 0$ and $\Phi$ is the scalar defining the electric piece of $k_{ab}$. Note the identity $\mathcal{D}_{(a} k^{(E)}_{b)c}-\mathcal{D}_ck^{(E)}_{ab}=0$ so that $k_{ab}$ can be replaced by $k_{ab}^{(B)}$ in Eq. \eqref{defIab}. 

These tensors satisfy the equations of motion 
\begin{align}
I^{a}_a =0, \qquad \mathcal{D}^b I_{ab} &= 0,\qquad 
     (\Box - 2)I_{ab} = 0,\label{eq:99b}\\
J_{a}^{a}= 0 ,\qquad \mathcal{D}^b J_{ab} &= 0,\qquad 
     (\Box - 2)J_{ab} = S_{ab} \label{BoxJab2},
\end{align} 
where the SDT source is 
\begin{align}
S_{ab}= &2\Big(I_{ab}-\mathcal{D}^c\sigma^{\text{inv}} \mathcal{D}_c\mathcal{D}_{\langle a}\mathcal{D}_{b \rangle}\sigma^{\text{inv}}+3 \mathcal{D}_c\mathcal{D}_{\langle a}\sigma^{\text{inv}} \mathcal{D}^c\mathcal{D}_{b\rangle}\sigma^{\text{inv}}+9 \sigma^{\text{inv}} \mathcal{D}_{\langle a}\mathcal{D}_{b\rangle}\sigma^{\text{inv}} \nonumber \\&+\frac{1}{2}k^{(B)ig} \mathcal{D}_{\langle a}\mathcal{D}_{b\rangle}\mathcal{D}_i\mathcal{D}_g\sigma^{\text{inv}}+\mathcal{D}_{\langle a}\mathcal{D}_{\vert i}\mathcal{D}_{g\vert}\sigma^{\text{inv}} (\mathcal{D}_{b\rangle}k^{(B)ig}-4 \mathcal{D}^ik_{b\rangle}^{(B)g})\nn\\&+\frac{1}{2}\mathcal{D}^i\mathcal{D}^g\sigma^{\text{inv}} \mathcal{D}_{\langle a}\mathcal{D}_{b\rangle}k_{gi}^{(B)}-4 \mathcal{D}^i\mathcal{D}^g\sigma^{\text{inv}} \mathcal{D}_i( \mathcal{D}_{(a}k_{b)g}^{(B)}- \mathcal{D}_gk_{ab}^{(B)})-6 k_{\langle a}^{(B)g} \mathcal{D}_{b\rangle}\mathcal{D}_g\sigma^{\text{inv}}\nonumber \\&+2 \mathcal{D}^g\sigma^{\text{inv}} ( \mathcal{D}_{(a}k_{b)g}^{(B)}- \mathcal{D}_gk_{ab}^{(B)})-6 \sigma^{\text{inv}} k_{ab}^{(B)} -\frac{1}{8} \mathcal D^{j}k^{(B)gi} \mathcal D_{j}\mathcal D_{\langle a}\mathcal D_{b\rangle}k_{gi}^{(B)}\nonumber \\&+\frac{3}{8} \mathcal D_{j}\mathcal D_{\langle a} k^{(B)gi} \mathcal D^{j}\mathcal D_{b\rangle}k_{gi}^{(B)}-\frac{5}{8} k^{(B)gi} \mathcal D_{\langle a}\mathcal D_{b\rangle}k_{gi}^{(B)}+\frac{5}{4} \mathcal D_{\langle a} k^{(B)gi} \mathcal D_{b\rangle }k_{gi}^{(B)}\nonumber \\&-  k^{(B)gi} \mathcal D_{i}\mathcal D_{( a} k^{(B)}_{b)g}+\frac{1}{2} k^{(B)gi} \mathcal D_{i}\mathcal D_{g}k_{ab}^{(B)}+2 \mathcal D^{i}k^{(B)g}_{\langle a} \mathcal D_{b\rangle}k_{gi}^{(B)}\nonumber \\&-\frac{1}{2}\mathcal D^{i}k^{(B)}_{g\langle a} \mathcal D^{g}k_{b\rangle i}^{(B)}+ \mathcal D_{i}k^{(B)g}_{\langle a} \mathcal D^{i}k^{(B)}_{b\rangle g}+\frac{9}{4} k^{(B)g}_{\langle a} k^{(B)}_{b\rangle g}\Big).\label{SourceSab}
\end{align} 

The tensor $I_{ab}$ is uniquely defined (up to a global factor) by imposing that it is invariant and that it satisfies Eq. \eqref{eq:99b}. The tensor $J_{ab}$ is defined such that it is invariant under proper supertranslations and logarithmic translations, SDT and such that $S_{ab}$ contains the minimal number of derivatives acting on Beig-Schmidt fields, which is four. It is ambiguous by shifts proportional to $I_{ab}$. We have checked that adding a non-linear SDT tensor in $\sigma$, $k_{ab}^{(B)}$ (see \cite{Compere:2011ve} for classification of such terms) will change the source $S_{ab}$ with higher derivative terms. 

As explained in \cite{Compere:2025bnf}, using the unit normal $n^a\partial_a = \partial_\tau$, one can convert the tensorial equations \eqref{eq:99b}-\eqref{BoxJab2} into the vectorial equations $D^a I_a = D^b J_b = 0$, $(\square-1)I_a =0$  and $(\square-1)J_a =S_a$ where 
\begin{align}
J_a &:= \cosh \tau n^b J_{ab}   
,\qquad I_a := \cosh \tau n^b I_{ab}, \label{defIa}
\end{align}
and $S_a := \cosh \tau  n^b S_{ab}$. The vector $I_a$ and the homogeneous part of $J_a$ are $n=0$ divergence-free vectors in the classification of \cite{Compere:2025bnf}, see that reference for a description of their properties. Using \cite{Compere:2011ve} the source $S_{ab}$ can be rewritten in terms of the stress-tensor of a boundary action. We perform this rewriting in Appendix \ref{app:boundaryaction}.

\subsection{Asymptotic behavior of the fields near the corners and their antipodal map}

The leading order fields $\sigma$, $\Phi$ and $k_{ab}^{(B)}$ obey homogeneous equations whose generic solutions are a combination of $p$-parity and $q$-parity harmonics with well-understood asymptotic behavior in the limit $\tau \to \pm\infty$ \cite{Compere:2025bnf}. Using the convergence of $\sigma^\pm$ as $\tau \to \pm\infty$ \cite{Compere:2023qoa}, the field $\sigma^\pm$ is a combination of $p$-parity $n=1$ scalar harmonics and $q$-parity $n=1$ $\ell=0,1$ harmonics. The field $\Phi$ is a combination of both $q$-parity and $p$-parity harmonics. In the notation of \cite{Compere:2025bnf} their asymptotic behaviors as $\tau \to \pm\infty$ read as
\begin{align}
\sigma^\pm &= \sigma^{(\mathbb{S})\pm} e^{-3\vert \tau \vert}  +o(e^{-3\vert \tau \vert})\label{sigmaAS},\\
\Phi &= \Phi^{(\mathbb{L})\pm} e^{\vert \tau \vert}-(\nabla^2+1)\Phi^{(\mathbb{L})\pm} e^{-\vert \tau \vert}+\left(\Phi^{(\mathbb{S})\pm}+\vert\tau\vert \nabla^2(\nabla^2+2)\Phi^{(\mathbb{L})\pm}\right)e^{-3\vert \tau \vert}+ o(e^{-3\vert \tau \vert}),
\end{align}
where $\sigma^{(\mathbb{S})\pm}(x^A)$, $\Phi^{(\mathbb{S})\pm}(x^A)$ and $\Phi^{(\mathbb{L})\pm}(x^A)$ are the free (subleading $\mathbb{S}$ or leading $\mathbb{L}$) asymptotic fields that obey the antipodal relationships
\begin{align}
    &\Upsilon^*\sigma^{(\mathbb{S})+}(x^A) = \sigma^{(\mathbb{S})-}(x^A),\\
    &\Upsilon^*\Phi^{(\mathbb{L})+}(x^A)-\Upsilon^*O^{(p)n=1}[\Phi^{(\mathbb{S})+}(x^A)]=-\big(\Phi^{(\mathbb{L})-}(x^A)-O^{(p)n=1}[\Phi^{(\mathbb{S})-}(x^A)]\big),\label{AntiPodPhiL}\\
    &
    \Upsilon^*\Phi^{(\mathbb{S})+}(x^A)-\Upsilon^*O^{(q)n=1}[\Phi^{(\mathbb{L})+}(x^A)]=+\big(\Phi^{(\mathbb{S})-}(x^A) -O^{(q)n=1}[\Phi^{(\mathbb{L})-}(x^A)]\big). 
\end{align}
We recall that the antipodal map acting on a function of $x^A$ is $\Upsilon^* f(\theta,\phi) = f(\pi-\theta,\phi+\pi)$. Here, the operators $O^{(p)n=1}$ and $O^{(q)n=1}$ acting on an arbitrary function $f(x^A)$ are defined as follows
\begin{align}
    O^{(p)n=1}[f(x^A)] := & \sum_{\ell=0}^{+\infty}\sum_{m=-\ell}^{+\ell}(-1)^{\ell}\frac{1}{\Gamma(\ell+3)\Gamma(2-\ell)}  Y_{\ell m}(x^A)\int_{S^2}d\Omega' \overline{Y_{\ell m}(x^{A'})}f(x^{A'}),
    \end{align}
    \begin{align}
    O^{(q)n=1}[f(x^A)] := &\sum_{\ell=0}^{+\infty}\sum_{m=-\ell}^{+\ell} \nabla^2(\nabla^2+2)(\frac{3}{4} -  H_{\ell} ) Y_{\ell m}(x^A)\int_{S^2}d\Omega' \overline{Y_{\ell m}(x^{A'})}f(x^{A'}),\label{defOq}
\end{align}
where $H_\ell = \sum_{k=1}^\ell k^{-1}$ are the harmonic numbers.
The operator $O^{(p)n=1}$ annihilates all $\ell\ge2$ harmonics and $O^{(q)n=1}$ annihilates all $\ell<2$ harmonics. Since $\Phi$ is defined up to $\ell=0,1$ harmonics, one can always choose $\Phi$ such that it contains only $\ell\ge2$ harmonics (see \cite{Compere:2023qoa}). This implies that the antipodal matching \eqref{AntiPodPhiL} becomes local 
\begin{align}
\Phi^{(\mathbb{L})+}(x^A)=-\Upsilon^*\Phi^{(\mathbb{L})-}(x^A). 
\end{align}
This is equivalent to the matching of supertranslation frames between $\scri^+$ and $\scri^-$ originally proposed in \cite{Strominger:2013jfa},  as demonstrated in Section \ref{sec:antipodalmatch}.  

The leading order Beig-Schmidt field $k^{(B)}_{ab}$ is a combination of SDT $n=-1$ magnetic-type tensors. Their asymptotic behavior as $\tau \to \pm\infty$ is provided in Eqs. (403)-(409) of \cite{Compere:2025bnf}. Explicitly, we have : 
\begin{align}
k^{(B)}_{\tau\tau} &= o(e^{-3\vert\tau\vert}),\\
k^{(B)}_{\tau A} &= \pm(2 \vert\tau\vert \nabla^B\Gamma^{(\mathbb{L},B)\pm}_{AB}+\nabla^B\Gamma^{(\mathbb{S},B)\pm}_{AB})e^{-\vert\tau\vert} + o(e^{-\vert\tau\vert}),\\
k^{(B)}_{AB} & =\frac{1}{4}\Big(2(\vert\tau\vert+1)\Gamma^{(\mathbb{L},B)\pm}_{AB} +\Gamma^{(\mathbb{S},B)\pm}_{AB}\Big)e^{\vert\tau\vert}+o(e^{\vert\tau\vert})\label{kabscri+-}. 
\end{align}
where $\Gamma^{(\mathbb{L},B)\pm}_{AB}(x^A)$ is the free leading asymptotic field of $q$-harmonics and $\Gamma^{(\mathbb{S},B)\pm}_{AB}(x^A)$ contains the free leading asymptotic field of $p$-harmonics and subleading contributions of $q$-harmonics. In \cite{Compere:2026jmk} (see also Eqs. \eqref{ktautau}-\eqref{kabscri+-bis}), it is shown that $\Gamma^{(\mathbb{L},B)\pm}_{AB}(x^A)=0$ as a result of asymptotic matching conditions given our hypotheses on the Bondi-Sachs expansion. Therefore $\Gamma^{(\mathbb{S},B)\pm}_{AB}(x^A)$ correspond to $p$-parity harmonics only. Taking that into account, the asymptotic data obeys the antipodal relationships
\begin{equation}
    \Gamma^{(\mathbb{S},B)+}_{AB}(x^A) = \Upsilon^*\Gamma^{(\mathbb{S},B)-}_{AB}(x^A).
\end{equation}
This determines the antipodal map of the dual mass aspect as demonstrated in Section \ref{sec:antipodalmatch}.

Let us now discuss the subleading order Beig-Schmidt fields. From the analysis \cite{Compere:2025bnf}, the asymptotic behavior of the homogeneous solutions are 
\begin{align}
V_\tau &= \pm 4 \nabla^A  V_A^{(\mathbb{L})\pm} e^{- \vert \tau \vert}\pm 4 (V^{(\mathbb S)\pm}-\vert\tau\vert \nabla^AV_A^{(\mathbb L,\tau)\pm})e^{- 3\vert \tau \vert}+o( e^{-3 \vert \tau \vert}), \label{Vtaubehav}\\
V_A &=V_A^{(\mathbb L)\pm}e^{\vert\tau\vert}+[ V_A^{(\mathbb L,\tau)\pm}\vert\tau\vert -V_A^{(\mathbb S)\pm}+V_A^{(\mathbb L)\pm}-2\nabla_A\nabla^BV_B^{(\mathbb L)\pm}]e^{-\vert\tau\vert}+o(e^{-\vert\tau\vert}),\label{VAbehav}
\end{align}
with 
\begin{align}
    V_A^{(\mathbb L,\tau)\pm} &= -2\Big((\nabla^2-1)\delta_{A}^B-2\nabla_A\nabla^B\Big)V_B^{(\mathbb L)\pm},\\
    V^{(\mathbb S)\pm} &= \nabla^AV_A^{(\mathbb{S})\pm}-3\nabla^AV_A^{(\mathbb{L})\pm}.
\end{align}
The arbitrary vectors on the sphere $V_A^{(\mathbb L)\pm}(x^A)$ and $V_A^{(\mathbb S)\pm}(x^A)$ obey the following antipodal matching conditions :
\begin{align}
     &V_A^{(\mathbb S)+}(x^A)- O^{(q)n=0}_V(V_A^{(\mathbb L)+}(x^A)) = +\Upsilon^*\Big(V_A^{(\mathbb S)-}(x^A) -  O^{(q)n=0}_V(V_A^{(\mathbb L)-}(x^A))\Big), \label{antipodalVa}\\
     &V_A^{(\mathbb L)+}(x^A)= -\Upsilon^*V_A^{(\mathbb L)-}(x^A). \label{antipodalVa2}
\end{align}
The operator $O^{(q)n=0}_V$ is explicitly given by \cite{Compere:2025bnf}: 
\begin{align}
    &O^{(q)n=0}_V[f_A(x^A)] =-\sum_{\ell=1}^{+\infty}\sum_{m=-\ell}^{+\ell} \Big[(-\ell)(\ell+1)(1 - 2 H_{\ell} )-2\Big]  \mathring V^{(E)\ell m}_A(x^A)\int_{S^2}d\Omega' \overline{\mathring V^{(E)\ell m}_C(x^{A'})}f^C(x^{A'})\nn\\&\qquad+\sum_{\ell=1}^{+\infty}\sum_{m=-\ell}^{+\ell} (-\ell)(\ell+1)(1 - 2 H_{\ell} )\mathring V^{(B)\ell m}_A(x^A)\int_{S^2}d\Omega' \overline{\mathring V^{(B)\ell m}_C(x^{A'})}f^C(x^{A'}),
\end{align}
where $f_A(x^A)$ is an arbitrary vector on the sphere and $H_{\ell}$ is the $\ell$-th harmonic number as defined below Eq. \eqref{defOq} above. We can rewrite this operator in a symmetric fashion under the exchange of $E$ and $B$ after using the completeness of the vector harmonic decomposition as 
\begin{align}\label{defOVq}
    &O^{(q)n=0}_V[f_A(x^A)] =-\sum_{\ell=1}^{+\infty}\sum_{m=-\ell}^{+\ell} \Big[(-\ell)(\ell+1)(1 - 2 H_{\ell} )-1\Big]  \mathring V^{(E)\ell m}_A(x^A)\int_{S^2}d\Omega' \overline{\mathring V^{(E)\ell m}_A(x^{A'})}f^A(x^{A'})\nn\\&\qquad+\sum_{\ell=1}^{+\infty}\sum_{m=-\ell}^{+\ell} \Big[(-\ell)(\ell+1)(1 - 2 H_{\ell} )-1\Big]\mathring V^{(B)\ell m}_A(x^A)\int_{S^2}d\Omega' \overline{\mathring V^{(B)\ell m}_A(x^{A'})}f^A(x^{A'})+f_{A}(x^A). 
\end{align}
As shown in \cite{Compere:2026jmk}, $I_a$ converges as $\tau\to\pm\infty$. Therefore, $V_A^{(\mathbb L)\pm}(x^A)=0$ and $I_a$ has $p$-parity. 

Using the asymptotics of the convergent definition of $\sigma$, the $p$-parity $I_{a}$ and $p$-parity $k^{(B)}_{AB}$, we get that the source of $J_a$ behaves as follows
\begin{align}
    S_{\tau}(\tau\to\pm\infty,x^A)&=\mp 16 \nabla^AS^{(0)\pm}_A(x^A) e^{-3\vert\tau\vert} + o(e^{-3\vert\tau\vert})\label{sourceBehav1taubis},\\
    S_{A}(\tau\to\pm\infty,x^A)&=  4S^{(0)\pm}_A(x^A)e^{-\vert\tau\vert} + o(e^{-\vert\tau\vert}).
\end{align}
The source $S_A^{(0)\pm}(x^A)$ will be explicitly provided in Eq. \eqref{SAAsympt}. 

Since $J_a$ obeys an inhomogeneous equation, one can split the solution into a homogeneous piece $\hat J_a$ and an inhomogeneous piece $J^{(S)}_a$. The ambiguities in the splitting are fixed by choosing the \emph{primitive inhomogeneous solution} as defined in Appendix \ref{sec:asymptBehav}, which allows to extract antipodal matching conditions from the homogeneous piece of the solution.  The asymptotic behavior of $J^{(S)}_a$ is found to be 
\begin{align}
    J^{(S)}_{\tau}(\tau\to\pm\infty,x^A) =&\pm 4 \nabla^A\mathcal{S}_A^{(p)\pm}e^{-\vert\tau\vert}+o\left(e^{-2\vert\tau\vert}\right),\\
    J^{(S)}_{A}(\tau\to\pm\infty,x^A)  =& \mathcal{S}_A^{(p)\pm}e^{\vert\tau\vert}+2 \Big[\Big(\vert\tau\vert+\frac{1}{2}\Big)S_{A}^{(0)\pm}(x^A)-S_{A}^{(0)E\pm}(x^A)-\mathcal{S}_A^{(q)\pm}\nn\\ &-\vert\tau\vert\Big((\nabla^2-1)\delta_{A}^B-2\nabla_A\nabla^B\Big)\mathcal{S}_B^{(p)\pm}\nn\\ &-\frac{1}{2}O^{(q)n=0}_V(\mathcal{S}_A^{(p)\pm})+\frac{1}{2}\mathcal{S}_A^{(p)\pm}-\nabla_A\nabla^B\mathcal{S}_B^{(p)\pm}\Big] e^{-\vert\tau\vert}+o\left(e^{-2\vert\tau\vert}\right),\label{SourceContribVA}
\end{align}
where $S_{A}^{(0)E\pm}(x^B)$ is the E-parity piece of $S_{A}^{(0)\pm}(x^B)$ and $\mathcal{S}_A^{(q)\pm}(x^B)$, $\mathcal{S}_A^{(p)\pm}(x^B)$ are defined from an integral over $dS_3$, see Eqs. \eqref{SpExplicit}-\eqref{SqExplicit}. The quantities $\mathcal{S}_A^{(q)\pm}$ vanish when $S_a$ is an $n=0$ $p$-parity vector and $\mathcal{S}_A^{(p)\pm}$ vanish when $S_a$ is $n=0$ q-parity vector. The homogeneous piece $\hat J_a$ admits the behavior \eqref{Vtaubehav}-\eqref{VAbehav} where the arbitrary functions satisfy the antipodal matching conditions \eqref{antipodalVa}.

\section{Matching between null and spatial infinity}
\label{sec:matching}
 
Asymptotically flat spacetimes have been defined by consistently matching their five asymptotic infinities such that they share a single Bondi-Metzner-Sachs (BMS) group of asymptotic symmetries and associated charges \cite{Compere:2023qoa} following the conjecture of \cite{Strominger:2013jfa}. This matching procedure allows to relate the charges defined at the corners of spatial infinity $i^0_\pm$ (defined as $\rho \to \infty \text{ at fixed $\tau$ then }  \tau \to \pm \infty$) to the asymptotic values of charges defined in the limit $u^\pm \to -\infty$ of null infinities, namely at $\mathcal I^+_-$ and $\mathcal I^-_+$. Note that the matching between null and timelike infinity at $\partial i^+=\mathcal I^+_+$ and $\partial i^- = \mathcal I^-_-$  in the limit $u^\pm \to + \infty$ has been partially derived at subleading order in \cite{Boschetti:2026gfd}. 

There is a bijection between the independent corner fields defined at $i^0_+/i^0_-$ and the conserved charges, see Section \ref{sec:charges}. We will already use in this section this property to denominate the corner fields as the conserved charges. 

There are many additional charges than just the BMS charges at spatial infinity. Our complete set of leading and subleading corner fields/conserved charges comprises 3 sets of leading fields and 3 sets of subleading fields defined for each spherical harmonic $(\ell,m)$, see Table \ref{tab:chargesnull}. The corresponding 3 sets of leading order charges will be defined in Eqs. \eqref{ch1}-\eqref{ch3} and the corresponding 3 sets of subleading order charges will be defined in Eqs. \eqref{Iacharges}-\eqref{Jacharges}.

We expect that a tower of charges can be defined at subsequent subleading orders but our analysis will be limited to subleading order in this paper.  

To proceed with the matching of charges between null infinities and spatial infinity, we need to relate the Beig-Schmidt fields to the Bondi-
Sachs fields at the corners $\scri^+_-$ and $\scri^-_+$. 

\subsection{Matching at the corners $i^0_+=\mathcal I^+_-$ and $i^0_-=\mathcal I^-_+$}
\label{matchIp}

We recall that the Bondi-Sachs fields take the asymptotic form \eqref{falloffIplus}. Accordingly, we start with the ansatz that the change of coordinates between Bondi coordinates and Beig-Schmidt coordinates in their overlapping range of validity takes the form 
\begin{align}
    u &= - e^{-\tau} \rho + u^{(1)}[\tau,y^A] + \frac{u^{(2)}[\tau,y^A]}{\rho}+ \frac{\log{\rho}~u^{(log,2)}[\tau,y^A]}{\rho}+o(\rho^{-1})\nonumber,\\
    r &= \rho \cosh\tau + r^{(1)}[\tau,y^A] + \frac{r^{(2)}[\tau,y^A]}{\rho}+ \frac{\log{\rho}~r^{(log,2)}[\tau,y^A]}{\rho}+o(\rho^{-1})\label{ChangeOfCord},\\
    x^A &= y^A + \frac{x^{(1)\,A}[\tau,y^A]}{\rho} + \frac{x^{(2)\,A}[\tau,y^A]}{\rho^2}+ \frac{\log{\rho}~x^{(log,2)\,A}[\tau,y^A]}{\rho^2}+o(\rho^{-2})\nonumber.
\end{align}

We view this change of coordinates as a global change of coordinates analogous to a change from Cartesian to spherical coordinates which is not a canonical transformation that affects the charges. In particular, we require that neither translation nor proper supertranslation is performed during that global coordinate change, i.e. that the supertranslation frame is unchanged. This imposes fixing a condition on $u^{(1)}[\tau,y^A]$. We require that 
\begin{equation}
    u^{(1)}[\tau,y^A] = o((e^{\tau})^0),
\end{equation}
as $\tau \to\pm\infty$ and we impose that $T^-(x^A) = - \Upsilon^* T^+(x^A)$. 

After imposing the Beig-Schmidt gauge conditions, we find a unique asymptotic expansion near the corners for all these coefficients up to a $p$-parity spi-supertranslation and a logarithmic translation, i.e., the coordinate change is determined at each corner up to one arbitrary function $w^\pm(x^A)$ on the sphere, which determines the even spi-supertranslation frame, and 4 arbitrary constants $L^{\pm \mu}$, $\mu=0,1,2,3$, which determine the logarithmic translation frame. 
The logarithmic translation frame is uniquely chosen such that $\sigma^\pm$ converge. Hence, the logarithmic translation frame differs between the corners, according to Eq. \eqref{etaH0}. The choice of the even spi-supertranslation frame is left arbitrary, as it is pure gauge and not relevant for the asymptotic matching.

The map between the two asymptotic solutions can then be asymptotically achieved at $\mathcal I^+_-$ and $\mathcal I^-_+$. The leading order Beig-Schmidt fields take the following asymptotic form in terms of the asymptotic Bondi-Sachs fields: 
\begin{align}
     \Phi &= \frac{1}{2} C^{\pm(0)} e^{\vert\tau\vert}-\frac{1}{2} (\nabla^2+1)C^{\pm(0)}e^{-\vert\tau\vert}+\left(w^{\pm}(y^A)+\frac{1}{2}\vert\tau\vert \nabla^2(\nabla^2+2)C^{\pm(0)}\right)e^{-3\vert\tau\vert}+ o(e^{-3\vert\tau\vert}),\label{MatchPhi}\\
    \sigma &= 2 m^{(0)\pm} e^{-3\vert\tau \vert}  + o(e^{-3\vert\tau\vert})\label{sigmascri+-},\\
   k_{\tau\tau} &= \left(8 \nabla^{A} \nabla^B C_{AB}^{\pm (0)}(\vert\tau\vert-\frac{3}{4})- 16 ~w^{\pm}(y^A) \right)  e^{-3\vert\tau\vert}+ o(e^{-3\vert\tau\vert}),\label{ktautau}\\
    k_{\tau A} &=\pm 2 \nabla^B C_{AB}^{\pm(0)} e^{-\vert\tau\vert} + o(e^{-\vert\tau\vert}),\\
    k_{AB} & = \frac{1}{2} C_{AB}^{\pm(0)} e^{\vert\tau\vert}+ o(e^{\vert\tau\vert})\label{kabscri+-bis}. 
\end{align}

The subleading proper-supertranslation-invariant Beig-Schmidt fields $I_a$ and $J_a$ take the following asymptotic form
\begin{align}
I_{\tau}&=o\left(e^{-2\vert\tau\vert}\right),\\
I_{ A} & = \pm 2 e^{- \vert\tau\vert} \mathcal{D}_A^{\pm(0)}+o\left(e^{- \vert\tau\vert}\right) ,\label{IetaabtauA+}\\
J_{\tau}&=2 \nabla^A\nabla^B\mathcal{C}^{\pm(1)}_{AB}e^{-\vert\tau\vert}+o\left(e^{-2\vert\tau\vert}\right),\\
J_{A}  &=\pm\frac{1}{2}\nabla^B\mathcal{C}^{\pm (1)}_{AB}e^{\vert\tau\vert}\nn\\&\pm2\Big(\vert\tau\vert (\nabla^B D_{AB}^{\pm(0)}+\frac{1}{2} \nabla_C \nabla_{\langle A} \nabla_{B\rangle} C^{\pm (1)\;BC})+N^{\pm(0)}_A\nn\\& -3 \mathcal{M}^{\pm(0)}_{AB}\nabla^BC^{\pm(0)}-C^{\pm(0)}\nabla^B\mathcal{M}^{\pm(0)}_{AB}-\frac{1}{4}C^{\pm(B)(0)\,B}_{A}\nabla^CC^{\pm(B)(0)}_{CB}-\frac{1}{16}\nabla_A (C^{\pm(B)(0)\,C}_{D}C^{\pm(B)(0)\,D}_{C})\nn\\&-(\log 2+1)\nabla^BD_{AB}^{\pm(0)}+\nabla_C \nabla_{\langle A} \nabla_{B\rangle}\mathcal{C}^{\pm(1)\,BC}-\frac{1}{2}\nabla_A \nabla^B\nabla^C \mathcal{C}^{\pm(1)}_{BC}\nn\\&+\frac{3}{2}\nabla_C \nabla_{\langle A} \nabla_{B\rangle}\Big((-E\pm P_i n_i)C^{\pm(B)(0)\,BC}\Big)-\frac{1}{4}\nabla_A \nabla^B\nabla^C \Big((-E\pm P_i n_i)C^{\pm(B)(0)}_{BC}\Big)\nn\\& \pm 2 P_i \mathcal{\tilde{M}}^{\pm(0)}\epsilon_{AB}\nabla^Bn_i\mp \frac{1}{4} P_i\nabla^B n_i C^{\pm(B)(0)}_{BA}+\frac{1}{4}(E\pm3P_in_i)\nabla^B C^{\pm(B)(0)}_{BA}\Big)e^{-\vert\tau\vert}+o\left(e^{- \vert\tau\vert}\right)\label{JetaABscri+-} ,
\end{align}
where $\mathcal{D}_A^{\pm(0)}$ is defined in Eq. \eqref{defDA} and $\mathcal{C}^{\pm (1)}_{AB}$ is defined in Eq. \eqref{defCAB1}. 

In order to perform the matching, we also need to compute the asymptotics of $S_a$. Let us define similarly to Eq. \eqref{defMAB} the leading order covariant mass tensor 
\begin{align}
 \mathcal M_{AB}^{\pm (0)} = m^{\pm (0)}\gamma_{AB} + \tilde{\mathcal M}^{\pm (0)} \epsilon_{AB} .\label{defMAB2}  
\end{align}
We note the identities 
\begin{align}
C^{\pm(1)BC}+2(-E\pm P_i n_i)C^{\pm(B)(0)\,BC} &=\mathcal C^{\pm(1)BC}+(-E\pm P_i n_i)C^{\pm(B)(0)\,BC}, \\
\pm6 P_i\tilde{\mathcal{M}}^{\pm(0)}\epsilon_{AB}\nabla^B n_i-2(E\mp P_in_i)\epsilon_{AB}\nabla^B\tilde{\mathcal{M}}^{\pm(0)} &=\nabla_C\nabla_{\langle A}\nabla_{B\rangle}\Big((-E\pm P_i n_i)C^{\pm(B)(0)\,BC}\Big). 
\end{align}
Using these identities and the definition \eqref{defDA} we deduce 
\begin{align}
&S_{\tau}=o\left(e^{-2\vert\tau\vert}\right),\\
&S_A= 4e^{-\vert\tau\vert}S^{(0)}_A + o\left(e^{- \vert\tau\vert}\right),
\end{align}
where 
\begin{align}
    S^{(0)}_A :=&\pm \left(\mathcal{D}_A^{\pm(0)}\mp 6 P_i (m^{\pm (0)}\nabla_A n_i +2\tilde{\mathcal{M}}^{\pm(0)}\epsilon_{AB}\nabla^B n_i)+ 2 (E\mp P_i n_i) (\nabla_A m^{\pm (0)}+2\epsilon_{AB}\nabla^B\tilde{\mathcal{M}}^{\pm(0)})\right)\nn\\=&\pm \left(\nabla^BD_{AB}^{\pm(0)}-\frac{1}{2}\nabla_C\nabla_{\langle A}\nabla_{B\rangle}\Big(C^{\pm(1)BC}+2(-E\pm P_i n_i)C^{\pm(B)(0)\,BC}\Big)\right)\nn\\=&\pm \left(\nabla^BD_{AB}^{\pm(0)}-\frac{1}{2}\nabla_C\nabla_{\langle A}\nabla_{B\rangle}\mathcal C^{\pm(1)BC}-\frac{1}{2}\nabla_C\nabla_{\langle A}\nabla_{B\rangle}\Big((-E\pm P_i n_i)C^{\pm(B)(0)\,BC}\Big)\right). \label{SAAsympt}
\end{align}

\subsection{Antipodal matching between $\mathcal I^+_-$ and $\mathcal I^-_+$}
\label{sec:antipodalmatch}

After having recognized how the asymptotic Bondi-Sachs fields as $u^\pm \to -\infty$ appear in the asymptotic expansion of the Beig-Schmidt fields as $\tau \to \pm \infty$, we can now use the properties of $p$ and $q$ harmonic solutions of the Beig-Schmidt fields on $dS_3$ that relate the $\tau\to\infty$ and $\tau \to - \infty$ behaviors to deduce the antipodal relationships of the asymptotic Bondi-Sachs fields between $\mathcal I^+_-$ and $\mathcal I^-_+$. We will proceed in sequential order (leading then subleading order) at the corners.

\begin{table}[!htbp]
    \centering
    \begin{tabular}{|r|c|c|c|c|}\hline 
        & Field & $dS_3$ type & $\Upsilon^*_{\mathcal H}$ eigenvalue & Corresponding field at $\tau\to\pm\infty$ \\ \hline\hline
Leading &  $\sigma$  & $p$ & $+$ & $m^{\pm(0)}$\\ \cline{2-5} 
        & $\Phi$ & $q$ & $-$ & $C^{\pm(0)}$ \\ \cline{2-5}
        &$k_{ab}^{(B)}$ & $p$ & $+$ & $C^{\pm (0)(B)}_{AB}$ $\Leftrightarrow$ $\tilde {\mathcal M}^{\pm (0)}$\\        \hline
Subleading & $I_{ab}$ & $p$ & $-$ & $\mathcal D_A^{\pm(0)}$ \\ \cline{2-5} 
        & $J_{ab}$  & $q$ & $+$ & $\mathcal N_A^{ (\log)\pm(0)}$ \\  \cline{3-5} 
        &           & $p$ & $-$ &  $\mathcal N_A^{ \text{inv}\pm(0)}$ \\ \hline 
      \end{tabular}
    \caption{Complete set of fields at spatial infinity leading to asymptotically conserved charges. For fields obeying inhomogeneous equations such as $J_{ab}$, the $\text{dS}_3$ type corresponds to $\hat J_{ab}$, the homogeneous piece of $J_{ab}$, after subtraction of the primitive inhomogeneous solution $J_{ab}^{(S)}$.}
    \label{tab:antipodal}
\end{table}

\subsubsection{Leading order}

As summarized in Table \ref{tab:antipodal}, the asymptotic Bondi mass aspect $m^{\pm (0)}$ and the asymptotic dual Bondi mass aspect $\tilde {\mathcal M}^{\pm (0)}$ are the asymptotic values of the $p$-parity part of the fields $\sigma$ and $k_{ab}^{(B)}$ on $dS_3$, respectively. Instead, the supertranslation frame $C^{\pm (0)}$ is the asymptotic value of the $q$-parity part of the field $\Phi$ on $dS_3$. It implies that 
\begin{align}
   \Upsilon^* m^{+(0)} &= m^{-(0)},\label{idm0} \\ 
   \Upsilon^* C_{AB}^{+(0)(E)} &= -  C_{AB}^{-(0)(E)} \qquad \Leftrightarrow \qquad  \Upsilon^* C^{+(0)} = -  C^{-(0)}, \label{idC0} \\ 
   \Upsilon^* C_{AB}^{+(0)(B)} &=   C_{AB}^{-(0)(B)}\qquad \Leftrightarrow \qquad  \Upsilon^* {\tilde {\mathcal M}}^{+(0)} =  -{\tilde {\mathcal M}}^{-(0)}. \label{idM0}
\end{align}
Using Eq. \eqref{defMAB2}, we can write Eqs. \eqref{idm0} and \eqref{idM0} as 
\begin{align}
\Upsilon^* { {\mathcal M}}^{+(0)}_{AB} =  {{\mathcal M}}^{-(0)}_{AB}    . \label{idMAB0}
\end{align}
Supertranslations act on $C_{AB}^{\pm (0)}$ as $\delta_T C_{AB}^{\pm (0)}= -2 D_{\langle A} D_{B\rangle} T^\pm$ where $T^-(x^A) = - \Upsilon^* T^+(x^A)$.

Let us comment on how our result fits within the literature. 
The leading order matching of $m^{\pm (0)}$ and $C^{\pm (0)}$ was originally conjectured by Strominger in \cite{Strominger:2013jfa} in the context of massless scattering without any massive asymptotic states, see also earlier work \cite{Ashtekar:1978zz,PhysRevLett.43.649,1979JMP....20.1362A,Herberthson:1992gcz} in different contexts.  The formal matching of $m^{\pm (0)}$ across spatial infinity was first performed by Troessaert \cite{Troessaert:2017jcm} in linearized gravity, see also \cite{1998JGP....24...83F,Compere:2011ve,Prabhu:2019daz,Magdy:2021rmi}. A matching in non-linear Einstein gravity under too restrictive assumptions (which remove the spacetime mass) was performed in \cite{Capone:2022gme}. The matching of $m^{\pm (0)}$ and $C^{\pm (0)}$ was performed in Einstein gravity in \cite{Compere:2023qoa} in an arbitrary supertranslation and logarithmic translation frame and for arbitrary scattering including massive states. The matching of $\tilde{\mathcal M}^{(0)}$ was conjectured in \cite{Kol:2019nkc} and formally proven in \cite{Compere:2026jmk}. Our result therefore reproduces the known results of the literature (demonstrated in full general relativity and with generic and physically motivated boundary conditions) at leading order.

\subsubsection{Subleading order}

As summarized in Table \ref{tab:antipodal}, the asymptotic fields $\mathcal D^{\pm (0)}_A$, $\mathcal{N}_A^{ (\log)\pm(0)}$ and $\mathcal{N}_A^{\text{inv}\pm (0)}$ are obtained from the asymptotic values of the $p$-parity part of the field $I_{ab}$ and either the $q$-parity or $p$-parity homogeneous part of the field $J_{ab}$ after subtraction of the primitive inhomogeneous solution, respectively. 

First, we read off $\mathcal D^{\pm (0)}_A$ from $I_{A}$ using Eq. \eqref{IetaabtauA+} and use the definition \eqref{defIa}. We obtain
\begin{align}
\Upsilon^{*} \mathcal D_A^{+ (0)} &= -\mathcal D_A^{-(0)},\label{cDABid} 
\end{align}
The subleading order matching of $\mathcal D_A^{\pm (0)}$ was obtained in \cite{Compere:2026jmk} and was shown to be consistent with alternative results \cite{Boschetti:2026ogm}.

Second, we first read off $\nabla^B\mathcal{C}^{\pm (1)}_{AB}$  from the leading $e^{\vert\tau\vert}$ behavior of $J_A$ in Eq. \eqref{JetaABscri+-}. This term identifies with the sum of $V_A^{(\mathbb L)\pm}$ from the homogeneous solution \eqref{VAbehav} and $\mathcal S^{(p)\pm}_A$ from the primitive inhomogeneous solution \eqref{SourceContribVA}. The source contribution $\mathcal S^{(p)\pm}_A$ can be expressed in terms of the source term $\mathcal S^{(0)\pm}_A$ as Eq. \eqref{SpLocalized} derived in Appendix \ref{sec:localization}.
We can therefore subtract this primitive inhomogeneous piece to $\nabla^B\mathcal{C}^{\pm (1)}_{AB}$ in order to isolate the $q$-parity homogeneous solution obeying an odd antipodal map \eqref{antipodalVa2}. This gives the antipodal map : 
\begin{equation}
    \Upsilon^{*} (\nabla^B\mathcal{C}^{+ (1)}_{AB}-2\mathcal{S}_A^{(p)+}) = \nabla^B\mathcal{C}^{- (1)}_{AB} + 2 \mathcal{S}_A^{(p)-} \label{nonlocalmapC1}
\end{equation}
where $\mathcal{S}_A^{(p)\pm} $ depend only on the p-parity piece of the source, i.e. $I_a$ and are related by Eq. \eqref{AntipodMatchalS}. This recovers the results obtained in \cite{Compere:2026jmk}. Using Eq. \eqref{IetaabtauA+} and the equation of motion $(\Box-1)I_a=0$, we deduce that $\mathcal{S}_A^{(p)\pm}$ only depends on $\mathcal{D}_A^{\pm(0)}$, i.e. $\mathcal{S}_A^{(p)\pm}(x^A)=\mathcal{S}_A^{(p)\pm}[\mathcal{D}_A^{\pm(0)}]$. Now, the source is related to the leading behavior of $I_a$ and therefore $\mathcal{D}_A^{\pm(0)}$ by the non-local operator over the sphere $K^{-1}$ (see Eq. \eqref{KOperator}) as demonstrated in Appendix \ref{sec:localization}.  In order to write a local antipodal identification, we now act with the operator $K[.]$ on the antipodal map \eqref{nonlocalmapC1} in order to cancel the operator $K^{-1}$ obtained in the map \eqref{SpLocalized}. The local antipodal map 
\begin{align}
\Upsilon^{*} \mathcal{N}_A^{ (\log)+(0)} &= \mathcal{N}_A^{(\log)- (0)},\label{cCABid} 
\end{align}
is then obtained after using the definition \eqref{defNAlog} and the property \eqref{propC}. This local form \eqref{cCABid} of the antipodal map \eqref{nonlocalmapC1} is a new result of this paper as compared with \cite{Compere:2026jmk}. 

Third, $\mathcal{N}_A^{\text{inv}\pm (0)}$ is extracted from $J_A$ using Eq. \eqref{JetaABscri+-}.
We first compute the asymptotics of $\hat J_a$ by removing the  contributions from the primitive inhomogeneous solution, with the source \eqref{SAAsympt}, using Eq. \eqref{SourceContribVA}. We can therefore identify both  $V_A^{(\mathbb S)\pm}$ and $V_A^{(\mathbb L)\pm}$. Then, using Eqs. \eqref{VAbehav}-\eqref{antipodalVa}, the quantity that $\mathcal{N}_A^{\text{inv}\pm (0)}$ that obeys the antipodal map is proportional to $V_A^{(\mathbb S)\pm}-O^{(q)n=0}_VV_A^{(\mathbb L)\pm}$. This leads precisely to the definition \eqref{Ninv0}-\eqref{curlyN0}, which has now been derived. The quantity $\mathcal{N}_A^{\text{inv}\pm (0)}$ satisfies by construction the odd antipodal matching condition:
\begin{align}
\Upsilon^{*}\mathcal{N}_A^{\text{inv}+(0)} &=-\mathcal{N}_A^{\text{inv}-(0)} . \label{NAinvid}
\end{align}
In Eqs. \eqref{Ninv0}-\eqref{curlyN0}, the $q$-parity source contribution $ S^{\pm(q)}_A$ only depends upon the initial data for $\sigma$ and $k_{ab}^{(B)}$ therefore formally depends upon $x^A$ through $\mathcal M_{AB}^{-(0)}$, i.e. $S^{\pm(q)}_A(x^B) = S^{\pm(q)}_A[\mathcal M_{AB}^{-(0)}(x^B)]$.

Note that the antipodal relations Eqs. \eqref{antipodalVa}-\eqref{antipodalVa2} for $V_A^{(\mathbb S)\pm}$ (resp. $V_A^{(\mathbb L)\pm}$) are even (resp. odd) but the antipodal relations for $\mathcal{N}_A^{\text{inv}\pm (0)}$ and $\mathcal{D}_A^{\pm (0)}$ (resp. $\mathcal{N}_A^{ (\log)\pm(0)}$) are odd (resp. even) because of the global $\pm$ factor in Eqs. \eqref{IetaabtauA+}-\eqref{JetaABscri+-}.

In the definition \eqref{Ninv0}, the field $\mathcal M_{AB}^{\pm (0)}$ is identified antipodally with a plus sign while $C^{\pm (0)}$ is identified antipodally with a minus sign, see Eqs. \eqref{idC0}-\eqref{idMAB0}. Their product is therefore identified antipodally with a minus sign and we deduce from Eqs. \eqref{cCABid} and \eqref{NAinvid} that 
\begin{align}
\Upsilon^{*} (\nabla^B D_{AB}^{+(0)} + \frac{1}{2} \nabla_C \nabla_{\langle A} \nabla_{B\rangle} C^{+ (1) BC}) &= \nabla^B D_{AB}^{-(0)} + \frac{1}{2} \nabla_C \nabla_{\langle A} \nabla_{B\rangle} C^{-(1) BC}, \label{idC4}\\ 
\Upsilon^{*}\mathcal{N}_A^{+(0)} &=-\mathcal{N}_A^{-(0)}.  \label{NAlaw}
\end{align}
Using the definition \eqref{NALog}, the antipodal matching condition \eqref{idC4} can be written as  
\begin{equation}
   \Upsilon^* (\nabla^B D_{AB}^{+(0)}-N_A^{+(\log)})=\nabla^B D_{AB}^{-(0)}-N_A^{-(\log)}.\label{NAlogMatch}
\end{equation}

The matching of $\mathcal{N}_A^{\text{inv}\pm(0)}$ or, equivalently, of $\mathcal{N}_A^{\pm(0)}$ is the main novel result of this paper. It generalizes the matching proved in \cite{Capone:2022gme} for the case $C^{(1)}_{AB}=C^{(B)(0)}_{AB}=D^{(0)}_{AB}=0$. Eq. \eqref{NAlaw} also reproduces the antipodal matching for the Lorentz charges when considering the $\ell=1$ harmonics of $\mathcal{N}_A^{\pm(0)}$.

\section{Asymptotically conserved charges at spatial infinity}
\label{sec:charges}

At spatial infinity, it is possible to construct conserved charges from first order and second order fields \cite{Compere:2011ve}. In \cite{Compere:2025bnf,Compere:2026jmk}, the concept of \emph{asymptotically conserved charge at the corners $i^0_\pm$ of spatial infinity} $i^0$ was introduced. A charge is asymptotically conserved at the corners $\tau \to \pm \infty$ of $i^0$ if its asymptotic values at $\tau\to\pm\infty$ are identical, even though its values at intermediate $\tau$ might differ. An asymptotically conserved charge at the corners of spatial infinity is necessary and sufficient to define an asymptotic conservation law between boundary fields at $\mathcal I^+_-$ and $\mathcal I^-_+$ after using the matching conditions at $\mathcal I^+_-= i^0_+$ and $\mathcal I^-_+= i^0_-$. Note that each charge defined from a $dS_3$ field at spatial infinity corresponds to two distinct charges at the corners after splitting the field into $p$-parity and $q$-parity parts. Such charges might trivially vanish in the absence of a given parity. The charges are furthermore conserved over all $dS_3$ if the $dS_3$ fields obey a homogeneous equation.

Let us discuss successively the leading order and subleading order charges. The complete set of charges defined at leading order in the large $\rho$ expansion is   
\begin{align}   Q_{\chi}^T[k^{(B)}_{ab}]&:= \int_{S^2} \frac{d^2x}{4 \pi} \sqrt{-q} n_a \bigg(\chi_{bc} \mathcal{D}^a k^{(B)bc}- k^{(B)bc}\mathcal{D}^a \chi_{bc} \bigg),\label{ch1}\\
    Q^S_{\omega}[\sigma] &:= \int_{S^2} \frac{d^2x}{4 \pi} \sqrt{-q} n_a \bigg(\omega \mathcal{D}^a \sigma- \sigma \mathcal{D}^a \omega\bigg),\label{ch2}\\
    Q^S_{H}[\Phi] &:= \int_{S^2} \frac{d^2x}{4 \pi} \sqrt{-q} n_a \bigg(H \mathcal{D}^a \Phi- \Phi \mathcal{D}^a H\bigg),\label{ch3}
\end{align}
where $\chi_{bc}$ is a SDT tensor obeying $(\Box-3)\chi_{bc}=0$, $\omega$ and $H$ are scalars obeying $(\Box+3)\omega=0$ and $(\Box+3)H=0$. At leading order, all charges are conserved because the tensors satisfy homogeneous equations, see Eqs. \eqref{BSEquationOrder1}.

We deduced from the matching conditions that $k_{ab}^{(B)}$ and $\sigma$ have a definite $p$-parity while $\Phi$ has both parities but only the $q$-parity leads to a canonical charge while the $p$-parity charge is pure gauge, see Table \ref{tab:antipodal}. By orthogonality of tensor harmonics, the charges are non-vanishing only if the symmetry generator has the opposite parity than the field considered. This defines the non-trivial conserved charges. 
The charges $Q_{\chi}^T[k^{(B)}_{ab}]$ are then the dual supertranslation charges \cite{Compere:2026jmk}, $Q^S_{\omega}[\sigma]$ are the supertranslation charges \cite{Compere:2011ve} and $Q^S_{H}[\Phi]$ are the super-logarithmic translation charges \cite{Fuentealba:2022xsz,girelli2026covariantformulationlogarithmicsupertranslations}.

At subleading order in the large $\rho$ expansion, the Beig-Schmidt expansion defines 2 $dS_3$ fields: $I_{ab}$ and $J_{ab}$. Only $I_{ab}$ obeys a homogeneous equation, see Eqs. \eqref{eq:99b}-\eqref{BoxJab2}. We deduced from the matching conditions that $I_{ab}$ has $p$-parity while $J_{ab}$ admits both parities, see Table \ref{tab:antipodal}. Hence, we have one set of conserved charges and 2 sets of asymptotically conserved charges at subleading order. Given our construction of the fields $I_{ab}$ and $J_{ab}$, see Eqs. \eqref{defIab}-\eqref{defJab}, the charges are supertranslation and logarithmic translation invariant. 

For any vectors $V_a$, $W_a$, the Klein-Gordon product
\begin{equation}
   Q^V_W[V_{a}]:=  \int_{S^2} \frac{d^2x}{8 \pi} \sqrt{-q} n_a \bigg(W^{b}\mathcal{D}^a V_{b}-V_{b} \mathcal{D}^a W^{b} \bigg)\label{Vacharges}
\end{equation}
is conserved if $V_{a}$ and $W_a$ satisfy the same equation of the form $(\Box-1)W_a=(\Box-1)V_a=0$. Moreover, for any vectors $V_a$, $W_a$, the second Klein-Gordon product
\begin{equation}
   Q^{V(2)}_W[V_{a}]:=  Q^V_W[(\square-1) V_{a}] + Q^V_{(\square-1) W}[V_{a}]\label{Vachargesv2}
\end{equation}
is conserved if $V_{a}$ and $W_a$ satisfy the same equation of the form $(\Box-1)^2W_a=(\Box-1)^2V_a=0$.

Let us recall $I_a := \cosh \tau n^b I_{ab}$, $J_a := \cosh \tau n^bJ_{ab}$ and the split $J_a = J^{(S)}_a + \hat J_a$ where $ \hat J_a$ is the homogeneous solution after subtraction of the primitive inhomogeneous solution $J^{(S)}_a$. We provide an algorithmic procedure to define $J^{(S)}_a$ asymptotically for large $\vert\tau\vert$ in Appendix \ref{sec:asymptBehav}. In principle, if we can define $J^{(S)}_a$ exactly, we can define the complete set of exactly conserved subleading charges as
\begin{align}
   Q^V_\chi[I_{a}] &=  \int_{S^2} \frac{d^2x}{8 \pi} \sqrt{-q} n_a \bigg(\chi^{b}\mathcal{D}^a I_{b}-I_{b} \mathcal{D}^a \chi^{b} \bigg)\label{Iacharges},\\
   Q^V_\chi[\hat J_{a}] &=  \int_{S^2} \frac{d^2x}{8 \pi} \sqrt{-q} n_a \bigg(\chi^{b}\mathcal{D}^a \hat J_{b}-\hat J_{b} \mathcal{D}^a \chi^{b} \bigg)\label{Jacharges},
\end{align}
where $\chi_a$ satisfies $(\Box-1)\chi_a=0$. Since $I_a$ has $p$-parity, the charge \eqref{Iacharges} is non-vanishing only when $\chi_a$ has $q$-parity. It defines the \emph{leading peeling-breaking charge} \cite{Compere:2026jmk}. The conserved charge \eqref{Jacharges}  for a $p$-parity $\chi_a$ defines the \emph{mixed tail/peeling-breaking  charge}. 

We see from Eq. \eqref{SourceSab} that $S_{ab}-2I_{ab}$ is even, i.e., it is a $q$-parity tensor. Since the $p$-parity piece of the source $S_a$ is $2 I_a$ which obeys $(\Box-1)I_a=0$, we have that $(\Box-1)^2J_a$ is only sourced by a $q$-parity term $S^{(q)}_a := S_a-2I_a$. Let $\Phi^{(p)}_a$ be a transverse $p$-parity vector that satisfies $(\Box-1)^2\Phi^{(p)}=0$ and $(\Box-1)\Phi^{(p)}\neq 0$ as defined in Appendix \ref{sec:inhom}. Using Eq. \eqref{PartialtauOmega}, we compute 
\begin{align}
    \lim_{\tau\to+\infty} Q^{V (2)}_{\Phi^{(p)}}[J_a]-\lim_{\tau\to-\infty} Q^{V (2)}_{\Phi^{(p)}}[J_a] &= -\frac{1}{8\pi}\int_{dS_3} d^3\phi \sqrt{-q}\,\Phi^{(p)}_a (\Box-1)^2J^a\nn\\&= -\frac{1}{8\pi}\int_{dS_3} d^3\phi \sqrt{-q}\,\Phi^{(p)}_a (\Box-1)S^{(q)a}. 
\end{align}
Since $S^{(q)}_a$ has the opposite parity on $dS_3$ than  $\Phi^{(p)}_a$, the right-hand side identically vanishes. The charge $Q^{V(2)}_{\Phi^{(p)}}[J_{a}] $ is therefore an \emph{asymptotically} conserved charge where the $q$-parity source does not spoil the conservation across spatial infinity. In Appendix \ref{sec:inhom}, we prove that this charge gives asymptotically 
\begin{align}
    \lim_{\tau\to+\infty} Q^{V\, (2)}_{\Phi^{(p)}}[J_a] &= \int_{S^2} \frac{d^2\Omega}{8\pi} Y^{A} \left(\mathcal N^{(\log)+}_{A}+F_A^{+}\right) \nn\\&= \int_{S^2} \frac{d^2\Omega}{8\pi}   (\Upsilon^* Y^{A} ) \left(\mathcal N^{(\log)-}_{A}+F_A^{-}\right)=\lim_{\tau\to-\infty} Q^{V\, (2)}_{\Phi^{(p)}}[J_a],
\end{align}
where $\mathcal N^{(\log)\pm}_{A}$ is defined in Eq. \eqref{defNAlog} and $F_A^\pm$ are the contributions from the $q$-parity source given in Eq. \eqref{defFA}. This construction provides a practical way to get the asymptotically conserved mixed tail/peeling-breaking charge and reproduce the antipodal matching \eqref{cCABid} since
\begin{equation}
    F_A^+ = \Upsilon^*F_A^-,
\end{equation}
as can be seen from Eq. \eqref{defFA} and the antipodal matching \eqref{idm0}-\eqref{idM0}.

In our earlier analysis \cite{Compere:2026jmk} we defined a distinct charge $Q^V_{\chi^{(p)}}[J_{a}]$, which is non-conserved at spatial infinity since it obeys a flux-balance law on $dS_3$. Now, in the limit $\tau \to\pm \infty$ this charge is exactly 
\begin{equation}
\lim_{\tau\to\pm\infty}Q^V_{\chi^{(p)}}[J_{a}]= Q^V_{\chi^{(p)}}[\hat J_{a}]\mp \frac{1}{16\pi}\int_{dS_3}d^3\phi\sqrt{-q}\chi^{(p)a}S_a,\label{SpatialLorentz}
\end{equation} 
where 
\begin{equation}
Q^V_{\chi^{(p)}}[\hat J_{a}]=\lim_{\tau\to\pm\infty}Q^{V (2)}_{\Phi^{(p)}}[J_a]-\int_{S^2} \frac{d^2\Omega}{8\pi} Y^{A}F_A^+,\label{SpatialLorentz2}
\end{equation} 
with $(\Box-1)\Phi^{(p)}_a=\chi^{(p)}_a$, see Appendix \ref{sec:localization}. In our current definition, the integral term over $dS_3$ has been written as a boundary term at $\tau \to \pm \infty$ is now included in the definition of the mixed tail/peeling-breaking charge. The localization of the $dS_3$ integral as the difference of two boundary terms at $i^0_\pm$ was made possible thanks to the existence of the second Klein-Gordon product \eqref{Vachargesv2} and the cancellation of the $q$-parity source terms for this charge.

Finally, the conserved charge \eqref{Jacharges} with a $q$-parity $\chi^a$ defines the \emph{super-Lorentz charge}. This definition is novel. It provides a general definition of super-Lorentz charges within General Relativity given our assumptions on boundary conditions. We only need to define the super-Lorentz and leading mixed tail/peeling-breaking  charges at the corners $i^0_\pm$ in order to obtain a conservation law between future and past null infinity. In other words, we do not need to define $\hat J_a$ everywhere on $dS_3$ which would allow to define the exactly conserved charge at spatial infinity but we can only define $\hat J_a$ asymptotically as $\tau \to \pm \infty$ in order to define the asymptotically conserved charge between the boundaries $i^0_\pm$. It is therefore sufficient to define $J^{(S)}_a$ for large $\vert\tau\vert$ and define the charge $Q^V_\chi[\hat J_{a}]$ at the corners.

In order to complete the matching between the charges at spatial infinity and the charges at $\scri^\pm$ performed in \cite{Compere:2026jmk}, we use the bijection between a $q$-parity $\chi^a$ and a vector over the sphere (see Eq. (19) in \cite{Compere:2026jmk}): 
\begin{align}
\chi^{(q)A}_Y(\tau\to+\infty,x^A) =- Y^A(x^A) e^{-\tau}+o(e^{-\tau}). 
\end{align}

Then, by direct evaluation of Eq. \eqref{Jacharges} (evaluating $\hat J_a$ at $\tau\to\pm\infty$) with $q$-parity $\chi^a$ using Eq. \eqref{JetaABscri+-} and subtracting the required fields from the primitive inhomogeneous solution, we find 
\begin{equation}
    Q^V_\chi[\hat J_{a}] = 
\int_{S^2} \frac{d^2\Omega}{8\pi} Y^{A}  \mathcal N^{\text{inv}+(0)}_{A}= -\int_{S^2} \frac{d^2\Omega}{8\pi}   (\Upsilon^* Y^{A} ) \mathcal N^{\text{inv}-(0)}_{A}.\label{MAtchNACharge} 
\end{equation}
where $\mathcal N^{\text{inv}+(0)}_{A}$ is defined in Eqs. \eqref{Ninv0}-\eqref{curlyN0}. This corresponds to the proper supertranslation invariant version of the charges \eqref{scri+superrotationscharges}-\eqref{scri-superrotationscharges} in the limit $u^\pm\to-\infty$.

For $\ell=1$, Eq. \eqref{Jacharges} or, equivalently, Eq. \eqref{MAtchNACharge} defines the Lorentz charges at spatial infinity. While it does not resemble the definition of \cite{Compere:2011ve}, both definitions match with the same Lorentz charges at $\scri^+_-$ and $\scri^-_+$ since they are both conserved charges obtained from $j_{ab}$ plus correction terms with identical normalization factors, see also \cite{Compere:2023qoa}.

\paragraph{Restricted phase space and conserved charges}

Let us close this section with one additional consideration. In this paragraph, we only consider a restricted phase space with stronger boundary conditions. We assume peeling at spatial infinity, which imposes $i_{ab}=0$. In addition, we assume the existence of a gauge in which $\sigma$ converges as $\tau\to+\infty$ and, independently, of a gauge in which it converges as $\tau\to-\infty$. This sets the charge $Q^V_\chi[I_a]=0$. The corresponding logarithmic translation frames are chosen according to Eq. \eqref{etaH0} to allow for a matching with either ingoing or outgoing radiative coordinates. 

These conditions are equivalent to 
\begin{align}
    \mathcal{D}_c \left( \mathcal{D}^c H_0~(\mathcal{D}_a \mathcal{D}_b \sigma + q_{ab} \sigma)\right) = 0.\label{Equsigma}\\
    (\mathcal{D}_{(a} k_{b)c}-\mathcal{D}_ck_{ab})\mathcal{D}^cH_0  =0\label{EqukabB}
\end{align}
The general solution is 
\begin{align}
    &\sigma =M\frac{1+2(\frac{H_0}{2M})^2}{\sqrt{1+(\frac{H_0}{2M})^2}} \pm H_0  \label{sigmaSol}\\
     &k_{ab}^{(B)}=0 \label{kabBSol}
\end{align}
where $M\ne 0 $ is the constant 
\begin{equation}
    \mathcal{D}_a H_0 \mathcal{D}^a H_0 + H_0^2 = - 4 M^2.
\end{equation}
The entire field $\sigma$ is fixed to be identical to the field of a boosted Schwarzschild black hole of mass $M$, energy $E$ and momentum $P_i$ in either incoming or outgoing  radiative frame after using Eq. \eqref{etaH0}. One can check that the solution \eqref{sigmaSol} satisfies $(\Box+3)\sigma=0$.  After performing the matching at $i^0_\pm$, the condition $Q^V_\chi[I_a]=0$ is equivalent to $\mathcal{D}_A^{\pm(0)}=0$ (see \cite{Compere:2026jmk} and Section \ref{sec:antipodalmatch}) while $i_{ab}=0$ imposes $\nabla^BD^{\pm(0)}_{AB}=\frac{1}{2}\nabla_C\nabla_{\langle A}\nabla_{B\rangle }C^{\pm(1)BC}$ (see asymptotics of $i_{ab}$ in eq. (D8) in \cite{Compere:2026jmk}). In \cite{Compere:2026jmk}, these conditions were shown to impose 
\begin{equation}\label{eq5}
       3 P_i m^{-(0)} \nabla_A n_i +  (E + P_i n_i) \nabla_A m^{-(0)}=0. 
\end{equation}
and 
\begin{equation}\label{eq6}
       3 P_i \mathcal{\tilde M}^{-(0)} \epsilon_{AB}\nabla^B n_i +  (E + P_i n_i)\epsilon_{AB} \nabla^B \mathcal{\tilde M}^{-(0)}=0. 
\end{equation}
which imposes $m^{-(0)}$ to be the Bondi mass of a single massive body $m^{-(0)}=\frac{m}{\gamma^3 (1+ v_i n_i)^3}$ and $\mathcal{\tilde M}^{-(0)}=0$. Then, injecting $\sigma$ in Eq. \eqref{SourceSab} we obtain $S_{ab}=0$. Hence, $J_{a}$ obeys a homogeneous equation, $(\Box-1)J_{a}=0$, and the charge $Q^V_\chi[\hat J_{a}]=Q^V_\chi[J_{a}]$ is exactly conserved at spatial infinity.

In summary, when assuming peeling at spatial infinity, we might still have violations of peeling at null infinity ($D_{AB}^{\pm (0)} \neq 0$). Yet, $J_{ab}$ obeys a homogeneous equation.

\section{Timelike infinities $i^\pm$}
\label{sec:timelike}

We will now derive the Beig-Schmidt expansion at timelike infinities $i^\pm$ and perform the matching with $\mathcal I^+_+$ and $\mathcal I^-_-$ at subleading order. 

\subsection{Beig-Schmidt expansion at timelike infinities}

Formally, the metric can be expanded at future timelike infinity by first performing a complex change of coordinates of the metric expanded at spatial infinity \eqref{BSExpansion} and by then restricting to real fields. The complex change of coordinates is given by Eq. (21) of \cite{Compere:2023qoa} : 
\begin{align}
    \begin{aligned}
\rho &\mapsto i\tau,
& \tau &\mapsto \rho-\frac{i\pi}{2},
& \partial_{\rho} &\mapsto -i\partial_{\tau},
& \partial_{\tau} &\mapsto \partial_{\rho},
\qquad q_{ab} \mapsto -h_{ab},
\\[4pt]
\sigma &\mapsto i\sigma^+,
& k_{ab} &\mapsto -ik^+_{ab},
& i_{ab} &\mapsto i^+_{ab},
& j_{ab} &\mapsto j^+_{ab}-i\frac{\pi}{2}i^+_{ab},
\end{aligned}
\end{align}
which leads to the expansion at $i^+$ as $\tau \to \infty$:
\begin{align} \label{BSexpansioni+}
\begin{aligned}
ds^{2}
={}&
\left(
-1-\frac{2\sigma^+}{\tau}
-\frac{(\sigma^+)^{2}}{\tau^{2}}
+o\!\left(\tau^{-2}\right)
\right)d\tau^{2}
+o\!\left(\tau^{-2}\right)\tau\,d\tau\,d\phi^{a}
\\
&\quad
+\tau^{2}
\left(
h_{ab}
+\tau^{-1}\left(k^+_{ab}-2\sigma^+ h_{ab}\right)
+\frac{\log\tau}{\tau^{2}}\,i^+_{ab}
+\tau^{-2}j^+_{ab}
+o\!\left(\tau^{-2}\right)
\right)
d\phi^{a}d\phi^{b}.
\end{aligned}
\end{align}
where $h_{ab}$ is the metric of the upper sheet of the two-sheet hyperboloid known as Euclidean $AdS_3$:
\begin{align}
 h_{ab}d\phi^a d\phi^b =d\rho^2+\sinh^2\rho (d\theta^2+\sin^2\theta d\phi^2). 
\end{align}
We will denote as $h^{ab}$ the inverse metric and $D_a$ the covariant derivative on the Euclidean hyperboloid. All indices $a,b,\dots$ of tensors defined at $i^+$ are raised and lowered with this metric. We again assume that 
\begin{equation}
 k^+_{ab}= -2 ( D_a  D_b - h_{ab})\Phi^+ + k_{ab}^{+(B)}
\label{kabPhii}
\end{equation}
where $\Phi^+$ is a scalar field on $EAdS_3$. The trace condition $k=0$ is equivalent to 
\begin{equation}
    (\Delta- 3)\Phi^+ = 0,
\label{EquationPhii}
\end{equation}
where $\Delta =  D_a D^a$ is the Laplacian on $EAdS_3$. 

The metric at past timelike infinity $i^-$ is formally the same as Eq. \eqref{BSexpansioni+} but with $\tau$ replaced with $-\tau$ and with fields $\sigma^-$, $k_{ab}^-$, $\dots$ instead of $\sigma^+$, $k_{ab}^+$, $\dots$ The boundary metric is the one of the lower sheet of Euclidean $AdS_3$ which is isometric to $h_{ab}$. This leads to the expansion at $i^-$ as $\tau \to -\infty$:
\begin{align} \label{BSexpansioni-}
\begin{aligned}
ds^{2}
={}&
\left(
-1+\frac{2\sigma^-}{\tau}
-\frac{(\sigma^-)^{2}}{\tau^{2}}
+o\!\left(\tau^{-2}\right)
\right)d\tau^{2}
+o\!\left(\tau^{-2}\right)\tau\,d\tau\,d\phi^{a}
\\
&\quad
+\tau^{2}
\left(
h_{ab}
-\tau^{-1}\left(k^-_{ab}-2\sigma^- h_{ab}\right)
+\frac{\log(-\tau)}{\tau^{2}}\,i^-_{ab}
+\tau^{-2}j^-_{ab}
+o\!\left(\tau^{-2}\right)
\right)
d\phi^{a}d\phi^{b}.
\end{aligned}
\end{align}

The main difference between spatial infinity and timelike infinities is the existence of sources at timelike infinity. Each individual body reaches $i^\pm$ at one point on the hyperboloid which is determined by its momentum. The Beig-Schmidt fields are therefore only determined in the sense of distributions. Outside of sources, we can perform the analytic continuation of the equations \eqref{generici0equ} obtained from Einstein's equations that govern the dynamics of Beig-Schmidt fields to obtain nested equations of the form
\begin{equation}
    (\Delta-((n+1)^2-1-r))\Psi_{i_1...i_r}=S_{i_1...i_r},\qquad D^{a}\Psi_{a\,i_2...i_r}=0,\qquad 
    h^{ab}\Psi_{ab\,i_3...i_r}=0.\label{generici0equi}
\end{equation}
Generic regular or singular homogeneous solutions for $r=0,1,2$ can be found after performing an analytic continuation of the solutions studied in \cite{Compere:2025bnf} for $dS_3$.

The solutions are built from the scalars satisfying the homogeneous $r=0$ solution to Eq. \eqref{generici0equi}, namely $(\Delta -n(n+2))\Psi=0$. There are two classes of solutions: the regular solutions $(R)$ (which diverge at the boundary $\rho\to \infty$ when $n \geq 1$), and the singular solutions $(S)$ (which are decaying at the boundary). A basis of such scalar harmonics reads as
\begin{align}  
&\psi_{n\ell m}^{(R)}(\rho,x^A) 
:=\left\{
    \begin{array}{ll}
       \frac{\sqrt{2\pi}}{(\sinh\rho)^\frac{1}{2}} Y_{\ell m}(\theta,\phi)\,  \left( P^{\ell +\frac{1}{2}}_{n+\frac{1}{2}}( \cosh{\rho})+\frac{2i}{\pi} Q^{\ell +\frac{1}{2}}_{n+\frac{1}{2}}( \cosh{\rho}) \right)& \mbox{for }\ell <(n+1) ;\\&\\
       \frac{1}{2\sinh\rho  }Y_{\ell m}(\theta,\phi)\, Q^{n+1}_\ell (\coth\rho)  & \mbox{for } \ell \ge(n+1),
    \end{array}
\right.\label{psi(p)nlm}
\intertext{and}
&\psi_{n\ell m}^{(S)}(\rho,x^A) 
:=\left\{
    \begin{array}{ll}
      \frac{i}{\sqrt{2\pi}(\sinh\rho)^\frac{1}{2}} Y_{\ell m}(\theta,\phi)\,  Q^{\ell +\frac{1}{2}}_{n+\frac{1}{2}}( \cosh{\rho}) & \mbox{for } \ell <(n+1); \\&\\
       \frac{1}{2\sinh\rho }Y_{\ell m}(\theta,\phi)\,  P^{n+1}_\ell (\coth\rho)  & \mbox{for } \ell \ge(n+1),
    \end{array}
    \right.\label{psi(q)nlm}
\end{align}
The associated Legendre functions are chosen to be Type 3 with branch cut on $]-\infty,1[$. The analysis in $dS_3$ of \cite{Compere:2025bnf} then extends straightforwardly though with much algebra to the case of fields in $EAdS_3$, where the $(p)/(q)$-parities are replaced by the regular/singular $(R)/(S)$ split.

\subsection{Matching at $\partial i^+=\mathcal I^+_+$ and $\partial i^-=\mathcal I^-_-$ }
\label{sec:matchi}

Matching conditions can be obtained at either corner $\partial i^+=\mathcal I^+_+$ and $\partial i^-=\mathcal I^-_-$ following the methods of \cite{Compere:2023qoa}. Given the presence of four corner regions, we need to slightly adapt our notation to accommodate for expansions close to timelike infinities. From now on we denote quantities evaluated in an asymptotic region $\mathcal{A}$ by $\vert_{\mathcal{A}}$. For example, the earlier notation $C_{AB}^{+(1)}$ is equivalent to $C_{AB}^{(1)}\vert_{\scri^+_-}$. Let us start by expanding the Bondi shear and Bondi aspects at $\scri^+_+$ and $\scri^-_-$ in the limit $u^\pm \to \infty$ as follows: 
\begin{align}
    &C_{AB}^\pm = C_{AB}^{(0)}\vert_{\partial i^\pm }-\frac{1}{u^\pm} C_{AB}^{(1)}\vert_{\partial i^\pm }+o(\frac{1}{u^\pm}),\\
    & D_{AB}^\pm=D_{AB}^{(0)}\vert_{\partial i^\pm },\\
    &m^{\pm}= m^{(0)}\vert_{\partial i^\pm}-\frac{1}{u^\pm } m^{(1)}\vert_{\partial i^\pm }+o(\frac{1}{u^\pm}),\label{mipmAsypt}\\
   &\mathcal{P}_A^\pm =  -N_A^{(-1)}\vert_{\partial i^\pm } u^\pm +N_A^{(\text{log})}\vert_{\partial i^\pm }\log(u^\pm)+N_A^{(0)}\vert_{\partial i^\pm }+o(1),
\end{align}
where we again decompose $C_{AB}^{(0)}\vert_{\partial i^\pm }$ using two scalars: 
\begin{equation}
  C_{AB}^{(0)}\vert_{\partial i^\pm }=\left(-2 \nabla_A \nabla_B + \gamma_{AB} \nabla^2\right)C^{(0)}\vert_{\partial i^\pm }+\epsilon_{C(A}\nabla_{B)}\nabla^C \Psi^{(0)}\vert_{\partial i^\pm }\label{CABODecompi} .
\end{equation}
The evolution equations at $\scri^\pm$ give the relations : 
\begin{align}
m^{(1)}\vert_{\partial i^\pm } &:=\frac{1}{4} \nabla^A \nabla^B C^{ (1)}_{AB}\vert_{\partial i^\pm },\label{EvolutionConstrainm1ipm}\\
N_A^{\pm (-1)}\vert_{\partial i^\pm }&:= -\nabla^B \mathcal M^{(0)}_{AB}\vert_{\partial i^\pm }, \\ 
    N_A^{ (\text{log})}\vert_{\partial i^\pm } &:=-\frac{1}{2} \nabla_C \nabla_{\langle A} \nabla_{B\rangle} C^{(1)\;BC}\vert_{\partial i^\pm }\label{NALogipm}. 
\end{align}

At $i^\pm$, the matching conditions between $\sigma^\pm$ and $m^{(0)}\vert_{\partial i^\pm }$ and between $\Phi^\pm$ and $C^{(0)}\vert_{\partial i^\pm }$ have been derived in \cite{Compere:2023qoa}. We get 
\begin{align}
    \Phi^\pm &= \frac{1}{2} C^{(0)}\vert_{\partial i^\pm} e^{\rho}+ o(e^{\rho}),\\
    \sigma^\pm &= - 2 m^{(0)}\vert_{\partial i^\pm} e^{-3\rho}  + o(e^{-3\rho})\label{sigmascri++},\\
   k_{\rho\rho}^{\pm (B)} &=  o(e^{-2\rho}),\\
    k_{\rho A}^{\pm (B)} &= -2 \nabla^B C_{AB}^{(B)(0)}\vert_{\partial i^\pm} e^{-\rho} + o(e^{-\rho}),\\
    k_{AB}^{\pm (B)} & = \frac{1}{2} C_{AB}^{(B)(0)}\vert_{\partial i^\pm} e^{\rho}+ o(1)\label{kabscri++bis}. 
\end{align}
Asymptotically, the fields $\sigma^\pm$ are combinations of singular harmonics. Since they can only be singular where the sources are located, their asymptotic behavior is uniquely determined by the sources (see \cite{Compere:2023qoa}). The field $\Phi^\pm$ is a combination of regular harmonics and pure-gauge terms. The matching of $k_{ab}^{\pm(B)}$ is new. Based on an analogous analysis in \cite{Compere:2026jmk}, it is built from $n=-1$ harmonics containing Legendre polynomials of the first kind, which are the singular harmonics in $EAdS_3$; see Eq. \eqref{psi(q)nlm}. Therefore, $k_{ab}^{\pm(B)}$ is singular. This implies that it must be sourced by a stress-energy tensor or vanish.

At subleading order, the matching between $i^\pm_{ab}$ and $N_A^{(\text{log})}\vert_{\partial i^\pm }$ and $D_{AB}^{(0)}\vert_{\partial i^\pm }$ was performed in  \cite{Boschetti:2026gfd}. Here, we notice that the matching is unchanged in the presence of $k_{ab}^{\pm (B)}$. It reads as
\begin{align}
i^\pm_{\rho\rho} &= -16\nabla^A(N_A^{(\text{log})}\vert_{\partial i^\pm}+\nabla_B D^{(0)B}_A\vert_{\partial i^\pm})e^{-4\rho}+o(e^{-4\rho}), \\ 
i^\pm_{\rho A} &=-4 (N_A^{(\text{log})}\vert_{\partial i^\pm}+\nabla_B D^{(0)B}_A\vert_{\partial i^\pm})e^{-2\rho}+o(e^{-3\rho}), \label{irhoA}\\
j_{\rho\rho}^\pm&= -16 m^{\pm (1)}\vert_{\partial i^\pm}e^{-2\rho}+o(e^{-2\rho}), \\ 
j_{\rho A} ^\pm&=\nabla^B C^{(1)}_{AB}\vert_{\partial i^\pm}+\bigg(-4 N_A^{(0)}-C^{(0)}_{AC} \nabla_B C^{(0)BC}+8\partial_A m^{(1)}+(4\rho-2)N_A^\text{(\text{log})}\nonumber \\ 
&-(-4 \log 2+6+4\rho)\nabla_B D^{(0)B}_A+\nabla^B C_{AB}^{(1)}\bigg)\vert_{\partial i^\pm}e^{-2\rho}+o(e^{-3\rho}).
\end{align}
This matching also reproduces the one obtained in \cite{Compere:2023qoa} in the case $D_{AB}^{(0)}\vert_{\partial i^\pm }=0$ and $\Psi^{(0)}\vert_{\partial i^\pm }=0$ up to one difference. 
In $j_{\rho A}$ we find $+8 \partial_A m^{(1)}$ instead of $+12 \partial_A m^{(1)}$ as written in \cite{Compere:2023qoa}\footnote{In \cite{Compere:2023qoa}, the definition of $m^{(1)}$ differs by a conventional minus sign (see Eq. \eqref{mipmAsypt} and Eqs. (240) and (243) in \cite{Compere:2023qoa}).}. This corrects a  typo in their Eq. (157).

From an analogous reasoning to the spatial infinity case, the asymptotic behavior $i_{ab}$ corresponds to Legendre P solutions which, in $EAdS_3$, are singular. As shown in \cite{Boschetti:2026gfd}, the asymptotic behavior of $i_{ab}$ is uniquely determined by the sources. Regarding $j_{ab}$, one can build an analogous SDT tensor $J_{ab}$, which is invariant under proper supertranslations. While we do not construct this tensor here, we can infer from the analysis at spatial infinity that the leading behavior will correspond to the Legendre Q solutions, hence regular solutions. Hence, $C^{(1)}_{AB}$ cannot be determined by the sources and must be computed from another formalism. Regarding $N^{(0)}_{A}$, we expect that it is a  singular solution and, similarly to $i_{ab}$ and $\sigma$, should be uniquely determined by the sources after solving Einstein's equations with a non-vanishing stress-energy tensor near $i^\pm$.

\section{Gravitational scattering}
\label{sec:comp}

Asymptotic particle states are classically defined as incoming and outgoing bodies reaching $i^\pm$. Effectively, any massive body is described by a point particle at $i^+/i^-$, i.e. there is an effective skeletonization at timelike infinity. After a quick review of the kinematics at $i^\pm$ and a review of the logarithmic shift of asymptotic trajectories due to gravitational interactions, we will derive the values of the corner fields at leading and subleading orders that can be deduced from $i^\pm$, following the work of Campiglia and Boschetti \cite{Boschetti:2026gfd}. This will allow us to discuss the soft theorems discussed in \cite{Laddha:2018myi,Laddha:2018vbn,Sahoo:2018lxl,Saha:2019tub,Sahoo:2021ctw,Boschetti:2026gfd,Boschetti:2026ogm}.  

\subsection{Kinematics at $i^\pm$}

Let us recall that Minkowski spacetime can be foliated close to $i^\pm$ by constant $\tau := \pm \sqrt{t^2-r^2}$ and $\rho=\text{arctanh}(r/(\pm t))$ slices. When including gravity, as described in Section \ref{sec:timelike}, it leads to the metric \eqref{BSexpansioni+} at $i^+$ and \eqref{BSexpansioni-} at $i^-$ where $h_{ab}d\phi^a d\phi^b$ is the unit metric on $EAdS_3$ with $D_a$ as its metric-compatible covariant derivative. In the following,   $\mu,\nu,\dots$ indices will be raised and lowered with the Minkowski metric $\eta_{\mu\nu}$ while $a,b,\dots$ indices will be raised and lowered with the unit hyperboloid metric $h_{ab}$. We introduce the notation $\cdot$ for the product of 4-vectors with the Minkowski metric. 

At leading order in the large $\tau \to \pm\infty$ expansion, the coordinates $(\tau,\phi^a)$ label spacetime points as $x^\mu = \tau v^\mu$ where $v^\mu(\phi^a)$ is the future directed unit normal to the hyperboloid which reads in asymptotic Minkowski coordinates as $v^\mu(x^a) := (\cosh\rho,\sinh\rho \vec{n})$. We can project a spacetime vector onto its normal and hyperboloidal components using $\delta^\mu_\nu =-v^\mu v_\nu +\partial^a v^\mu \partial_a v_\nu$. At large radius we have $v^\mu=\frac{1}{2}e^\rho n^\mu+O(e^{-\rho})$ where $n^\mu := (1,\vec{n})$.

Let us consider the scattering of $n^-$ incoming massive bodies at $i^-$ to $n^+$ outgoing massive bodies at $i^+$. 
Following standard conventions, the momentum of the body $i$ is defined as $p^\mu_{(i)} = m_{(i)} v^\mu_{(i)}$ with $v_\mu^{(i)} v^\mu_{(i)} =-1$ if the body is outgoing ($i \in \text{out}$ or $i \in +$) and $p^\mu_{(i)} = - m_{(i)} v^\mu_{(i)}$ if the body is incoming ($i \in \text{in}$ or $i \in -$). Massless incoming and outgoing particles are also labelled with $i$. The total momentum is 
\begin{align}
P^\mu=\sum_{i \in \text{out}}p_{(i)}^\mu = -\sum_{i \in \text{in}}p_{(i)}^\mu  \label{MomentumEnergyConservation}   
\end{align}
where the summation includes both massive and massless particles. Let $\tau_{(i)}$ be the proper time of the massive body $i$, $\rho_{(i)}$ its rapidity and $\gamma_{(i)}:= \cosh \rho_{(i)}=1/\sqrt{1-v_{(i)}^i v_{(i)}^i}$ its Lorentz boost factor. In an asymptotically Minkowskian frame, its velocity can be written as 
\begin{align}
v_{(i)}^\mu = (\cosh \rho_{(i)},\sinh \rho_{(i)} \vec{n}(x^A_{(i)}))    
\end{align}
where $x^A_{(i)}$ are the incidence angles of the body on the sphere at future timelike infinity. A free particle originates from the antipodally related point $\Upsilon^* x^A_{(i)}$ on $S^2$ at $i^-$ with respect to its location on the sphere $x^A_{(i)}$ at $i^+$. Using  $\vec{n}(\Upsilon^* x^A_{(i)})=-\vec{n}(x^A_{(i)})$ and the convention $p^0_{(i)}<0$ for an incoming particle, the four-momentum of an incoming particle is indeed opposite to the four-momentum of an outgoing particle with the same velocity.

\subsection{Logarithmic deviation vector}

The asymptotic trajectories at large $\tau_{(i)} \to \pm \infty$ (in the approach to $i^\pm$) read as \cite{Eder:1983hm}
\begin{align}
 x^\mu_{(i)}(\tau_{(i)}) = \tau_{(i)} v_{(i)}^\mu + \log \vert \tau_{(i)} \vert c_{(i)}^{\pm \mu}(v_{(i)})+O(\tau_{(i)}^0)     
\end{align}
where $c_{(i)}^{\pm \mu}(v_{(i)})$ is the logarithmic deviation vector at $i^\pm$, which is described in detail in \cite{Boschetti:2026gfd} (see also \cite{Bini:2018ywr,Sahoo:2018lxl}). 

Let us first consider future timelike infinity. The logarithmic deviation vector can be obtained by solving for the coupled motion of $n$ point particles in the presence of gravity at leading order in the large $\tau$ expansion in a Beig-Schmidt patch $(\tau,\phi^a)$ near timelike infinity. Following \cite{Boschetti:2025tru} and Eq. (84) of \cite{Compere:2023qoa}, the result is 
\begin{align}
c_{(i)}^{+ \mu}(v_{(i)}) &=( \partial^a v^\mu \partial_a \sigma^+-v^\mu\sigma^+)\Bigg\vert_{\hspace{-4pt}\footnotesize\begin{array}{l} x^\mu \mapsto \tau v^\mu_{(i)} \\ v^\mu \mapsto v^\mu_{(i)} \end{array}}.  
\end{align}
The Newtonian potential is 
\begin{align}\label{sigmaval}
\sigma^+ =-L^{+\mu}v_\mu(\phi^a) +\sum_{j \in \text{out}}m_{(j)}\left(2 \chi_{(j)}- \frac{2\chi_{(j)}^2-1}{\sqrt{\chi_{(j)}^2-1}} \right) ,
\end{align}
where $\chi_{(i)}(\phi^a ; v_{(i)}) := -v^\mu(\phi^a)v_\mu(v_{(i)}) = \gamma_{(i)}(\cosh \rho - v_{(i)}^i n_i(\phi^A) \sinh\rho)$ and we defined the four-vector $L^{+\mu} :=(L^{+0},L^{+i})$. Here the field $-L^{+\mu}v_\mu(\phi^a)$ sets the logarithmic translation frame at $i^+$. It is the analogue of Eq. \eqref{HlogSolutions} but at future timelike infinity.   In the outgoing radiative logarithmic frame $L^{+\mu}=0$, the field $\sigma$ is convergent at $\partial i^+$ which allows to match to an outgoing Bondi frame  \cite{Compere:2023qoa}. In a distinct logarithmic frame we have simply $c_{(i)}^{+ \mu}(v_{(i)})=c_{(i)}^{+ \mu}(v_{(i)})\vert_{L^{+\mu}=0} - L^{+\mu}$ given our definition of $L^{+\mu}$. Let us denote the four-vector $L^+ := L^{+\mu}\partial_\mu$. The Newtonian potential can be written alternatively as 
\begin{align}
\sigma^+(\phi^a) &=  -(L^{+} +2 P_{\text{massive}}) \cdot v(\phi^a)-\mathop{\sum_{j \in \text{out}}}_{m_{(j)}\neq 0} \frac{2 (v(\phi^a) \cdot p_{(j)})^2-m_{(j)}^2}{\sqrt{(v(\phi^a) \cdot p_{(j)})^2-m_{(j)}^2}} \\ 
&=   -(L^{+} +2 P) \cdot v(\phi^a)-\mathop{\sum_{j \in \text{out}}} \frac{2 (v(\phi^a) \cdot p_{(j)})^2-m_{(j)}^2}{\sqrt{(v(\phi^a) \cdot p_{(j)})^2-m_{(j)}^2}} \label{sigmap}
\end{align}
where $P_\text{massive}=\sum_{j \in \text{out}} m_{(j)}v^\mu (v_{(j)}) \partial_\mu $ is the total 4-momentum from outgoing massive particles while $P=P^\mu \partial_\mu$ is the total momentum including massless particles. We used $v \cdot p_{(i)}=-\vert v \cdot p_{(i)}\vert$ for outgoing particles. In harmonic gauge, the Newtonian potential is identified as Eq. \eqref{sigmap} with $L^+ =-2P$. This fixes the logarithmic frame of harmonic gauge. In a generic gauge this leads to 
\begin{align}
c_{(i)}^{+\mu} =- L^{+\mu} -2 P^\mu -\mathop{\sum_{\substack{j \in \text{out} \\ j \neq i}}}\frac{(2 (v_{(i)} \cdot p_{(j)})^3-3m_{(j)}^2 v_{(i)} \cdot p_{(j)})p_{(j)}^\mu-m_{(j)}^4 v^\mu_{(i)}}{((v_{(i)}\cdot p_{(j)})^2-m_{(j)}^2)^{3/2}}  
\end{align}
where the sum is over all massless and massive outgoing particles excluding the particle considered as self-field is excluded. In outgoing radiative gauge, $c_{(i)}^{\text{rad}+\mu} :=c_{(i)}^{+\mu}\vert_{L^{+\mu}=0}$.

The reasoning is analogous at past timelike infinity with result 
\begin{align}
c_{(i)}^{-\mu} &=-L^{-\mu} +2 P^\mu +\mathop{\sum_{\substack{j \in \text{in} \\ j \neq i}}}\frac{(2 (v_{(i)} \cdot p_{(j)})^3-3m_{(j)}^2 v_{(i)} \cdot p_{(j)})p_{(j)}^\mu-m_{(j)}^4 v^\mu_{(i)}}{((v_{(i)}\cdot p_{(j)})^2-m_{(j)}^2)^{3/2}} \\ 
&=- L^{-\mu}+2 P^\mu_\text{massive} +\mathop{\sum_{j \in \text{in},\, j \neq i}}_{m_{(j)}\neq 0}\frac{(2 (v_{(i)} \cdot p_{(j)})^3-3m_{(j)}^2 v_{(i)} \cdot p_{(j)})p_{(j)}^\mu-m_{(j)}^4 v^\mu_{(i)}}{((v_{(i)}\cdot p_{(j)})^2-m_{(j)}^2)^{3/2}}
\end{align}
where $L^{-\mu} :=(L^{-0},L^{-i})$ sets the logarithmic translation frame at $i^-$ and we used $v_{(i)}\cdot p_{(j)}>0$ for $j \in \text{in}$. Harmonic gauge corresponds to $L^{-\mu}= 2 P^{\mu}$.  In incoming radiative gauge, we have simply $c_{(i)}^{\text{rad}-\mu} :=c_{(i)}^{-\mu}\vert_{L^{-\mu}=0}$.

Logarithmic translations act globally on both $i^+$ and $i^-$ as $\delta L^+ = \delta L^-$ (with our definition of $L^-$ ) as a consequence of matching conditions between infinities \cite{Compere:2023qoa,Boschetti:2025tru}. The shift of the logarithmic translation frame between outgoing and incoming frame is dictated by the total energy-momentum at spatial infinity, see Eq. \eqref{etaH0}. It reads as 
\begin{align}
 L^{+\mu}- L^{-\mu} =-4 P^{\mu}
\end{align}
where $P^\mu:= (E,P^i)$ is the conserved total momentum. The logarithmic translation frames are summarized in the following Table \ref{logframes} identical to one provided in \cite{Boschetti:2026gfd}.

\begin{table}
\begin{center}
\begin{tabular}{|l||c | c|}
\hline
Logarithmic translation frame & $L^{+\mu}$ &  $L^{-\mu}$  \\  \hline\hline
Outgoing radiative & $0$ & $4P^\mu$ \\ \hline
Harmonic & $-2P^\mu$ & $2P^\mu $ \\ \hline
Incoming radiative & $-4P^\mu$ & $0$  \\\hline
\hline
\end{tabular}
\caption{Value of $L^{\pm \mu}$ in three different logarithmic translation frames.}\label{logframes}
\end{center}
\end{table}


\subsection{Corner fields at $\partial i^\pm$}

After solving for the Beig-Schmidt field at $i^\pm$, one can read off their boundary value as $\rho \to \infty$ and use the matching conditions at $\partial i^\pm$ derived in Section \ref{sec:matchi} to read off the asymptotic values of the combination of the Bondi shear and aspects at $u^\pm \to +\infty$ that are entirely determined by the behavior at $i^\pm$. (Note that part of the asymptotic values of the Bondi shear and aspects is not determined by the solution at $i^\pm$.)

The solution for $\sigma^+$ \eqref{sigmaval} and the matching conditions \eqref{sigmascri++} directly give
\begin{align}
    m^{(0)}\vert_{\partial i^+} = - \mathop{\sum_{i\in \text{out}}}_{m_{(i)}\ne 0}\frac{m_{(i)}^4}{(p_{(i)} \cdot n)^3},\qquad      m^{(0)}\vert_{\partial i^-} = + \mathop{\sum_{i\in \text{in}}}_{m_{(i)}\ne 0}\frac{m_{(i)}^4}{(p_{(i)} \cdot n^-)^3}.\label{valsm0}
\end{align}
where $n^-=n^{-\mu}\partial_\mu$ with $n^{-\mu}:=(1,-\vec{n})$. We also define $n^{+\mu} := n^{\mu}=(1,\vec{n})$.  The result at $i^-$ can be fixed by considering the compatibility of the expression with the matching at spatial infinity \eqref{idm0} in the $G \mapsto 0$ limit when there is no interaction\footnote{Note that in \cite{Campiglia:2014yka}, the spherical coordinates at $i^-$ are antipodally related to our spherical coordinates at $i^-$. Moreover, $m^{(0)}\vert_{\partial i^-}$ is defined with a relative global minus sign. This explains the minus sign difference and substitution $n \mapsto n^-$ of our expression with respect to Eq. (5.9) of \cite{Campiglia:2014yka}.}.

The field $k_{ab}$ is not sourced at $i^\pm$ in the case of gravitational scattering. Therefore, the irregular part of $k_{ab}$ vanishes. Matching the fields at the corners $\partial i^\pm$ yields 
\begin{align}
C_{AB}^{(B)(0)}\vert_{\partial i^\pm}=0.    \label{CABB0}
\end{align} 

The $\rho A$ components of the field $i_{ab}$ were solved in \cite{Boschetti:2026gfd}. After switching their convention for $\rho$ as our $\sinh\rho$, their Eq. (4.24) combined with the matching at the corners $\partial i^\pm$ of $i_{ab}$ (see Eq. \eqref{irhoA}) gives \footnote{We use the convention $c_{[\mu}p_{\nu]}=\frac{1}{2}(c_{\mu}p_{\nu}-p_{\mu}c_{\nu})$ while, in \cite{Boschetti:2025tru}, they use $c_{[\mu}p_{\nu]}=c_{\mu}p_{\nu}-p_{\mu}c_{\nu}$.}
\begin{align}
\nabla^B D^{(0)}_{AB}\vert_{\partial i^\pm }+ N_A^{(\log)}\vert_{\partial i^\pm }=- 6\mathop{\sum_{i \in \pm}}_{m_{(i)}\neq 0} m_{(i)}^4\frac{\partial_A n^{\pm\mu} n^{\pm \nu} c^{\text{rad}\pm}_{(i)[\mu}p^{(i)}_{\nu]}}{(p_{(i)}\cdot n^\pm)^4} .\label{DplusNtimelike}
\end{align}

Using Eq. (39) and (42) in \cite{Boschetti:2026ogm}, this can be rewritten as 
\begin{align}
    &\nabla^B D^{(0)}_{AB}\vert_{\partial i^\pm }+ N_A^{(\log)}\vert_{\partial i^\pm }=2  \mathop{\sum_{i\in \pm}}_{m_{(i)}\ne0} \tilde{s}_A[p_{(i)},c_{(i)}^{\text{rad}\pm},n^\pm]\label{BCMAtch1}
\end{align}
where, for a four-momentum $p^\mu$ of mass $m^2 =-p_\mu p^\mu$ and a four-vector $b^\mu$, $\tilde{s}_A[p,b,n]$ is defined by 
\begin{equation}
    \tilde{s}_A[p,b,n]:=-3m^4\frac{\partial_A n^{\mu} n^{ \nu} b_{[\mu}p_{\nu]}}{(p\cdot n)^4} =-\frac{1}{2}(n^\mu b_\mu\nabla_A+3 b_\mu\nabla_A n^\mu) \frac{m^4}{(p \cdot n)^3}.\label{sAdef}
\end{equation}
Eq. \eqref{BCMAtch1} reproduces Eqs. (68) and (69) in \cite{Boschetti:2026ogm}. 

\subsection{Soft theorems: Comparison with Boschetti-Campiglia}

Let us now revisit the leading classical soft graviton theorem and the logarithmic correction to the subleading classical soft graviton theorem by analyzing the work of  \cite{Boschetti:2026gfd,Boschetti:2026ogm}.

In this section, we consider the stress-energy tensor near $\scri^\pm$ to be the stress-energy tensor for massless particles, which takes the form \eqref{TAB} with 
\begin{align}\label{hypoT}
T_{rr}^{\pm (0)}=T_{rA}^{\pm (0)}=T_{\langle A B\rangle}^{\pm (1)}=0,    
\end{align}
while the other coefficients are compactly supported on $\scri^\pm$. As before, such stress-energy tensor does not contribute to the antipodal relationships at spatial infinity at leading and subleading orders. In order to be consistent with the literature, we also assume the absence of a leading-order magnetic component of the shear at spatial infinity:  
\begin{equation}
C_{AB}^{(B)(0)}\vert_{i^0_\pm} = 0. \label{CABBzero}
\end{equation}

\paragraph{Leading soft graviton theorem} Integrating the evolution equation \eqref{FluxBalancedLawsm} and using the simplification \eqref{hypoT} yields the following relations between the values of the Bondi mass aspect at the corners:
\begin{align}
    &m^{(0)}\vert_{\scri^+_+}-m^{(0)}\vert_{\scri^+_-}=\frac{1}{4} \nabla^A\nabla^B[C_{AB}^{(0)}\vert_{\scri^+_+}-C_{AB}^{(0)}\vert_{\scri^+_-}]-4\pi \int_{-\infty}^\infty du \rho_{\text{massless}}\label{eqm1},\\
   & m^{(0)}\vert_{\scri^-_+}-m^{(0)}\vert_{\scri^-_-}=\frac{1}{4} \nabla^A\nabla^B[C_{AB}^{(0)}\vert_{\scri^-_+}-C_{AB}^{(0)}\vert_{\scri^-_-}] +4\pi \int_{-\infty}^\infty dv \rho_{\text{massless}}\label{eqm2},
\end{align}
where the local densities of radiated energy at $\scri^+$ and at $\scri^-$ read as 
\begin{align}
  \rho_{\text{massless}}(u,x^A)=  T^{(0)}_{uu}+\frac{1}{32\pi} N_{AB}^+N^{+\,AB} ,\\
 \rho_{\text{massless}}(v,x^A) = T^{(0)}_{vv}+ \frac{1}{32\pi} N_{AB}^-N^{-\,AB} .
\end{align}

In general, there are tails in the news tensors ($N^\pm_{AB} \sim (u^\pm)^{-2}$ terms as $u^\pm \to \pm \infty$ and higher subleading terms) which are related between future and past null infinity, as deduced from the antipodal map \eqref{idC4} and, as we expect,  higher subleading antipodal maps. We therefore postulate 
\begin{align}
   \int_{-\infty}^\infty du \rho_{\text{massless}} = \mathop{\sum_{i\in \text{out}}}_{m_{(i)}=0} E_{(i)} \delta(\phi,\phi_{(i)})+\frac{1}{4\pi}f^+_\text{tail}(\phi,\{\phi_{(j)}\}),\label{rho1}\\
    \int_{-\infty}^\infty dv \rho_{\text{massless}} = -\mathop{\sum_{i\in \text{in}}}_{m_{(i)}=0} E_{(i)} \Upsilon^*\delta(\phi,\phi_{(i)})+\frac{1}{4\pi}f^-_\text{tail}(\phi,\{\phi_{(j)}\}),\label{rho2}
\end{align}
where $f^\pm_\text{tail}(\phi,\{\phi_{(j)}\})$ are possible contributions from tails, originating from gravitational interactions. We do not see any reason to set them to zero at this stage. The first terms in Eqs. \eqref{rho1}-\eqref{rho2} are standard in the literature and correspond to the energy of outgoing hard particles \cite{Strominger:2013jfa,Strominger:2014pwa}. The additional terms $f^\pm_\text{tail}(\phi,\{\phi_{(j)}\})$ are typically omitted when treating asymptotic states as free, as in the original derivation of the leading soft theorem from conservation laws \cite{Strominger:2013jfa}. Instead, we consider them as contributions, and we will demonstrate that they cancel in the classical soft graviton theorem under an appropriate hypothesis.
The tails in the news tensor scale as $G^2$ as they originate from interactions and therefore $f^\pm_\text{tail}$ scales as $G^4$. 

We now add Eq. \eqref{eqm1} with the equation obtained from acting with the antipodal map on \eqref{eqm2}. Using the identity \eqref{idm0} at spatial infinity and the substitutions \eqref{rho1}-\eqref{rho2} we obtain 
\begin{align}
m^{(0)}\vert_{\mathcal I^+_+}-\Upsilon^* m^{(0)}\vert_{\mathcal I^-_-} &= \frac{1}{4} \nabla^A\nabla^B[C_{AB}^{(0)}\vert_{\scri^+_+}-C_{AB}^{(0)}\vert_{\scri^+_-}]+\Upsilon^* \frac{1}{4} \nabla^A\nabla^B[C_{AB}^{(0)}\vert_{\scri^-_+}-C_{AB}^{(0)}\vert_{\scri^-_-}] \nonumber\\
&\hspace{-2cm}- 4\pi\mathop{\sum_{i }}_{m_{(i)}=0} E_{(i)} \delta(\phi,\phi_{(i)})-f^+_\text{tail}(\phi,\{\phi_{(j)}\}) + \Upsilon^*f^-_\text{tail}(\phi,\{\phi_{(j)}\})\label{leadSoft}
\end{align}
where the sum over $i$ is over both the incoming and outgoing particles. Note that the contribution at spatial infinity of $C_{AB}^{(0)}$ does not cancel out on the right-hand side since the antipodal identification of $C_{AB}^{(0)}$ contains a minus sign, see Eq. \eqref{idC0}. We can now substitute the massive contributions $m^{(0)}\vert_{\mathcal I^+_+}$ and $m^{(0)}\vert_{\mathcal I^-_-}$ using Eq. \eqref{valsm0}. We can furthermore reabsorb the massless contributions in the qualitative form of massive contributions using the identity
\begin{equation}\label{idm}
\lim_{m_{(i)}\to0}\frac{m_{(i)}^4}{(p_{(i)} \cdot n)^3}=-4\pi E_{(i)} \delta(\phi,\phi_{(i)}). 
\end{equation}
We finally obtain that the sum of the incoming and outgoing displacement memories is determined in terms of the Bondi aspect as 
\begin{align}
&\nabla^A\nabla^B[C_{AB}^{(0)}\vert_{\scri^+_+}-C_{AB}^{(0)}\vert_{\scri^+_-}]+\Upsilon^*\nabla^A\nabla^B[C_{AB}^{(0)}\vert_{\scri^-_+}-C_{AB}^{(0)}\vert_{\scri^-_-}]=-4 \sum_{i}\frac{m_{(i)}^4}{(p_{(i)} \cdot n)^3}\nonumber\\ 
&+4[f^+_\text{tail}(\phi,\{\phi_{(j)}\})-\Upsilon^*f^-_\text{tail}(\phi,\{\phi_{(j)}\})]. \label{leadingsoft}
\end{align}
This is simply the reformulation of the leading classical soft graviton theorem \cite{Weinberg:1965nx} derived in \cite{Boschetti:2026gfd} when we take the hypothesis that the tail contributions exactly compensate: 
\begin{align}
f^+_\text{tail}(\phi,\{\phi_{(j)}\})=\Upsilon^*f^-_\text{tail}(\phi,\{\phi_{(j)}\}).\label{tailc}
\end{align}
We did not derive \eqref{tailc} but we think that it is a reasonable hypothesis given the proven leading order matching of the tail \eqref{idC4} and the expected similar matching of subleading tails. The hypothesis is also consistent with the conservation of energy-momentum \eqref{MomentumEnergyConservation} obtained by integrating Eq. \eqref{leadingsoft} times $n^\mu$ over the sphere.

\paragraph{Logarithmic subleading soft theorem}
Now, let us perform the analogous derivation, which was done in \cite{Boschetti:2026gfd}, for the logarithmic soft theorem. 

Let us start with the antipodal matching condition \eqref{cDABid} which, upon using Eqs. \eqref{defDA}, \eqref{NALog} and \eqref{CABBzero}, can be rewritten as 
\begin{align}
  &  \nabla^B D^{(0)}_{AB}\vert_{\scri^{+} }+ N_A^{(\log)}\vert_{\scri^{+}_- }+2 \Big(3 P_i \nabla_A n_i+P_\mu n^\mu \nabla_A\Big)m\vert_{\scri^{+}_- }\nn\\&= -\Upsilon^*\nabla^B D^{(0)}_{AB}\vert_{\scri^{-} }- \Upsilon^*N_A^{(\log)}\vert_{\scri^{-}_+ }-2\Big(3 P_i \nabla_A n_i+P_\mu n^\mu\nabla_A\Big)\Upsilon^*m\vert_{\scri^{-}_+ }.\label{CurlyDAMatch}
\end{align}
Using the matching at $i^\pm$ \eqref{BCMAtch1}, along with the matching of the asymptotic Bondi mass aspect \eqref{idm0}, we get 
\begin{align}
  &  N_A^{(\log)}\vert_{\scri^+_+} - N_A^{(\log)}\vert_{\scri^+_-}-\Upsilon^*[N_A^{(\log)}\vert_{\scri^-_+} - N_A^{(\log)}\vert_{\scri^-_-}]=2  \mathop{\sum_{i\in+}}_{m_{(i)}\ne0} \tilde{s}_A[p_{(i)},c_{(i)}^{\text{rad}+},n^+]\nn\\&\hspace{1cm}+2  \mathop{\sum_{i\in -}}_{m_{(i)}\ne0} \tilde{s}_A[p_{(i)},c_{(i)}^{\text{rad}-},n^+]+4 \Big(3 P_i \nabla_A n_i+P_\mu n^\mu\nabla_A\Big)\Upsilon^*m\vert_{\scri^{-}_+}. \label{NAeq}
\end{align}
We note that the contribution of $N^{(\text{log})}_A$ at spatial infinity does not cancel in the left-hand side since the antipodal relationship \eqref{NAlogMatch} has a positive sign. We now express $m\vert_{\scri^{-}_+}$ using Eq. \eqref{eqm2} and \eqref{rho2} and use Eq.  \eqref{valsm0} to substitute $m^{(0)}\vert_{\mathcal I^-_-}$ to obtain
\begin{align}
&4 \Big(3 P_i \nabla_A n_i+P_\mu n^\mu\nabla_A\Big)\Upsilon^*m\vert_{\scri^{-}_+}=  -8 \sum_{i\in -}\tilde{s}_A[p_i,P,n] \nonumber\\
&\qquad +4 \Big(3 P_i \nabla_A n_i+P_\mu n^\mu\nabla_A\Big)\Upsilon^*\Big(\frac{1}{4}\nabla^A\nabla^B[C_{AB}^{(0)}\vert_{\scri^-_+}-C_{AB}^{(0)}\vert_{\scri^-_-}]+f^-_\text{tail}(\phi,\{\phi_{(j)}\})\Big)\label{eq:45}
\end{align}
where the properties \eqref{sAdef} and \eqref{idm} were used. Substituting Eq. \eqref{eq:45} into the last term of Eq. \eqref{NAeq} finally provides the rewriting of the logarithmic classical soft theorem as written in Eq. (5.16) of \cite{Boschetti:2026gfd} up to the tail contribution.  This tail contribution identically vanishes if one assumes that 
\begin{align}
f^-_\text{tail}(\phi,\{\phi_{(j)}\})=\frac{C_\text{tail}(\phi_{(j)})}{(P \cdot n^-)^3} 
\end{align}
where $P^\mu$ is the total four-momentum and $C_\text{tail}(\phi_{(j)})$ is a constant that may depend upon the directions of the incoming bodies. This is our prescription for determining the effective tail factor. 

\subsection{Soft theorems: Comparison with Laddha-Saha-Sahoo-Sen}

The total displacement memory accumulated over $\scri^-$ and $\scri^+$ takes the form \cite{1974SvA....18...17Z,PhysRevD.46.4304,1987Natur.327..123B}
\begin{align}
&\nabla^A\nabla^B[C_{AB}^{(0)}\vert_{\scri^+_+}-C_{AB}^{(0)}\vert_{\scri^+_-}]+\Upsilon^*  \nabla^A\nabla^B[C_{AB}^{(0)}\vert_{\scri^-_+}-C_{AB}^{(0)}\vert_{\scri^-_-}]=-4 \sum_{i}\frac{m_{(i)}^4}{(p_{(i)} \cdot n)^3}. 
\end{align}
In the Laddha-Saha-Sahoo-Sen (LSSS) framework \cite{Laddha:2018myi,Laddha:2018vbn,Sahoo:2018lxl,Saha:2019tub,Sahoo:2021ctw}, the leading behavior of the shear obeys 
\begin{align}
&\nabla^A\nabla^B[C_{AB}^{(0)}\vert_{\scri^+_+}-C_{AB}^{(0)}\vert_{\scri^+_-}]=-4 \sum_{i}\frac{m_{(i)}^4}{(p_{(i)} \cdot n)^3}.\label{LSSS1}
\end{align}
which is compatible with the absence of incoming displacement memory
\begin{equation}
\nabla^A\nabla^B[C_{AB}^{(0)}\vert_{\scri^-_+}-C_{AB}^{(0)}\vert_{\scri^-_-}]=0\label{noinc}.
\end{equation}
Eq. \eqref{LSSS1} is also compatible with our formulation of the soft graviton theorem \eqref{leadingsoft} after assuming Eq. \eqref{noinc} and the cancellation of tails \eqref{tailc}. 

In the absence of incoming displacement memory \eqref{noinc} and in the absence of tail contributions in \eqref{leadingsoft}, the classical logarithmic subleading soft graviton theorem \eqref{NAeq} (with the substitution \eqref{eq:45}) gives
\begin{align}
  &  N_A^{(\log)}\vert_{\scri^+_+} - N_A^{(\log)}\vert_{\scri^+_-}-\Upsilon^*[N_A^{(\log)}\vert_{\scri^-_+} - N_A^{(\log)}\vert_{\scri^-_-}]=2  \mathop{\sum_{i\in+}}_{m_{(i)}\ne0} \tilde{s}_A[p_{(i)},c_{(i)}^{\text{rad}+},n]\nn\\&\hspace{1cm}+2  \mathop{\sum_{i\in -}}_{m_{(i)}\ne0} \tilde{s}_A[p_{(i)},c_{(i)}^{\text{rad}-},n] -8 \sum_{i\in -}\tilde{s}_A[p_i,P,n]  .\label{NAeq2}
\end{align}

At subleading order, the results of LSSS reformulated in \cite{Boschetti:2026ogm} give in our notation
\begin{align}
     N_A^{(\log)}\vert_{\scri^+_+} =2  \mathop{\sum_{i\in+}}_{m_{(i)}\ne0} \tilde{s}_A[p_{(i)},c_{(i)}^{\text{rad}+},n]-4\sum_{i\in -}\tilde{s}_A[p_{(i)},P,n],\label{LSSSResults1}\\
     N_A^{(\log)}\vert_{\scri^+_-} =-2  \mathop{\sum_{i\in -}}_{m_{(i)}\ne0} \tilde{s}_A[p_{(i)},c_{(i)}^{\text{rad}-},n]+4\sum_{i\in -}\tilde{s}_A[p_{(i)},P,n].\label{LSSSResults2}
\end{align}
Using the antipodal matching \eqref{NAlogMatch}, \eqref{NAeq} and \eqref{BCMAtch1}-\eqref{LSSSResults1}-\eqref{LSSSResults2}, we find 
\begin{align}
N_A^\text{(log)}\vert_{\mathcal I^-_+} =0,\qquad N_A^\text{(log)}\vert_{\mathcal I^-_-} =0,\label{NoIncomingRadiation}
\end{align} 
which is compatible with the non-incoming radiation condition $N_{AB}^- =0$. This result which combines our matching conditions with the results of \cite{Saha:2019tub} and \cite{Boschetti:2026gfd} (see also Eq. \eqref{BCMAtch1}) is compatible with the two postulated antipodal matching conditions in \cite{Boschetti:2026ogm}. More precisely, if we consider (\eqref{idC4}$+$\eqref{cDABid}) and (\eqref{idC4}$-$\eqref{cDABid}), we get : 
\begin{align}
    \Upsilon^*\nabla^B D_{AB}^{+(0)} = \frac{1}{2} \nabla_C \nabla_{\langle A} \nabla_{B\rangle} C^{- (1) BC} + 6 P_i \mathcal{M}^{-(0)}_{AB}\nabla^B n_i+2 (E+ P_i n_i) \nabla^B\mathcal{M}^{-(0)}_{AB}\\
    \nabla^B D_{AB}^{-(0)} = \frac{1}{2} \Upsilon^*(\nabla_C \nabla_{\langle A} \nabla_{B\rangle} C^{+ (1) BC}) + 6 P_i \mathcal{M}^{-(0)}_{AB}\nabla^B n_i+ 2 (E+ P_i n_i) \nabla^B\mathcal{M}^{-(0)}_{AB}
\end{align}
which correspond exactly to Eqs. (73) and (74) in \cite{Boschetti:2026ogm}.

\subsection{Violation of peeling: Comparison with Damour}
Let us review the comparison performed in  \cite{Boschetti:2026ogm}, using our notation, and prove that our results are compatible with Damour's results \cite{Damour:1985cm}. 
Let us still assume the absence of incoming displacement memory \eqref{noinc} and we assume the LSSS formulae \eqref{LSSSResults1}-\eqref{LSSSResults2}. Substituting Eqs. \eqref{LSSSResults1}-\eqref{NoIncomingRadiation} into Eq. \eqref{BCMAtch1}, we find the leading order violation of peeling terms at $\scri^+$ and $\scri^-$:
\begin{align}
   &\nabla^B D_{AB}\vert_{\scri^{+} } = 4 \sum_{i\in -}\tilde{s}_A[p_{(i)},P,n]\label{eq1},\\
   &\nabla^B D_{AB}\vert_{\scri^{-} } = 2 \mathop{\sum_{i\in -}}_{m_{(i)}\ne0} \tilde{s}_A[p_{(i)},c_{(i)}^{\text{rad}-},n^-].\label{eq2}
\end{align}
These can be reexpressed in terms of the Weyl tensor  \eqref{Weyl1}-\eqref{Weyl2} as 
\begin{align}
\Psi_0\vert_{\scri^+} &= \frac{2}{r^4} \sum_{i\in -}\tilde{s}_{AB}[p_{(i)},P,n]m_1^A m_1^B+o(r^{-4}),\\ 
\Psi_4\vert_{\scri^-} &=\frac{1}{r^4}\Big(\mathop{\sum_{i\in -}}_{m_{(i)}\ne0} \tilde{s}_{AB}[p_{(i)},c_{(i)}^{\text{rad}-},n^-]\Big)\bar m_1^A \bar m_1^B+o(r^{-4}), 
\end{align}
where $\tilde s_{AB}$ is explicitly given by  \cite{Boschetti:2026ogm}
\begin{align}
 \tilde s_{AB}[p,b,n] &=\left( 4 \frac{p_{\langle A}p_{B \rangle}p_C}{(p \cdot n)^3}\partial^C n^\mu n^\nu-\frac{2p_C p^C p_{\langle A} \partial_{B \rangle} n^\mu n^\nu}{(p \cdot n)^3} +\frac{6}{(p \cdot n)^2} p_{\langle A} \partial_{B \rangle} n^\mu p^\nu\right)b_{[\mu}p_{\nu ]},  \label{saB} 
\end{align}
with $p_A(x^B) := p \cdot \partial_A n$. By explicit computation, we find
\begin{align}
\nabla^B \tilde s_{AB}=\tilde s_A +3 \nabla_A n^\mu (\nabla^2 +1)n^\nu  b_{[\mu}p_{\nu ]}      
\end{align}
where $\tilde s_A$ is defined in Eq. \eqref{sAdef}. Therefore, $\sum_{i \in -}\nabla^B \tilde s_{AB}[p,P,n] = \sum_{i \in -}\tilde s_A$ as $\sum_{i \in -} P_{[\mu}p_{\nu ]} =0$. Also, $\mathop{\sum_{i\in -}}_{,\,m_{(i)}\ne0}\nabla^B \tilde s_{AB}[p,P,n] = \mathop{\sum_{i\in -}}_{,\,m_{(i)}\ne0}\tilde s_A$ as $\mathop{\sum_{i\in -}}_{,\,m_{(i)}\ne0} c^{\text{rad}-}_{[\mu}p_{\nu ]} =0$ \cite{Boschetti:2026ogm}.

In the Newtonian limit, incoming $p_{(i)}$ take the form $p^\mu_{(i)}\approx -(m_{(i)},\vec{p}_{(i)})$ while outgoing $p_{(i)}$ take the form $p^\mu_{(i)}\approx (m_{(i)},\vec{p}_{(i)})$. The total momentum is at leading Newtonian order $P^\mu \approx (M,\vec{0})$. The radiative logarithmic deviation vector takes the form $c_{(i)}^{\text{rad}\pm \mu}= (\pm 2M, \vec{c}^\text{rad  N}_{(i)})$ where $\vec{c}^\text{rad N}_{(i)}= \sum_{j \neq i} \frac{m_{(j)}}{\vert \vec{v}_{(j)}-\vec{v}_{(i)}\vert^3}(\vec{v}_{(j)} - \vec{v}_{(i)})$. At leading Newtonian order, only the third term in Eq. \eqref{saB} contributes and gives at $\scri^\pm$
\begin{align}
 \tilde s_{AB}[p,b,n^\pm]=\pm 3 (\vec{p} \cdot \partial_{\langle A}\vec{n})[\partial_{B \rangle}(\vec{n} \cdot \vec{b})-(\partial_{B\rangle} \vec{n} \cdot \vec{p})\frac{b^0}{m}]. 
\end{align}

At $\scri^+$, we obtain 
\begin{align}
\Psi_0\vert_{\scri^+} &= -\frac{3M}{r^4}A_{ij}^-  e^i_{\langle A} e^j_{B \rangle} m_1^A m_1^B+o(r^{-4})
\end{align}
where $A_{ij}^- := 2 \sum_{k \in -} m_{(k)}v^{\langle i}_{(k)}v^{j \rangle}_{(k)}$. This exactly reproduces the leading violation of peeling at $\scri^+$ in the mass-quadrupole approximation obtained in Eq. (4.42) of \cite{Damour:1985cm}, as also checked in \cite{Boschetti:2026ogm}. 

At $\scri^-$, we obtain 
\begin{align}
\Psi_4\vert_{\scri^-} &=\frac{1}{r^4}\frac{3}{2} B_{ij}^- e^i_{\langle A} e^j_{B \rangle} \bar m_1^A \bar m_1^B+o(r^{-4}).     \label{eq2bis}
\end{align}
where $B_{ij}^- :=  2\sum_{i \in -} m_{(i)}v^{\langle i}_{(i)}c^{\text{rad N}j \rangle}_{(i)}$. After changing conventions (Damour's $\ell^\mu$ is our $\sqrt{2}\ell^\mu$ and his $n^\mu$ is our $\frac{1}{\sqrt{2}}n^\mu$ with $m_1^A$ identical), this second result corresponds the violation of peeling at $\scri^-$ obtained in the literature \cite{PhysRevD.19.3495,Damour:1985cm}, as proven in \cite{Boschetti:2026ogm}.

\section{Conclusion}
\label{sec:ccl}

We derived the conservation law of the super-Lorentz charge at spatial infinity. It completes the other two conservation laws at subleading order (the conservation of leading tails and the conservation of leading violation of peeling) already derived in \cite{Compere:2026jmk}. Our result \eqref{curlyN0}-\eqref{NAlaw} significantly differs from the original proposal of 2014 \cite{Kapec:2014opa} that motivated this work, as our result takes into account the tail effects in the shear as well as the effects of breaking of peeling and the shift of logarithmic translation frame at spatial infinity proportional to the total spacetime momentum; it also allows for a magnetic leading shear at spatial infinity, besides correcting the global minus sign of the original conjecture as already pointed out in \cite{Compere:2023qoa}.

Our systematic construction of conserved charges at the corners of spatial infinity extends earlier constructions in the literature where some of these charges have been defined under more constrained assumptions on the news and the matter fields
\cite{Flanagan:2015pxa,Blanchet:2023pce,Compere:2018ylh,Compere:2022zdz,Grant:2021hga}. The conservation of super-Lorentz charges is in a sense a conservation law of the subleading multipolar structure of the gravitational field \cite{Compere:2022zdz,kmec2026quasilocalcelestialchargesmultipoles}. 

We checked that our results are compatible with the classical leading soft graviton theorem and the logarithmic soft graviton theorem, with some caveats that we highlighted, namely, the existence of tail effects, which were not explicitly included in the past literature. Our demonstrated antipodal matching condition \eqref{idC4} is compatible with the results of Sahoo, Saha, Sen and Laddha \cite{Laddha:2018myi,Laddha:2018vbn,Sahoo:2018lxl,Saha:2019tub,Sahoo:2021ctw} and demonstrates under our assumptions the conjectured antipodal map given in Eq. (73)-(74) of \cite{Boschetti:2026ogm}.  Our results combined with the ones of Sahoo, Saha, Sen and Laddha are compatible with the seminal result of Damour on the violation of peeling at $\scri^+$ \cite{Damour:1985cm}.  We expect that the non-local feature of the matching of the angular momentum aspect will also appear in a related form in the loop corrections to the subleading soft graviton theorem \cite{He_2014,Donnay_2022,alessio20242pmwaveformloopcorrected}.

In the alternative approach of Kehrberger and Masaood \cite{Kehrberger:2024aak}, a linear solution around Schwarzschild is found which admits $C^{(1)+}_{AB}\ne 0$ while $C^{(1)-}_{AB}= 0$. The hypotheses of that paper match with our Hypotheses 1 and 2 as their class of spacetimes admits a polyhomogeneous expansion of the form \eqref{BondiSachsMetric}-\eqref{hABdef}-\eqref{BondihExp} at $\scri^+$ and at $\scri^-$ and the shear admits a leading tail of the form \eqref{CABasymptoticsscri}. It would be interesting to verify that the antipodal matching conditions that we derived are obeyed in their analysis.  
 
Given our all-order analysis \cite{Compere:2025bnf} we expect that subsubleading and higher subleading charges can be constructed following the same methods. It would be interesting to obtain an algebra of these charges, and compare it with a proposed symmetry algebra for Einstein gravity \cite{Strominger:2021mtt}. 

\section*{Acknowledgements}
We acknowledge discussions with Gianni Boschetti, Miguel Campiglia, Dima Fontaine, Kevin Nguyen and Ana Raclariu. We gratefully thank our referee Thibault Damour for his useful comments and suggestions. We acknowledge Ana-Maria Raclariu, Laurent Freidel and Domenico Salluce for helping to spot a typo in an earlier version of the manuscript. G.C. would like to thank the organizers of the IFPU Focus Week, "Peeling the gravitational onion” in
Trieste. G.C. is Research Director of the FNRS.  



\appendix
\numberwithin{equation}{section}

\section{Asymptotic behavior of sourced fields at second order in Beig-Schmidt expansion}
\label{sec:asymptBehav}
\subsection{Generic case with convergent source}
In this appendix, we study the asymptotic behavior of transverse vectors $V_a$ that satisfy the equation 
\begin{equation}
    (\Box-1)V_a=S_a\label{GeneralVaEqu} . 
\end{equation}
The source $S_{a}$ is transverse and admits the following asymptotic behavior 
\begin{align}
    S_{\tau}(\tau\to\pm\infty,x^A)&= \mp 16 \nabla^AS^{(0)\pm}_A(x^A)e^{-3\vert\tau\vert} + o(e^{-4\vert\tau\vert}),\label{sourceBehav1tau}\\
    S_{ A}(\tau\to\pm\infty,x^A)&= 4 S^{(0)\pm}_A(x^A)e^{-\vert\tau\vert} + o(e^{-2\vert\tau\vert}).
\end{align}
This exactly corresponds to the equation satisfied by $J_a$ and the behavior of its source. 
We are interested in subtracting contributions from the source while keeping control of the behavior at $\tau\to\pm \infty$. 
All subsequent computations will be based on our earlier analysis  \cite{Compere:2025bnf}. 

To find the general solution to Eq. \eqref{GeneralVaEqu}, we define the following scalars: 
\begin{align}
    \eta^E = \cosh\tau n^a V_{a}, \qquad \eta^B = \cosh\tau n^a  \text{Curl}(V_a),\\
    \kappa^E = \cosh\tau n^a S_{a}, \qquad \kappa^B = \cosh\tau n^a  \text{Curl}(S_a).
\end{align}
Eq. \eqref{GeneralVaEqu} can then equivalently be written as 
\begin{equation}
    \Box  \eta^D = \kappa^D,\qquad D=E,B,  \label{scalarSDTEquation}
\end{equation}
where the sources $\kappa^D$ ($D=E,B$) admit the following asymptotic behavior 
\begin{equation}
    \kappa^D(\tau\to\pm\infty,x^A) = \kappa^{D\pm(0)}(x^A) e^{-2\vert\tau\vert} + o(e^{-3\vert\tau\vert}).\label{kappaAsympt}
\end{equation}
The solution to Eq. \eqref{scalarSDTEquation} is (see \cite{Compere:2025bnf} with $n=0$):
\begin{align}
  \eta^D &= \sum_{\ell,m} \bigg[ A^{(D)}_{\ell m}(\tau; \bar\tau)\psi^{(p)}_{0\ell m}(\tau,x^A)  
   +B^{(D)}_{\ell m}(\tau; \bar\tau) \psi^{(q)}_{0\ell m}(\tau,x^A)\bigg] \label{ScalNHSol},\quad D=E,B,
\end{align}
where $\bar\tau$ is the time corresponding to a Cauchy surface and 
\begin{align}\label{ASscalar}
    A^{(D)}_{\ell m}(\tau ; \bar\tau)&:= a_{\ell m}^{(D)}+\int_{\bar\tau}^\tau d\tau'  \cosh^2\tau'\int_{S^2} d^2\Omega \,\kappa^D(\tau',x^A)\overline{\psi^{(q)}_{0\ell m}(\tau',x^A)} , \\
\label{BSscalar}
    B^{(D)}_{\ell m}(\tau; \bar\tau)&:=b_{\ell m}^{(D)}-\int_{\bar\tau}^\tau  d\tau' \cosh^2\tau'\int_{S^2} d^2\Omega \,\kappa^D(\tau',x^A)\overline{\psi^{(p)}_{0\ell m}(\tau',x^A)}.
\end{align}

Plugging Eq. \eqref{kappaAsympt} into Eqs. \eqref{ASscalar}-\eqref{BSscalar} and using the asymptotics of the scalar harmonics derived in \cite{Compere:2025bnf} we get  
\begin{align}
    A_{\ell m}^{(D)}(\tau\to\pm\infty;\bar\tau) &=-(\pm1)^{\ell+1}\frac{\vert\tau\vert }{4} \int_{S^2} d^2\Omega \, \kappa^{D\pm(0)}(x^A)
     \overline{Y_{\ell m}(x^{A})}+\tilde A^{(D)\pm}_{\ell m}(\bar \tau) + o(e^{-\vert\tau\vert}),\label{Atopminfty}\\
    B_{\ell m}^{(D)}(\tau\to\pm\infty;\bar\tau) &= \tilde B^{(D)\pm}_{\ell m}(\bar \tau) -\frac{(\pm1)^{\ell}}{4}e^{-2\vert\tau\vert} \int_{S^2}d^2\Omega  \kappa^{D\pm(0)}(x^A) \overline{Y_{\ell m}(x^{A})} + o(e^{-3\vert\tau\vert}),\label{Btopminfty}
\end{align}
where $\tilde A^{(D)\pm}_{\ell m}(\bar \tau)$ and $\tilde B^{(D)\pm}_{\ell m}(\bar \tau)$ are integration constants that depend upon $a^{(D)}_{\ell m}$, $b^{(D)}_{\ell m}$. They are related by 
\begin{align}
    &\tilde A^{(D)+}_{\ell m}(\bar \tau)-\tilde A^{(D)-}_{\ell m}(\bar \tau)=\text{FP}\int_{\text{dS}_3} d^3\phi\,\sqrt{-q} \kappa^D(\phi^a) \overline{\psi^{(q)}_{0\ell m}(\phi^a)} \label{DiffAD}\\
    &\tilde B^{(D)+}_{\ell m}(\bar \tau)-\tilde B^{(D)-}_{\ell m}(\bar \tau) =-\int_{\text{dS}_3} d^3\phi\,\sqrt{-q}  \kappa^D(\phi^a) \overline{\psi^{(p)}_{0\ell m}(\phi^a)},\label{DiffBD}
\end{align}
where the Finite Part (FP) removes the linearly divergent piece of the integral\footnote{For $f(\tau\to\pm\infty)=f_0^\pm+o(e^{-\vert\tau\vert})$, we use the prescription $FP \int_{-\infty}^{\infty} f(\tau) d\tau  = \int_{0}^{\infty} (f(\tau) -f_0^+) d\tau+\int_{-\infty}^{0} (f(\tau) -f_0^-) d\tau$.}. An important property is that the right-hand sides of Eqs. \eqref{DiffAD}-\eqref{DiffBD} do not depend upon $\bar\tau$. This allows us to rewrite the constants as follows 
\begin{align}
     &\tilde A^{(D)+}_{\ell m}(\bar \tau)=\frac{1}{2}\text{FP}\int_{\text{dS}_3} d^3\phi\,\sqrt{-q}  \kappa^D(\phi^a) \overline{\psi^{(q)}_{0\ell m}(\phi^a)} + \tilde A^{(D)}_{\ell m}(\bar \tau)\label{ApParityplus},\\&\tilde A^{(D)-}_{\ell m}(\bar \tau)=-\frac{1}{2}\text{FP}\int_{\text{dS}_3} d^3\phi\,\sqrt{-q} \kappa^D(\phi^a) \overline{\psi^{(q)}_{0\ell m}(\phi^a)} + \tilde A^{(D)}_{\ell m}(\bar \tau),\label{ApParityminus}\\
    &\tilde B^{(D)+}_{\ell m}(\bar \tau) =-\frac{1}{2}\int_{\text{dS}_3} d^3\phi\,\sqrt{-q}  \kappa^D(\phi^a) \overline{\psi^{(p)}_{0\ell m}(\phi^a)}+\tilde B^{(D)}_{\ell m}(\bar \tau),\label{BpParityplus}\\
    &\tilde B^{(D)-}_{\ell m}(\bar \tau) =\frac{1}{2}\int_{\text{dS}_3} d^3\phi\,\sqrt{-q}  \kappa^D(\phi^a) \overline{\psi^{(p)}_{0\ell m}(\phi^a)}+\tilde B^{(D)}_{\ell m}(\bar \tau)\label{BpParityminus},
\end{align}
where $\tilde A^{(D)}_{\ell m}(\bar \tau)$ and $\tilde B^{(D)}_{\ell m}(\bar \tau)$ are constant. The finite part prescription is defined in a specified Lorentz frame. We do study its transformation laws under a change of frame.

We will study asymptotic solutions that have the same parity as the source regardless of its parity. This can be achieved by defining the inhomogeneous solution where $\tilde A^{(D)}_{\ell m}(\bar \tau)=\tilde B^{(D)}_{\ell m}(\bar \tau)=0$. We call the resulting inhomogeneous solution the \emph{primitive inhomogeneous solution}.  The parity properties of $\kappa^D$ will then imply parity properties of the so-defined primitive inhomogeneous solution $\eta^D$. 

We can check that, if the source has $p$-parity, i.e. the source is odd, then the constants \eqref{ApParityplus}-\eqref{ApParityminus} vanish and $\tilde B^{(D)+}_{\ell m}(\bar \tau) =-\tilde B^{(D)-}_{\ell m}(\bar \tau) $ (see \eqref{BpParityplus}-\eqref{BpParityminus}). This implies, in Eqs. \eqref{Atopminfty}-\eqref{Btopminfty}, that 
\begin{align}
    A_{\ell m}^{(D)}(\tau\to+\infty;\bar\tau)=A_{\ell m}^{(D)}(\tau\to-\infty;\bar\tau) \\ \intertext{and} B_{\ell m}^{(D)}(\tau\to+\infty;\bar\tau)=-B_{\ell m}^{(D)}(\tau\to-\infty;\bar\tau).
\end{align}
 Injecting these asymptotic parity properties in Eq. \eqref{ScalNHSol}, we can check that the primitive inhomogeneous solution $\eta^D$ has (asymptotically) $p$-parity, i.e. 
 \begin{equation}
     \eta^D(\tau\to+\infty,x^A)=-\Upsilon^*\eta^D(\tau\to-\infty,x^A).
 \end{equation}
 The same reasoning applies for $q$-parity sources. 

From the solution \eqref{ScalNHSol}, we can recover the vector solution to Eq. \eqref{GeneralVaEqu} using the decomposition (see \cite{Compere:2025bnf}): 
\begin{align}
   & V_a =\sum_{\ell=1}^{+\infty}\sum_{m=-\ell}^\ell \bigg\{- \frac{ A^{(B)}_{\ell m}(\tau;\bar\tau)}{\sqrt{\ell(\ell+1)}}V^{(B,p)0\ell m}_a - \frac{ B^{(B)}_{\ell m}(\tau;\bar\tau)}{\sqrt{\ell(\ell+1)}} V^{(B,q)0\ell m}_a \nn\\&\;\;- \epsilon_{a}^{\,bc}\mathcal{D}_b \left(\frac{ A^{(E)}_{\ell m}(\tau;\bar\tau) }{\sqrt{\ell(\ell+1)}} V^{(B,p)0\ell m}_c\right)- \epsilon_{a}^{\,bc}\mathcal{D}_b \left(\frac{ B^{(E)}_{\ell m}(\tau;\bar\tau) }{\sqrt{\ell(\ell+1)}} V^{(B,q)0\ell m}_c\right) \bigg\} .\label{VTNHdecomp}
\end{align}
where $V^{(D,c)0\ell m}_a$ ($D=E,B$ and $c=p,q$) are the $n=0$ transverse vector harmonics defined in \cite{Compere:2025bnf}. 
Taking the limit $\tau\to\pm\infty$ in \eqref{VTNHdecomp} and setting $\tilde A^{(D)}_{\ell m}(\bar \tau)=0$ and $\tilde B^{(D)}_{\ell m}(\bar \tau)=0$, we find the behavior of the primitive inhomogeneous solution $V_a^{(S)}$: 
\begin{align}
   V^{(S)}_{\tau}(\tau\to\pm\infty,x^A) =&\pm 4 \nabla^A\mathcal{S}_A^{(p)\pm}e^{-\vert\tau\vert}+o\left(e^{-2\vert\tau\vert}\right),\\
    V_{A}^{(S)}(\tau\to\pm\infty,x^A)  =& \mathcal{S}_A^{(p)\pm}e^{\vert\tau\vert}+2 \Big[\Big(\vert\tau\vert+\frac{1}{2}\Big)S_{A}^{(0)\pm}(x^A)-S_{A}^{(0)E\pm}(x^A)-\mathcal{S}_A^{(q)\pm}\nn\\ &-\vert\tau\vert\Big((\nabla^2-1)\delta_{A}^B-2\nabla_A\nabla^B\Big)\mathcal{S}_B^{(p)\pm}\nn\\ &-\frac{1}{2}O^{(q)n=0}_V(\mathcal{S}_A^{(p)\pm})+\frac{1}{2}\mathcal{S}_A^{(p)\pm}-\nabla_A\nabla^B\mathcal{S}_B^{(p)\pm}\Big] e^{-\vert\tau\vert}+ o( e^{-\vert\tau\vert})
\end{align}
where $S_{A}^{(0)E\pm}(x^A)$ is the $E$ parity piece of $S_{A}^{(0)\pm}(x^A)$. $\mathcal{S}_A^{(p)\pm}$ and $\mathcal{S}_A^{(q)\pm}$ are explicitly defined as : 
\begin{align}
    \mathcal S_A^{(p)\pm}(x^B)
    =\frac{1}{4}
    \sum_{\ell=1}^{+\infty}\sum_{m=-\ell}^{\ell}
    \bigg[
    &(\pm1)^\ell
    \mathring V_A^{(E)\ell m}(x^B)
    \int_{\mathrm{dS}_3} d^3\phi\,\sqrt{-q(\phi^b)}\,
    S_a(\phi^b)\,
    \overline{V^{(E,p)0\ell m\,a}(\phi^b)}
    \nonumber\\
    &+
    (\pm1)^{\ell+1}
    \mathring V_A^{(B)\ell m}(x^B)
    \int_{\mathrm{dS}_3} d^3\phi\,\sqrt{-q(\phi^b)}\,
    S_a(\phi^b)\,
    \overline{V^{(B,p)0\ell m\,a}(\phi^b)}
    \bigg],
    \label{SpExplicit}
\end{align}
and
\begin{align}
    \mathcal S_A^{(q)\pm}(x^B)
    =\frac{1}{4}\,\mathrm{FP}
    \sum_{\ell=1}^{+\infty}\sum_{m=-\ell}^{\ell}
    \bigg[
    &\frac{(\pm1)^{\ell+1}}{\sqrt{\ell(\ell+1)}}
    \mathring V_A^{(E)\ell m}(x^B)
    \int_{\mathrm{dS}_3} d^3\phi\,\sqrt{-q(\phi^b)}\,
    \kappa^{E}(\phi^b)\,
    \overline{\psi^{(q)}_{0\ell m}(\phi^b)}
    \nonumber\\
    &-
    \frac{(\pm1)^\ell}{\sqrt{\ell(\ell+1)}}
    \mathring V_A^{(B)\ell m}(x^B)
    \int_{\mathrm{dS}_3}d^3\phi\,\sqrt{-q(\phi^b)}\,
    \kappa^{B}(\phi^b)\,
    \overline{\psi^{(q)}_{0\ell m}(\phi^b)}\,
    \bigg].
    \label{SqExplicit}
\end{align}
Equivalently, these expressions can be written as  bulk-to-boundary maps
\begin{align}
    \mathcal S_A^{(p)\pm}(x^B)
    &=
    \int_{\mathrm{dS}_3} d^3\phi\,\sqrt{-q(\phi^b)}\,
    \mathcal P_A^{(p)\pm\,a}(\phi^b;x^B)\,
    S_a(\phi^b),
    \label{SpCompact}
    \\
    \mathcal S_A^{(q)\pm}(x^B)
    &=
    \mathrm{FP}
    \int_{\mathrm{dS}_3} d^3\phi\,\sqrt{-q(\phi^b)}\,
    \mathcal P_A^{(q)\pm\,a}(\phi^b;x^B)\,
    S_a(\phi^b),
    \label{SqCompact}
\end{align}
where the $(p)$-parity and $(q)$-parity bulk/boundary operators are defined by
\begin{align}
    \mathcal P_A^{(p)\pm\,a}(\phi^b;x^B)
    :=
    \frac{1}{4}
    \sum_{\ell=1}^{+\infty}\sum_{m=-\ell}^{\ell}
    \bigg[
    &(\pm1)^\ell
    \mathring V_A^{(E)\ell m}(x^B)\,
    \overline{V^{(E,p)0\ell m\,a}(\phi^b)}
    \nonumber\\
    &+
    (\pm1)^{\ell+1}
    \mathring V_A^{(B)\ell m}(x^B)\,
    \overline{V^{(B,p)0\ell m\,a}(\phi^b)}
    \bigg],
    \label{PpDef}
\end{align}
and
\begin{align}
    \mathcal P_A^{(q)\pm\,a}(\phi^b;x^B)
    :=
    \frac{1}{4}
    \sum_{\ell=1}^{+\infty}\sum_{m=-\ell}^{\ell}
    \frac{1}{\sqrt{\ell(\ell+1)}}
    \bigg[
    &(\pm1)^{\ell+1}
    \mathring V_A^{(E)\ell m}(x^B)\,\overline{\psi^{(q)}_{0\ell m}(\phi^b)}\,
    \cosh\tau\,n^a
    \nonumber\\
    &+
    (\pm1)^\ell
    \mathring V_A^{(B)\ell m}(x^B)\,
    \cosh\tau\, \overline{\psi^{(q)}_{0\ell m}(\phi^b)}
    \epsilon^{abc}n_b\,
    \mathcal D_c
    \bigg].
    \label{PqDef}
\end{align}

As a direct consequence of the properties of vector harmonics on the sphere under the antipodal map, the quantities $\mathcal{S}_A^{(p)+}$ and $\mathcal{S}_A^{(q)+}$ defined in \eqref{SpExplicit}-\eqref{SqExplicit} satisfy the antipodal matching conditions : 
\begin{equation}
    \mathcal{S}_A^{(p)+} = \Upsilon^*\mathcal{S}_A^{(p)-},\qquad  \mathcal{S}_A^{(q)+} = -\Upsilon^*\mathcal{S}_A^{(q)-}\label{AntipodMatchalS}
\end{equation}
Hence, we have defined a particular inhomogeneous solution, the primitive inhomogeneous solution $V_a^{(S)}$,  with well-defined asymptotic properties as $\tau \to \pm \infty$. By construction the subtracted field $\hat V_a := V_a-V_a^{(S)}$ obeys the homogeneous wave equation $(\square-1)\hat V_a=0$.

\subsection{Localization of the $dS_3$ source term onto the sphere} 
\label{sec:localization}
We now consider the particular case in which the transverse source $S_a$ is itself an $n=0$, $p$-parity solution of the homogeneous vector equation,
\begin{equation}
(\Box-1)S_a=0,
\qquad
\mathcal D^a S_a=0.
\label{homogeneousSourceEq}
\end{equation}
Using the harmonic decomposition performed in \cite{Compere:2025bnf}, the most general such source can be expanded as
\begin{equation}
S_a
=
\sum_{\ell=1}^{+\infty}\sum_{m=-\ell}^{\ell}
\left(
s_{\ell m}^{E}\,
V_a^{(E,p)0\ell m}
+
s_{\ell m}^{B}\,
V_a^{(B,p)0\ell m}
\right).
\label{pSourceHarmonicDecomposition}
\end{equation}
We denote $L_\ell:=\ell(\ell+1)$ and recall the definition of the normalized electric- and magnetic-parity vector harmonics on the unit sphere,
\begin{equation}
\mathring V_A^{(E)\ell m}
:=
\frac{1}{\sqrt{L_\ell}}\nabla_A Y_{\ell m},
\qquad
\mathring V_A^{(B)\ell m}
:=
\frac{1}{\sqrt{L_\ell}}
\epsilon_A{}^B\nabla_B Y_{\ell m}.
\label{sphereVectorHarmonics}
\end{equation}

The scalar quantities associated with the source,
\begin{equation}
\kappa^E=\cosh\tau\,n^aS_a,
\qquad
\kappa^B=\cosh\tau\,n^a\text{Curl}(S_a),
\end{equation}
are particularly simple for the harmonic decomposition
\eqref{pSourceHarmonicDecomposition}. Indeed, using the definitions and
duality relations of the vector harmonics of
\cite{Compere:2025bnf}, one finds
\begin{equation}
\kappa^E
=
\sum_{\ell,m}
\sqrt{L_\ell} s_{\ell m}^{E}\,
\psi^{(p)0\ell m},
\qquad
\kappa^B
=
-\sum_{\ell,m}
\sqrt{L_\ell} s_{\ell m}^{B}\,
\psi^{(p)0\ell m}.
\label{kappaForpSource}
\end{equation}
The $n=0$ $p$-parity scalar harmonics obey
\begin{equation}
\int_{\mathrm{dS}_3}
d^3\phi\,\sqrt{-q}\,
\psi^{(p)0\ell m}(\phi^b)
\overline{\psi^{(p)0\ell' m'}(\phi^b)}
=
\frac{1}{L_\ell}
\delta_{\ell\ell'}\delta_{mm'}.
\label{pScalarL2Norm}
\end{equation}
It follows that
\begin{align}
s_{\ell m}^{E}=\sqrt{L_\ell}\int_{\mathrm{dS}_3}
d^3\phi\,\sqrt{-q}\,
\kappa^E
\overline{\psi^{(p)0\ell m}},\qquad
s_{\ell m}^{B}=-\sqrt{L_\ell}\int_{\mathrm{dS}_3}
d^3\phi\,\sqrt{-q}\,
\kappa^B
\overline{\psi^{(p)0\ell m}}.
\end{align}

Using the orthogonality of vector harmonics over the sphere, the contribution $\mathcal S_A^{(p)\pm}$ \eqref{SpExplicit} to the leading
asymptotic behavior of the primitive inhomogeneous solution therefore
reduces to
\begin{equation}
\mathcal S_A^{(p)\pm}
=
-\frac{1}{4}
\sum_{\ell=1}^{+\infty}\sum_{m=-\ell}^{\ell}
\frac{1}{L_\ell}
\left[
(\pm1)^\ell s_{\ell m}^{E}
\mathring V_A^{(E)\ell m}
-
(\pm1)^{\ell+1}s_{\ell m}^{B}
\mathring V_A^{(B)\ell m}
\right].
\label{SpHomogeneousSource}
\end{equation}

This bulk expression can be entirely localized on the asymptotic
sphere. Indeed, the $n=0$ $p$-parity vector harmonics behave as
\begin{align}
V_A^{(E,p)0\ell m}(\tau\rightarrow\pm\infty)
&=
-(\pm1)^\ell
\mathring V_A^{(E)\ell m}
e^{-|\tau|}
+o(e^{-|\tau|}),
\\
V_A^{(B,p)0\ell m}(\tau\rightarrow\pm\infty)
&=
-(\pm1)^{\ell+1}
\mathring V_A^{(B)\ell m}
e^{-|\tau|}
+o(e^{-|\tau|}).
\end{align}
Comparing with our convention
\begin{equation}
S_A(\tau\rightarrow\pm\infty,x^B)
=
4S_A^{(0)\pm}(x^B)e^{-|\tau|}
+o(e^{-|\tau|}),
\end{equation}
we obtain
\begin{equation}
S_A^{(0)\pm}
=
-\frac14
\sum_{\ell=1}^{+\infty}\sum_{m=-\ell}^{\ell}
\left[
(\pm1)^\ell s_{\ell m}^{E}
\mathring V_A^{(E)\ell m}
+
(\pm1)^{\ell+1}s_{\ell m}^{B}
\mathring V_A^{(B)\ell m}
\right].
\label{S0HomogeneousSource}
\end{equation}
Writing the electric-magnetic decomposition of the corner source as
\begin{equation}
S_A^{(0)\pm}
=
S_A^{(0)E\pm}
+
S_A^{(0)B\pm},
\end{equation}
comparison of Eqs.~\eqref{SpHomogeneousSource} and
\eqref{S0HomogeneousSource} gives, mode by mode,
\begin{equation}
\mathcal S_A^{(p)\pm}
=
\frac{1}{L_\ell}
\left(
S_A^{(0)E\pm}
-
S_A^{(0)B\pm}
\right).
\label{SpModeByMode}
\end{equation}

The last relation can be written without reference to the harmonic
decomposition. We define
\begin{equation}
K[f_A] := (\nabla^2 - 1)f_A-2\nabla_A \nabla^B f_B . \label{KOperator}
\end{equation}
Then the vector spherical harmonics satisfy
\begin{equation}
K[\mathring V_A^{(E)\ell m}]
=
L_\ell
\mathring V_A^{(E)\ell m},\qquad K[\mathring V_A^{(B)\ell m}]
=
-L_\ell
\mathring V_A^{(B)\ell m}
\end{equation}

The operator $K[.]$ is invertible in the space of $\ell \geq 1$ harmonics. We thus obtain the localized expression
\begin{equation}
\mathcal S_A^{(p)\pm}
=
K^{-1}[S_A^{(0)\pm}].
\label{SpLocalized}
\end{equation}

Hence, although $\mathcal S_A^{(p)\pm}$ is originally defined through
an integral of the source over the entire $\mathrm{dS}_3$
hyperboloid, when the source is itself an $n=0$ $p$-parity
homogeneous transverse vector its dependence localizes to the corner
data $S_A^{(0)\pm}$, up to the non-local operator
$K^{-1}$ on the celestial sphere.

\section{Inhomogeneous solution to field equations at spatial infinity} 
\label{sec:inhom}

\subsection{Construction of a basis}
\label{sec:constrBasis}
We want to construct the generic solution to the following field equations on $dS_3$: 
\begin{equation}
    (\Box-1)^2\Phi_a = 0,\qquad \mathcal D^a\Phi_a=0.\label{GenericPhiaEq}
\end{equation}
This is done by first solving the scalar equation  \begin{equation}
    \Box^2 \Phi =0. 
\end{equation}
It admits the following generic solution: 
\begin{align}
  \Phi &= \sum_{\ell\ge1,m} \bigg[ \bar A_{\ell m } \psi^{(p)}_{0\ell m}(\tau,x^A)  
   +\bar B_{\ell m }\psi^{(q)}_{0\ell m}(\tau,x^A)+\bar C_{\ell m } \phi^{(p)}_{0\ell m}(\tau,x^A)  
   +\bar D_{\ell m }\phi^{(q)}_{0\ell m}(\tau,x^A)\bigg] \label{ScalBox2Sol},
\end{align}
where $\psi^{(p)}_{0\ell m}(\tau,x^A)$ and $\phi^{(q)}_{0\ell m}(\tau,x^A)$ obey the homogeneous equations $\Box\psi^{(p)}_{0\ell m}=\Box\psi^{(q)}_{0\ell m}=0$ (see \cite{Compere:2025bnf} for an explicit construction) and $\phi^{(p)}_{0\ell m}(\tau,x^A)$ and $\phi^{(q)}_{0\ell m}(\tau,x^A)$ obey
\begin{equation}
    \Box\phi^{(p)}_{0\ell m}=\psi^{(p)}_{0\ell m},\qquad\Box\phi^{(q)}_{0\ell m}=\psi^{(q)}_{0\ell m}.
\end{equation}
Explicitly, $\phi^{(p)}_{0\ell m}$ and $\phi^{(q)}_{0\ell m}$ are given by 
\begin{align}
    \phi^{(p)}_{0\ell m}(\tau,x^A) &=\phi^{(p)}_{0\ell }(\tau)Y_{\ell m}(x^A), \qquad \phi^{(p)}_{0\ell }(\tau):= \frac{1}{2\ell(\ell+1)}P_{\ell-1}(\tanh\tau),\\
    \phi^{(q)}_{0\ell m}(\tau,x^A) &= \phi^{(q)}_{0\ell }(\tau)Y_{\ell m}(x^A), \qquad \phi^{(q)}_{0\ell }(\tau):=\frac{1}{2}Q_{\ell-1}(\tanh\tau),
\end{align}
and satisfy the antipodal matching conditions 
\begin{equation}
\Upsilon^*_{\mathcal{H}}\phi^{(p)}_{0\ell m}(\tau,x^A)=-\phi^{(p)}_{0\ell m}(\tau,x^A),\qquad \Upsilon^*_{\mathcal{H}}\phi^{(q)}_{0\ell m}(\tau,x^A)=\phi^{(q)}_{0\ell m}(\tau,x^A).
\end{equation}
Asymptotically, these functions behave as 
\begin{align}
     &\phi^{(p)}_{0\ell m}(\tau\to\pm\infty,x^A) = (\pm1)^{\ell+1}\frac{1}{2\ell(\ell+1)} Y_{\ell m}-(\pm1)^{\ell+1}\frac{\ell(\ell-1)}{2\ell(\ell+1)}Y_{\ell m} e^{-2\vert\tau\vert}+o(e^{-2\vert\tau\vert}),\\& \phi^{(q)}_{0\ell m}(\tau\to\pm\infty,x^A) = (\pm1)^{\ell}\vert\tau\vert\frac{1}{2} Y_{\ell m}-(\pm1)^{\ell}\frac{1}{2} H_{\ell-1}Y_{\ell m}\nn\\&\qquad\qquad\qquad\qquad\qquad+(\pm1)^{\ell}(-\vert\tau\vert+H_{\ell-1} -1)\frac{1}{2}\ell(\ell-1)Y_{\ell m}e^{-2\vert\tau\vert}+o(e^{-2\vert\tau\vert}).
\end{align}

Given functions $f,\tilde f$ on $\mathbb R$, we define the $dS_3$ vector transverse electric parity (E) harmonics for $\ell \geq 1$ as 
\begin{align}
   V_a^{(E)\:\ell m} (f)=  \left(
\begin{array}{c}
\sqrt{\ell(\ell+1)} \sech\tau f(\tau)Y_{\ell m}(\theta,\phi) \\\\
-\cosh\tau \left(\partial_\tau + \tanh\tau\right)f(\tau)\mathring{V}^{(E)\ell m}_\theta(\theta, \phi)\\\\
-\cosh\tau\left(\partial_\tau + \tanh\tau\right)f(\tau)\mathring{V}^{(E)\ell m}_\phi(\theta, \phi) \\
\end{array}
\right),\label{VE}
\end{align} 
and the transverse magnetic parity (B) harmonics for $\ell \geq 1$ as
\begin{align}
   V_a^{(B)\:\ell m}(\tilde{f})=  \left(
\begin{array}{c}
0 \\\\
\cosh\tau\tilde{f}(\tau) \mathring{V}^{(B)\ell m}_\theta(\theta, \phi)\\\\
\cosh\tau\tilde{f}(\tau)\mathring{V}^{(B)\ell m}_\phi(\theta, \phi) \\
\end{array}
\right).\label{VB}
\end{align}
A generic solution to  Eq. \eqref{GenericPhiaEq} can then be decomposed as 
\begin{align}
    \Phi_a =  \sum_{\ell \ge 1,m} \bigg(&a_{\ell m}^{(E,p)}  \Psi^{(E,p)\ell m}_a +a_{\ell m}^{(B,p)}  \Psi^{(B,p)\ell m}_a
    +b_{\ell m}^{(E,q)}  \Psi^{(E,q)\ell m}_a +b_{\ell m}^{(B,q)}  \Psi^{(B,q)\ell m}_a \nn\\&+c_{\ell m}^{(E,p)}  \Phi^{(E,p)\ell m}_a +c_{\ell m}^{(B,p)}  \Phi^{(B,p)\ell m}_a
    +d_{\ell m}^{(E,q)}  \Phi^{(E,q)\ell m}_a +d_{\ell m}^{(B,q)}  \Phi^{(B,q)\ell m} _a\bigg),\label{expVT}
\end{align}
where 
\begin{align}
    \Psi_a^{(E,p)\;0\ell m} & := V_a^{(E)\:\ell m}(f=\psi_{0\ell}^{(p)}),\qquad 
    \Psi_a^{(B,p)\;0\ell m}  := V_a^{(B)\:\ell m}(\tilde{f}=\psi_{0\ell}^{(p)}),\\\Phi_a^{(E,p)\;0\ell m} & := V_a^{(E)\:\ell m}(f=\phi_{0\ell}^{(p)}),\qquad 
    \Phi_a^{(B,p)\;0\ell m}  := V_a^{(B)\:\ell m}(\tilde{f}=\phi_{0\ell}^{(p)}),
\end{align}
with $\psi_{0\ell}^{(p)}:=\psi_{0\ell m}^{(p)}/Y_{\ell m}$  and
\begin{align}
    \Psi_a^{(E,q)\;0\ell m} & :=- V_a^{(E)\:\ell m}(f=\psi_{0\ell}^{(q)}),\qquad  
    \Psi_a^{(B,q)\;0\ell m}  := V_a^{(B)\:\ell m}(\tilde{f}=\psi_{0\ell}^{(q)}),\\\Phi_a^{(E,q)\;0\ell m} & :=- V_a^{(E)\:\ell m}(f=\phi_{0\ell}^{(q)}),\qquad 
    \Phi_a^{(B,q)\;0\ell m} := V_a^{(B)\:\ell m}(\tilde{f}=\phi_{0\ell}^{(q)}),
\end{align}
with $\psi_{0\ell}^{(q)}:=\psi_{0\ell m}^{(q)}/Y_{\ell m}$.

\subsection{Derivation of the local antipodal matching from a vector inner product}
\label{sec:vectorOmega4Matching}
Let us define $L:=\Box-1$. Let $V_a$ be a transverse vector field on $dS_3$ satisfying
\begin{equation}
    LV_a=S_a,
    \qquad
    LS_a=0.
    \label{eq:vectorSourcedEq}
\end{equation}
It follows immediately that $L^2V_a=0$. Therefore, $V_a$ and $LV_a$ form a Jordan chain of order 2 since $L^2$ acts nilpotently on the span of $V_a$ and $LV_a$. 

For two transverse vector fields $X_a$ and $Y_a$, we recall the
Klein-Gordon product defined in Eq. \eqref{Vacharges}:
\begin{equation}
    \Omega^V[X,Y]
    :=
    \int_{S^2(\tau)}\frac{d^2\Omega}{8\pi }\,\cosh^2\tau\,
    n^b\left(
        X^a\mathcal D_b Y_a
        -
        Y^a\mathcal D_b X_a
    \right),
    \label{eq:vectorOmega2}
\end{equation}
and the second Klein-Gordon product defined in Eq. \eqref{Vachargesv2}: 
\begin{equation}
    \Omega^{V \,(2)}[X,Y]
    :=
    \Omega^V[X,LY]
    +
    \Omega^V[LX,Y],
    \label{eq:vectorOmega4}
\end{equation}
which satisfies 
\begin{equation}
    \partial_\tau\Omega^{V \,(2)}[X,Y] = - \int_{S^2(\tau)}\frac{d^2\Omega}{8\pi}\,\cosh^2\tau\,
    \left(
        X^aL^2 Y_a
        -
        Y^aL^2 X_a
    \right).\label{PartialtauOmega}
\end{equation}
For fields satisfying $L^2X=L^2Y=0$, this product is conserved, $    \partial_\tau\Omega^{V \,(2)}[X,Y]=0$.

We now choose a generalized $p$-parity transverse vector field $\Phi^{(p)}_a$ satisfying (see section \ref{sec:constrBasis})
\begin{equation}
    L^2\Phi^{(p)}_a=0,\qquad \Upsilon_{\mathcal H}^* \Phi^{(p)}_a = \Phi^{(p)}_a
\end{equation}
and define
\begin{equation}
    \Psi^{(p)}_a:=L\Phi^{(p)}_a,
    \qquad
    L\Psi^{(p)}_a=0.
    \label{eq:PsiFromPhi}
\end{equation}
The conserved quantity of interest is therefore
\begin{equation}
    \Omega^{V \,(2)}[\Phi^{(p)},V]
    =
    \Omega^V[\Phi^{(p)},S]
    +
    \Omega^V[\Psi^{(p)},V].
    \label{eq:Omega4Decomposition}
\end{equation}

Let us assume that the solution $V_a$ behaves as
\begin{equation}
    V_A(\tau\to\pm\infty,x^A)
    =
    V_A^{(\mathbb L)\pm}(x^A)e^{|\tau|}
    +
    O\!\left(
        |\tau|e^{-|\tau|}
    \right).
    \label{eq:VGenericLeading}
\end{equation}
The divergence-free condition gives
\begin{equation}
    V_\tau(\tau\to\pm\infty,x^A)
    =
    \pm4
    \nabla^AV_A^{(\mathbb L)\pm}
    e^{-|\tau|}
    +
    O\!\left(
        |\tau|e^{-3|\tau|}
    \right).
    \label{eq:VtauGenericLeading}
\end{equation}
Correspondingly, the homogeneous source $S_a$ is taken to behave as
\begin{align}
    S_A(\tau\to\pm\infty,x^A)
    &=
    4S_A^{(0)\pm}(x^A)e^{-|\tau|}
    +
    o(e^{-|\tau|}),
    \label{eq:IGenericA}
    \\
    S_\tau(\tau\to\pm\infty,x^A)
    &=
    \mp16
    \nabla^AS_A^{(0)\pm}(x^A)e^{-3|\tau|}
    +
    o(e^{-3|\tau|}).
    \label{eq:IGenericTau}
\end{align}

For the generalized $p$-parity transverse vector $\Phi_a^{(p)}$, we define $Y_A^\pm$ by 
\begin{equation}
    \Phi_A^{(p)}(\tau\to\pm\infty,x^A)
    =
   -\frac{1}{8} Y_A^\pm(x^A)e^{|\tau|}
    +
    O\!\left(
        |\tau|e^{-|\tau|}
    \right).
    \label{eq:PhiGenericLeading}
\end{equation}
Its divergence-free condition then implies
\begin{equation}
    \Phi^{(p)}_\tau(\tau\to\pm\infty,x^A)
    =
    \mp\frac{1}{2} 
    \nabla^AY_A^\pm
    e^{-|\tau|}
    +
    O\!\left(
        |\tau|e^{-3|\tau|}
    \right).
    \label{eq:PhiTauGeneric}
\end{equation}
Since $\Phi_a^{(p)}$ is antipodally even, its leading boundary data obey
\begin{equation}
    Y_A^+=\Upsilon^*Y_A^-.
\end{equation}
The homogeneous field $\Psi^{(p)}_a=L\Phi^{(p)}_a$ has the asymptotic behavior
\begin{align}
    \Psi^{(p)}_A(\tau\to\pm\infty,x^A)
    &=
    +\frac{1}{2}
    \left[
        2\nabla_A\nabla^BY_B^\pm
        -
        (\nabla^2-1)Y_A^\pm
    \right]
    e^{-|\tau|}
    +
    o(e^{-|\tau|}),
     \\ 
    &= -\frac{1}{2}  \,  K[Y_A^\pm] 
        e^{-|\tau|}
    +
    o(e^{-|\tau|}),\label{eq:PsiGenericLeading}
\end{align}
where we recognized the operator $K[]$ acting on vectors over the sphere defined in Eq. \eqref{KOperator}. 
The $\tau$-component of $\Psi^{(p)}$ is $O(e^{-3|\tau|})$ and will not contribute to the
finite part of the charge.

Evaluating $\Omega^{V \,(2)}[\Phi^{(p)},V]$ at $\tau\to\pm\infty$, we obtain : 
\begin{equation}
    \Omega^{V \,(2)}[\Phi^{(p)},V]\vert_{\tau\to+\infty}
    =
    \frac{1}{8\pi}
    \int_{S^2}d^2\Omega\,
    Y^{+A}
    \Big[S_A^{(0)+}-
       K[V_A^{(\mathbb L)+}]
    \Big].
    \label{eq:Omega4PlusGeneric}
\end{equation}
and 
\begin{equation}
    \Omega^{V \,(2)}[\Phi^{(p)},V]\vert_{\tau\to-\infty}
    =
   - \frac{1}{8\pi}
    \int_{S^2}d^2\Omega\,
    Y^{-A}
    \Big[S_A^{(0)-}-
        K[V_A^{(\mathbb L)-}]
    \Big].
    \label{eq:Omega4MinusGeneric}
\end{equation}
As $ \Omega^{V \,(2)}[\Phi^{(p)},V]$ is asymptotically conserved, we have  $\Omega^{V \,(2)}[\Phi^{(p)},V]\vert_{\tau\to-\infty}= \Omega^{V \,(2)}[\Phi^{(p)},V]\vert_{\tau\to+\infty}$, which implies :
\begin{equation}
   S_A^{(0)+}-K[V_A^{(\mathbb L)+}]
    =
    -
    \Upsilon^*
    \left(
        S_A^{(0)-}-K[V_A^{(\mathbb L)-}]
    \right).
    \label{eq:GenericLocalVectorMatching}
\end{equation}

\paragraph{Specialization to the gravitational data.}
We now use the specific asymptotics
\begin{equation}
    V_A^{(\mathbb L)\pm}
    =
    \pm\frac12
    \nabla^B\mathcal C_{AB}^{\pm(1)}.
    \label{eq:VLPhysical}
\end{equation}
The source behaves as
\begin{align}
    S_A(\tau\to\pm\infty,x^A)
    =
    \pm4
    \Big(&\mathcal{D}_A^{\pm(0)}\mp 6 P_i (m^{\pm (0)}\nabla_A n_i +2\tilde{\mathcal{M}}^{\pm(0)}\epsilon_{AB}\nabla^B n_i)\nn\\&+ 2 (E\mp P_i n_i) (\nabla_A m^{\pm (0)}+2\epsilon_{AB}\nabla^B\tilde{\mathcal{M}}^{\pm(0)})\Big)
    e^{-|\tau|}
    +
    o(e^{-|\tau|}),
\end{align}
which, compared with Eq.~\eqref{eq:IGenericA}, gives
\begin{equation}
    S_A^{(0)\pm}
    =
    \pm
    \Big(\mathcal{D}_A^{\pm(0)}\mp 6 P_i (m^{\pm (0)}\nabla_A n_i +2\tilde{\mathcal{M}}^{\pm(0)}\epsilon_{AB}\nabla^B n_i)+ 2 (E\mp P_i n_i) (\nabla_A m^{\pm (0)}+2\epsilon_{AB}\nabla^B\tilde{\mathcal{M}}^{\pm(0)})\Big).
    \label{eq:I0Physical}
\end{equation}

Using Eq.~\eqref{eq:VLPhysical}, we find
\begin{align}
      2\nabla_A\nabla^BV_B^{(\mathbb L)\pm}
    -
    (\nabla^2-1)V_A^{(\mathbb L)\pm}
    &=
    \pm
    \left[
        \nabla_A\nabla^B\nabla^C
        \mathcal C_{BC}^{\pm(1)}
        -\frac12
        (\nabla^2-1)
        \nabla^B\mathcal C_{AB}^{\pm(1)}
    \right] \nonumber\\
    &  = \pm
    \nabla_C\nabla_{\langle A}\nabla_{B\rangle}
    \mathcal C^{\pm(1)BC}.
    \label{eq:LocalOperatorOnVL}
\end{align}
Hence Eq.~\eqref{eq:GenericLocalVectorMatching} reduces to
\begin{equation}
    \Omega^{V \,(2)}[\Phi^{(p)},V]\vert_{\tau\to+\infty}
    =
    \frac{1}{8\pi}
    \int_{S^2}d^2\Omega\,
    Y^{+A}
    \Big[\mathcal N_A^{\text{(log)}+ (0)}+F_{A}^+
    \Big],
\label{eq:Omega4PlusGenericbis}
\end{equation}
and 
\begin{equation}
    \Omega^{V \,(2)}[\Phi^{(p)},V]\vert_{\tau\to-\infty}
    =
    \frac{1}{8\pi}
    \int_{S^2}d^2\Omega\,
    Y^{-A}
    \Big[\mathcal N_A^{\text{(log)}- (0)}+F_{A}^-
    \Big],
    \label{eq:Omega4MinusGenericbis}
\end{equation}
where $\mathcal  N_A^{\text{(log)}\pm (0)}$ is defined Eq. \eqref{defNAlog} and $F^\pm_A$ is defined as 
\begin{equation}
    F^\pm_A=\mp 6 P_i (m^{\pm (0)}\nabla_A n_i +2\tilde{\mathcal{M}}^{\pm(0)}\epsilon_{AB}\nabla^B n_i)+ 2 (E\mp P_i n_i) (\nabla_A m^{\pm (0)}+2\epsilon_{AB}\nabla^B\tilde{\mathcal{M}}^{\pm(0)}).\label{defFA}
\end{equation}
In the cases where the charge is asymptotically conserved (as in the main text), Eq.~\eqref{eq:GenericLocalVectorMatching} becomes  
\begin{equation}
    \mathcal  N_A^{\text{(log)}+ (0)}+F_{A}^+
    =
    \Upsilon^*
    \left[
        \mathcal  N_A^{\text{(log)}- (0)}+F_{A}^-
    \right].
    \label{eq:CalDCalCMatchingFromOmega4}
\end{equation}


\section{Quadratic source as stress-energy tensor of a boundary action}
\label{app:boundaryaction}

In this appendix we prove that the quadratic source \eqref{SourceSab} can be rewritten as a local operator acting on the stress-energy tensor of a quadratic boundary action on $dS_3$ of the boundary fields $\sigma^\text{inv}$ and $k^{(B)}_{ab}$. The proof is based on the formalism and observations set in \cite{Compere:2011ve} which we complete here in order to obtain a complete proof including the mixed $\sigma-k$ sector. In this appendix only we shall denote $\sigma^\text{inv}$ simply as $\sigma$ and $k^{(B)}_{ab}$ simply as $k_{ab}$ in order to avoid notational clutter. 

We consider the quadratic boundary action defined on an arbitrary (off-shell) metric $q_{ab}$: 
\begin{equation}
  I_{\partial}[\sigma, k_{ab} ; q_{ab}]
  :=I_{\sigma\sigma}[\sigma ; q_{ab}]+I_{\sigma k}[\sigma,k_{ab} ;q_{ab}]
    +I_{kk}[k_{ab} ;q_{ab}],
  \label{eq:complete-boundary-action-short}
\end{equation}
where
\begin{subequations}\label{eq:three-actions-short}
\begin{align}
 I_{\sigma\sigma}
 &=\int_{dS_3}d^3\phi\,\sqrt{-q}\,
   \left(-\frac12\mathcal D_a\sigma\mathcal D^a\sigma
         +\frac32\sigma^2\right),
 \label{eq:Iss-short}\\
 I_{\sigma k}
 &=-\frac12\int_{dS_3}d^3\phi\,\sqrt{-q}\,
   \mathcal G_{ab}[q] \,\sigma k^{ab},
 \label{eq:Isk-short}\\
 I_{kk}
 &=I_k^{\rm DN}
   -\frac1{16}\int_{dS_3}d^3\phi\,\sqrt{-q}\,
   \mathcal G_{ab}[q] k^a{}_c k^{bc}.
 \label{eq:Ikk-short}
\end{align}
\end{subequations}
Here
\begin{equation}
 \mathcal G_{ab}[q]
 :=\mathcal R_{ab}-\frac12\mathcal R q_{ab}+q_{ab},
 \label{eq:cosmological-Einstein-short}
\end{equation}
is the boundary vacuum Einstein equation with positive cosmological constant $1$ and the massive spin-two action (already considered by Deser and Nepomechie \cite{Deser:1983mm} and Higuchi \cite{Higuchi:1989gz}) is
\begin{equation}
\begin{aligned}
 I_k^{\rm DN}=\frac14\int_{dS_3}d^3\phi\,\sqrt{-q}\,\Big[&
 -\frac14\mathcal D_ak_{bc}\mathcal D^ak^{bc}
 +\frac12\mathcal D_ak_{bc}\mathcal D^bk^{ac}
 +\frac14\mathcal D_ak\mathcal D^ak
 -\frac12\mathcal D_ak\mathcal D_bk^{ab}
 \\
 &+k_{ab}k^{ab}-\frac12k^2
 -\frac{M^2_k}{4}\big(k_{ab}k^{ab}-k^2\big)\Big],
\end{aligned}
\label{eq:IPM-short}
\end{equation}
where $k:=q^{ab}k_{ab}$ and the mass is set to $M^2_k := 1$. 
The terms proportional to $\mathcal G_{ab}$ identically vanish once $q_{ab}$ is set to be the unit metric on the hyperboloid, but their metric variations are nonzero and give rise to a non-vanishing stress-energy tensor. We define
\begin{equation}
 T^{\partial}_{ab}
 :=-\frac{2}{\sqrt{-q}}\frac{\delta I_{\partial}}{\delta q^{ab}}
   \bigg|_{dS_3,\,\mathrm{EOM}}
 =T^{(\sigma\sigma)}_{ab}+T^{(\sigma k)}_{ab}+T^{(kk)}_{ab}.
 \label{eq:Tboundary-short}
\end{equation}
where the stress-energy tensor is evaluated on-shell, for $q_{ab}$ the unit metric on the hyperboloid and using the equations of motion of the boundary fields: $(\Box+3)\sigma=0$, $k=0$, $\mathcal D^bk_{ab}=0$ and
$(\Box-3)k_{ab}=0$. The three contributions are
\begin{subequations}\label{eq:three-stresses-short}
\begin{align}
 T^{(\sigma\sigma)}_{ab}
 &=\mathcal D_a\sigma\mathcal D_b\sigma
   +q_{ab}\left(\frac32\sigma^2
   -\frac12\mathcal D_c\sigma\mathcal D^c\sigma\right),
 \label{eq:Tss-short}\\
 T^{(\sigma k)}_{ab}
 &=-\kappa^{[\sigma,k,I]}_{ab},
 \label{eq:Tsk-short}\\
 T^{(kk)}_{ab}
 &=-\frac14\left(Y^{(2)}_{ab}+\kappa^{[k,k,I]}_{ab}\right).
 \label{eq:Tkk-short}
\end{align}
\end{subequations}
The tensors entering these expressions in the notation of \cite{Compere:2011ve} are
\begin{align}
 \kappa^{[\sigma,k,I]}_{ab}
 ={}&\sigma k_{ab}
 +\mathcal D_c\mathcal D_{(a}\sigma\,k_{b)}{}^c
 -\frac12q_{ab}\mathcal D_c\mathcal D_d\sigma\,k^{cd}
 \nonumber\\[-1mm]
 &-\mathcal D^c\sigma\mathcal D_ck_{ab}
 +\mathcal D^c\sigma\mathcal D_{(a}k_{b)c},
 \label{eq:kappask-short}\\
 \kappa^{[k,k,I]}_{ab}
 ={}&\frac34 k_{cd}k^{cd}q_{ab}-k_{ac}k_b{}^c
 +\frac14\mathcal D_ck_{de}\mathcal D^ck^{de}q_{ab}
 -\frac12\mathcal D_ak_{cd}\mathcal D_bk^{cd},
 \label{eq:kappakk-short}\\
 B^{(1)}_{ab}:={}&\frac12\epsilon_a{}^{cd}\mathcal D_ck_{db},
 \label{eq:B1-short}\\
 Y^{(2)}_{ab}={}&-4B^{(1)}_{c(a}B^{(1)c}{}_{b)}
 -2\epsilon_{cd(a}\mathcal D^ck_{b)}{}^eB^{(1)d}{}_e
 -2\epsilon_{cd(a}\mathcal D_{b)}B^{(1)c}{}_e k^{de}.
 \label{eq:Y2-short}
\end{align}
It is useful to shift the invariant second-order tensor $J_{ab}$ defined in the main text in Eq. \eqref{defJab} by the local SDT
quadratic tensor $Y^{(2)}_{ab}$,
\begin{equation}
  \widetilde J_{ab}:=J_{ab}-Y^{(2)}_{ab}.
  \label{eq:Jtilde-short}
\end{equation}

With this definition, the shifted tensor obeys the compact equation
\begin{equation}
 (\Box-2)\widetilde J_{ab}
 =2I_{ab}+4\big(\mathscr C^2T^{\partial}\big)_{ab}.
 \label{eq:Jtilde-source-short}
\end{equation}
where the symmetrized curl of a symmetric tensor  $X_{ab}$ is denoted as 
\begin{equation}
  (\mathscr C X)_{ab}:=\epsilon_{cd(a}\mathcal D^cX_{b)}{}^d.
  \label{eq:symcurl-short}
\end{equation}
The first term on the right-hand side of Eq. \eqref{eq:Jtilde-source-short} is the $p$-parity source while the second term is the $q$-parity source.





\providecommand{\href}[2]{#2}\begingroup\raggedright\endgroup

\end{document}